\documentclass[11pt]{article}

\usepackage{amssymb,amsmath,amsfonts,amssymb}
\usepackage{graphics,graphicx,color}
\hsize \textwidth
\advance \hsize by -\marginparwidth
\evensidemargin \oddsidemargin
\advance\hoffset by 5mm

\makeatletter \@addtoreset{equation}{section}

\@addtoreset{equation}{section}

\makeatother

\renewcommand\thefigure{\thesection.\@arabic\c@figure}
\renewcommand\thetable{\thesection.\@arabic\c@table}

\def\@abssec#1{\vspace{.05in}\footnotesize \parindent .2in
{\bf #1. }\ignorespaces}
\def\proof{\par{\it Proof}. \ignorespaces}
\def\endproof{{\ \vbox{\hrule\hbox{%
   \vrule height1.3ex\hskip0.8ex\vrule}\hrule
  }}\par}

\newtheorem{theorem}{Theorem}[section]
\newtheorem{lemma}[theorem]{Lemma}
\newtheorem{proposition}[theorem]{Proposition}

\newtheorem{remark}[theorem]{Remark}

\def \Rm {\mathbb R}

\newcommand{\eps}{\varepsilon}

\newcommand{\disp}{\displaystyle}
\newcommand{\pdr}[2]{\dfrac{\partial{#1}}{\partial{#2}}}

\newcommand{\pdrt}[3]{\dfrac{\partial^2{#1}}{\partial{#2}{\partial{#3}}}}

\newcommand{\bk}{\mathbf k} 
\newcommand{\bx}{\mathbf x} 
\newcommand{\vu}{\mathbf u}\newcommand{\vy}{\mathbf y}
\newcommand{\vz}{\mathbf z}
\newcommand{\bz}{\mathbf z} \newcommand{\vx}{\mathbf x}

\newcommand{\obm}{\mathbf m}

\newcommand{\vc}{{\bf c}}
\newcommand{\bp}{\mathbf p} 
\newcommand{\bu}{\mathbf u} \newcommand{\bv}{\mathbf v}

\newcommand{\vb}{{\bf b}}

\newcommand{\vhatk}{\hat{\bk}}

\newcommand{\commentout}[1]{}
\newcommand{\bes}{\begin{displaymath}}
\newcommand{\ees}{\end{displaymath}}
\newcommand{\ba}{\begin{eqnarray}}
\newcommand{\ea}{\end{eqnarray}}
\newcommand{\bas}{\begin{eqnarray*}}
\newcommand{\eas}{\end{eqnarray*}}

\newcommand{\bl}{{\bml}}

\newcommand{\bze}{{\bf 0}}

\newcommand{\bt}{\beta}

\newcommand{\bby}{{\bf y}}
\newcommand{\by}{{\bmy}}

\newcommand{\bbl}{{\bf l}}

\newcommand{\R}{{\mathbb R}}
\newcommand{\bbP}{{\mathbb P}}
\newcommand{\E}{{\mathbb E}}

\newcommand{\bbk}{{\bf k}}

\newcommand{\vphi}{\varphi}

\newcommand{\si}{\sigma}
\newcommand{\ga}{\gamma}
\newcommand{\al}{\alpha}

\newcommand{\la}{\lambda}

\newcommand{\bbx}{{\bf x}}
\newcommand{\ep}{\epsilon}

\newcommand{\bE}{\mathbb{E}}

\newcommand{\vv}{{\bf v}}

\newcommand{\Om}{\Omega}
\newcommand{\om}{\omega}

\newcommand{\bbR}{\mathbb R^d}

\newcommand{\nwc}{\newcommand}

\nwc{\beq}{\begin{eqnarray}} \nwc{\beqn}{\begin{eqnarray*}}
\nwc{\beqast}{\begin{eqnarray*}} \nwc{\bm}{\boldmath}
\nwc{\eeq}{\end{eqnarray}} \nwc{\eeqn}{\end{eqnarray*}}
\nwc{\eeqast}{\end{eqnarray*}}

\nwc{\veps}{\varepsilon} \nwc{\ie}{\mbox{e}} \nwc{\ibi}{\mbox{i}}

\nwc{\m}{\mbox} \nwc{\re}{\hbox{Re}}
\nwc{\lamb}{\lambda_\varepsilon} \nwc{\ls}{\stackrel{<}{\sim}}
\nwc{\gs}{\stackrel{>}{\sim}} \nwc{\ubm}{\unboldmath}
\nwc{\cls}{{\cal L}^s} \nwc{\mt}{\bar{t}} \nwc{\bla}{\m{\bm
$\lambda$\ubm}} \nwc{\bxsi}{\m{\bm $\xi$\ubm}} \nwc{\bpsi}{\m{\bm
$\psi$\ubm}} \nwc{\bmeta}{\m{\bm $\eta$\ubm}} \nwc{\bma}{\m{\bm
$a$\ubm}} \nwc{\bmb}{\m{\bm $b$\ubm}} \nwc{\bmc}{\m{\bm $c$\ubm}}
\nwc{\bmd}{\m{\bm $d$\ubm}} \nwc{\bme}{\m{\bm $e$\ubm}}\nwc{\bmL}{\m{\bm $L$\ubm}}
\nwc{\bmf}{\m{\bm $f$\ubm}} \nwc{\bmg}{\m{\bm $g$\ubm}}
\nwc{\bmh}{\m{\bm $h$\ubm}} \nwc{\bmi}{\m{\bm $i$\ubm}}
\nwc{\bmj}{\m{\bm $j$\ubm}} \nwc{\bmk}{\m{\bm $k$\ubm}}
\nwc{\bml}{\m{\bm $l$\ubm}} \nwc{\bmn}{\m{\bm $n$\ubm}}
\nwc{\bmo}{\m{\bm $o$\ubm}} \nwc{\bmp}{\m{\bm $p$\ubm}}
\nwc{\bmq}{\m{\bm $q$\ubm}} \nwc{\bmr}{\m{\bm $r$\ubm}}
\nwc{\bms}{\m{\bm $s$\ubm}} \nwc{\bmt}{\m{\bm $t$\ubm}}
\nwc{\bmu}{\m{\bm $u$\ubm}} \nwc{\bmv}{\m{\bm $v$\ubm}}
\nwc{\bmw}{\m{\bm $w$\ubm}} \nwc{\bmx}{\m{\bm $x$\ubm}}
\nwc{\bmxt}{\m{\bm $x$\ubm}^\varepsilon (t)} \nwc{\bmy}{\m{\bm
$y$\ubm}} \nwc{\bmz}{\m{\bm $z$\ubm}}

\nwc{\bmX}{\m{\bm $X$\ubm}} \nwc{\bmR}{\m{\bm $R$\ubm}}
\nwc{\bmU}{\m{\bm $U$\ubm}} \nwc{\bmE}{\m{\bm $E$\ubm}}
\nwc{\bmF}{\m{\bm $F$\ubm}} \nwc{\bmH}{\m{\bm $H$\ubm}}
\nwc{\bmI}{\m{\bm $I$\ubm}} \nwc{\bmP}{\m{\bm $P$\ubm}}
\nwc{\bmM}{\m{\bm $M$\ubm}} \nwc{\bmJ}{\m{\bm $J$\ubm}}
\nwc{\bmA}{\m{\bm $A$\ubm}} \nwc{\bmD}{\m{\bm $D$\ubm}}
\nwc{\bS}{{\bf S}}
\nwc{\bmtheta}{\m{\bm $\theta$\ubm}} \nwc{\bmnu}{\m{\bm
$\nu$\ubm}} \nwc{\bmomega}{\m{\bm $\omega$\ubm}}
\nwc{\bmsigma}{\m{\bm $\sigma$\ubm}} \nwc{\bmnabla}{\m{\bm
$\nabla$\ubm}} \nwc{\bmLambda}{\m{\bm $\Lambda$\ubm}}
\nwc{\bmlambda}{\m{\bm $\lambda$\ubm}} \nwc{\bmtau}{\m{\bm
$\tau$\ubm}} \nwc{\bmPhi}{\m{\bm $\Phi$\ubm}} \nwc{\bmphi}{\m{\bm
$\phi$\ubm}} \nwc{\bmGamma}{\m{\bm $\Gamma$\ubm}}
\nwc{\bmgamma}{\m{\bm $\gamma$\ubm}}

\nwc{\ca}{{\cal A}}
\nwc{\ii}{{\mbox{i}}} \nwc{\cao}{{\cal A}^{-1}} \nwc{\cb}{{\cal
B}} \nwc{\cc}{{\cal C}} \nwc{\cd}{{\cal D}} \nwc{\bone}{{\bf 1}}
\nwc{\ce}{{\cal E}} \nwc{\cf}{{\cal F}} \nwc{\cg}{{\cal G}}
\nwc{\vV}{{\bf V}} \nwc{\ch}{{\cal H}} \nwc{\ci}{{\cal I}}
\nwc{\cj}{{\cal J}} \nwc{\ck}{{\cal K}} \nwc{\cl}{{\cal L}}
\nwc{\cle}{{\cal L}^\varepsilon} \nwc{\clu}{{\cal L}{\cal U}}
\nwc{\cm}{{\cal M}} \nwc{\cn}{{\cal N}} \nwc{\co}{{\cal O}}
\nwc{\cp}{{\cal P}} \nwc{\cpt}{{\cal P}^\varepsilon_t}
\nwc{\cq}{{\cal Q}} \nwc{\calr}{{\cal R}} \nwc{\cs}{{\cal S}}
\nwc{\ct}{{\cal T}} \nwc{\cu}{{\cal U}} \nwc{\cv}{{\cal V}}
\nwc{\cw}{{\cal W}} \nwc{\cy}{{\cal Y}} \nwc{\cz}{{\cal Z}}
\nwc{\ob}{{\cal B(\Om)}}
\nwc{\bbE}{\mathbb{E}}
\nwc{\bbZ}{\mathbb{Z}^d}

\nwc{\uP}{{\em \bf Proof: }} \nwc{\uT}{\underline{Theorem:}}

\renewcommand{\arraystretch}{1.5}

\title{Diffusion in a weakly random Hamiltonian flow}

\author{Tomasz Komorowski
\thanks{Institute of Mathematics,
UMCS, pl. Marii Curie Sk\l odowskiej 1, 20-031 Lublin,
Poland; e-mail: komorow@hektor.umcs.lublin.pl}
\and {Lenya Ryzhik}
\thanks{Department of Mathematics,
University of Chicago, Chicago, IL 60637, USA; e-mail:
{ryzhik@math.uchicago.edu}}}

\begin{document}

\maketitle

\begin{abstract}
We consider the motion of a particle governed by a weakly random
Hamiltonian flow. We identify temporal and spatial scales on which the
particle trajectory converges to a spatial Brownian motion. The
main technical issue in the proof is to obtain error estimates for
the convergence of the solution of the stochastic acceleration
problem to a momentum diffusion. We also apply our results to the
system of  random geometric acoustics equations and show that the
energy density of the acoustic waves undergoes a spatial
diffusion.
\end{abstract}


\renewcommand{\thefootnote}{\fnsymbol{footnote}}
\renewcommand{\thefootnote}{\arabic{footnote}}

\renewcommand{\arraystretch}{1.1}





\section{Introduction}
\label{sec:intro}

The long time, large distance behavior of a massive particle in a
weakly random time-independent potential field is described by the
momentum diffusion: the particle momentum undergoes the Brownian
motion on the energy sphere. This intuitive result has been first
proved in \cite{KP} in dimensions higher than two, and later extended
to two dimensions with the Poisson distribution of scatterers in
\cite{DGL}. On the other hand, the long time limit of a momentum
diffusion is the standard spatial Brownian motion.  Hence, a natural
question arises if it is possible to obtain such a Brownian motion
directly as the limiting description in the original problem of a
particle in a quenched random potential. This necessitates the control
of the particle behavior over times longer than those when the
momentum diffusion holds.

We consider a particle that moves in an isotropic weakly
random Hamiltonian flow with the Hamiltonian of the form
$H_\delta(\vx,\bk)=H_0(k)+\sqrt{\delta}H_1(\vx,k)$, $k=|\bk|$, and
$\vx,\bk\in\Rm^d$ with $d\ge3$:
\begin{eqnarray}\label{intro-1}
\frac{dX^\delta}{dt}=\nabla_\bk H_\delta,~~~
\frac{dK^\delta}{dt}=-\nabla_\vx H_\delta,~~X^\delta(0)=0,~~K^\delta(0)=\bk_0.
\end{eqnarray}
Here $H_0(k)$ is the background Hamiltonian and $H_1(\vx,k)$ is a
random perturbation.  As we have mentioned, it has been shown in
\cite{KP} that, when $\disp
H_\delta(\vx,\bk)=\frac{k^2}{2}+\sqrt{\delta} V(\vx)$, and under
certain mixing assumptions on the random potential $V(\vx)$, the
momentum process $K^\delta(t/\delta)$ converges to a diffusion process
$K(t)$ on the sphere $k=k_0$ and the rescaled spatial component
$\tilde X^\delta(t)=\delta X^\delta(t/\delta)$ converges to $
X(t)=\int_0^tK(s)ds.$ This is the momentum diffusion mentioned above.
Another special case,
\begin{equation}\label{intro-waves-hamilt}
H_\delta(\vx,\bk)=
(c_0+\sqrt{\delta}c_1(\vx))|\bk|,
\end{equation}
arises in the geometrical optics limit of wave propagation. Here $c_0$
is the background sound speed, and $c_1(\vx)$ is a random
perturbation.  This case has been considered in \cite{bakoryz}, where it has
been shown that, once again, $K^\delta(t/\delta)$ converges to a
diffusion process $K(t)$ on the sphere $\{k=k_0\}$ while $\tilde X^\delta(t)=\delta
X^\delta(t/\delta)$ converges to
$X(t)=c_0\int_0^t \hat K(s)ds$, $\hat K(t):=K(t)/|K(t)|.$

We show in this paper that this analysis may be pushed beyond the
time of the momentum diffusion, and that under certain assumptions
concerning the mixing properties of $H_1$ in the spatial variable
there exists $\alpha_0>0$ so that the process
$\delta^{1+\alpha}X^\delta(t/\delta^{1+2\alpha})$ converges to the
standard Brownian motion in $\Rm^d$ for all
$\alpha\in(0,\alpha_0)$. The main difficulty of the proof is to
obtain error estimates in the convergence of $K^\delta(\cdot)$ to
the momentum diffusion on time scales of the order $\delta^{-1}$.
The error estimates allow us to push the analysis to times much
longer than $\delta^{-1}$ where the momentum diffusion converges
to the standard Brownian motion.  The method of the proof is a
modification of the cut-off technique used in \cite{bakoryz} and
\cite{KP}.

A similar question arises in the semi-classical limit of the quantum
mechanics and high frequency wave propagation. The Wigner transform
\cite{GMMP}, or the phase space energy density of the solution of the
Schr\"odinger equation, is approximated in a weakly random medium by
the solution of a deterministic linear Boltzmann equation
\cite{Erdos-Yau}. This behavior is also conjectured for the acoustic
and other waves in a weakly random medium
\cite{RPK-WM}. As in the momentum diffusion model for a particle,
the long time limit of the Boltzmann equation is the spatial
diffusion equation. It has been recently shown in \cite{ESY} that,
indeed, one may push the analysis of \cite{Erdos-Yau} beyond the
times on which the Boltzmann equation holds and obtain the
diffusive behavior of the energy density of the solutions of the
Schr\"odinger equation in the weak coupling limit.

We also apply our results to the problem of multiple scattering of the
acoustic waves. Our approach is different from that of \cite{ESY}
mentioned above: we first consider the random geometrical optics
approximation of the wave phase space energy density. The rays in the
phase space satisfy the Hamiltonian equations (\ref{intro-1}) with the
Hamiltonian given by (\ref{intro-waves-hamilt}). Therefore, the
aforementioned convergence result of the solutions of (\ref{intro-1})
to the standard Brownian motion, combined with the error estimates on
the geometrical optics approximation of the Wigner distribution of the
solutions of the wave equation, allows us to establish rigorously the
diffusive behavior of the wave energy density. To the best of our knowledge,
this is the first result of such kind for classical waves.

This paper is organized as follows. Section \ref{prelim} contains
the main results on the convergence of solutions of
(\ref{intro-1}) to the Brownian motion as well as the assumptions
on the random medium. Sections \ref{sec:liouv} and \ref{sec:wd-sd}
contain the proof of our main results, Theorems \ref{main-thm},
\ref{thm-wave-space-main} and \ref{thm-liouv-diff}: the error
estimates on the passage to the momentum diffusion and the passage
from the momentum diffusion to the spatial diffusion,
respectively. Section \ref{sec:waves} discusses the application to
the wave equation. Finally, Appendix \ref{sec:append} contains the
proof of a technical result that is stated in Lemma \ref{lmA1}. We
note that all constants appearing throughout the paper do not
depend on $\delta\in(0,1]$ unless otherwise specified.

{\bf Acknowledgment.}
The research of TK was partially supported by KBN grant 2PO3A03123.
The work of LR was partially supported by an NSF grant DMS-0203537,
an ONR grant N00014-02-1-0089 and an Alfred P. Sloan Fellowship.

\section{The main result and preliminaries}
\label{prelim}

\subsection{The notation}

As we will avoid the singular point $\bk=0$, we denote
$\R^d_*:=\R^d\setminus\{\bze\}$ and $\R^{2d}_*:=\R^d\times\R^d_*$.
Also $\mathbb S_R^{d-1}(\bx)$ ($\mathbb B_R(\bx)$) shall stand for a
sphere (open ball) in $\R^d$ of radius $R>0$ centered at $\bx$.  We
shall drop writing either $\bx$, or $R$ in the notation of the sphere
(ball) in the particular cases when either $\bx=0$, or $R=1$.  For a
fixed $M>0$ we define the spherical shell $ A(M):=[\bk\in
\R^{d}_*:M^{-1}\le|\bk|\le M]$ in the $\bk$-space, and ${\cal
A}(M):=\R^d\times A(M)$ in the whole phase space. Given a vector
$\vv\in\R^d_*$ we denote by $\hat\vv:=\vv/|\vv|\in\mathbb S^{d-1}$ the
unit vector in the direction of $\vv$. For any set $A$ we shall denote by
$A^c$ its complement.

For any non-negative integers $p,q,r$, positive times $T>T_*\ge0$ and
a function $G:[T_*,T]\times \R^{2d}_*\to\R$ that has $p$, $q$ and $r$
derivatives in the respective variables we define
\begin{equation}\label{62701}
\|G\|^{[T_*,T]}_{p,q,r}:=\sum\sup\limits_{(t,\bx,\bk)\in[T_*,T]\times\R^{2d}}|
\partial_t^\al\partial_\bx^\beta\partial_\bk^\gamma G(t,\bx,\bk)|.
\end{equation}
The summation range covers all integers $0\le \al\le p$ and all
integer valued multi-indices $|\beta|\le q$ and $|\gamma|\le r$.  In
the special case when $T_*=0$, $T=+\infty$ we write
$\|G\|_{p,q,r}=\|G\|^{[0,+\infty)}_{p,q,r}$.  We denote by
$C^{p,q,r}_b([0,+\infty)\times \R^{2d}_*)$ the space of all functions
$G$ with $\|G\|_{p,q,r}<+\infty$.  We shall also consider spaces
of bounded and a suitable number of times continuously differentiable
functions $C^{p,q}_b( \R^{2d}_*)$ and $C^{p}_b(\R^{d}_*)$ with the
respective norms $\|\,\cdot\,\|_{p,q}$ and $\|\,\cdot\,\|_{p}$.

\subsection{The background Hamiltonian}

We assume that the background Hamiltonian $H_0(k)$ is isotropic, that
is, it depends only on $k=|\bk|$, and is uniform in space.  Moreover,
we assume that $H_0:[0,+\infty)\to\R$ is a strictly increasing function
satisfying $H_0(0)\ge 0$ and such that it is of $C^3$-class of
regularity in $(0,+\infty)$ with $H_0'(k)>0$ for all $k>0$, and let
\begin{equation}\label{100901}
    h^*(M):=\max\limits_{k\in[M^{-1},M]}(H_0'(k)+|H_0'\!'(k)|+|H_0'''(k)|),\quad
     h_*(M):=\min\limits_{k\in[M^{-1},M]}H_0'(k).
\end{equation}
Two examples of such Hamiltonians are the quantum Hamiltonian $H_0(k)=k^2/2$
and the acoustic wave Hamiltonian $H_0(k)=c_0k$.



\subsection{The random medium}\label{sec-rand-assump}

Let $(\Om,\Sigma,\mathbb P)$ be a probability space, and let $\bbE$
denote the expectation with respect to $\bbP$.  We denote by
$\|X\|_{L^p(\Om)}$ the $L^p$-norm of a given random variable
$X:\Om\to\R$, $p\in[1,+\infty]$. Let $
H_1:\R^d\times[0,+\infty)\times\Om\rightarrow\R$ be a random field
that is measurable and strictly stationary in the first variable. This
means that for any shift $\bx\in\R^d$, $k\in[0,+\infty)$,
and a collection of points $\bx_1,\ldots,\bx_n\in\R^d$ the laws of
$(H_1(\bx_1+\bx,k),\ldots,H_1(\bx_n+\bx,k))$ and
$(H_1(\bx_1,k),\ldots,H_1(\bx_n,k))$ are identical. In addition, we
assume that $\mathbb E H_1(\bx,k)=0$ for all $k\ge0$, $\vx\in\Rm^d$,
the realizations of $H_1(\bx,k)$ are $\bbP$--a.s. $C^2$-smooth in
$(\bx,k)\in \R^d\times(0,+\infty)$ and they satisfy
\begin{equation}\label{d-i}
D_{i,j}(M):=\max\limits_{|\al|=i}\,\mathop{\mbox{ess-sup}}
\limits_{(\bx,k,\om)\in \R^d\times[M^{-1},M]\times\Om}
|\partial_\bbx^\al\partial_k^j
H_1(\bbx,k;\om)|<+\infty,\quad i,j=0,1,2.
\end{equation}
We define
$\tilde D(M):=\sum_{0\le i+j\le 2} D_{i,j}(M)$.

We suppose further that the random field is strongly mixing in the
uniform sense.  More precisely, for any $R>0$ we let ${\cal C}_{R}^i$
and ${\cal C}_{R}^e$ be the $\si$--algebras generated by random
variables $H_1(\bbx,k)$ with $k\in[0,+\infty)$, $\bbx\in \mathbb B_R$
and $\bbx\in \mathbb B_R^c$ respectively. The uniform mixing
coefficient between the $\si$--algebras is
\[
\phi(\rho):=\sup[\,|\mathbb P(B)-\mathbb P(B|A)|:\,R>0,\,A\in
{\cal C}_{R}^i,\,B\in {\cal C}_{R+\rho}^e\,],
\]
for all $\rho>0$. We suppose that $\phi(\rho)$ decays faster than any
power: for each $p>0$
\begin{equation}\label{DR}
h_p:=\sup\limits_{\rho\ge0}\rho^p\phi(\rho)<+\infty.
\end{equation}
The two-point spatial correlation function of the random field $H_1$
is $ R(\bby,k):=\bbE[H_1(\bby,k)H_1(\bze,k)].  $
\commentout{
Note that stationarity of $H_1(\cdot,k)$ implies that $
\nabla_\by R(\bze,k)=\bze$. We also assume that
\begin{equation}\label{011104}
\nabla_\by R_1(\bze,k)=\bze.
\end{equation}
} 
Note  that (\ref{DR}) implies that for each $p>0$
\begin{equation}\label{53102-intro}
h_{p}(M):=\,\sum\limits_{i=0}^4\sum\limits_{|\al|=i}\sup\limits_{(\bby,k)
\in\R^d\times[M^{-1},M]}(1+|\bby|^2)^{p/2}
|\partial_\bby^\al
 R(\bby,k)|<+\infty,\quad
\,M>0.
\end{equation}
We also assume that the correlation function $R(\vy,l)$ is of the
$C^\infty$-class for a fixed $l>0$, is sufficiently smooth in $l$,
and that  for any fixed $l>0$
\begin{equation}\label{non-trivial}
\hbox{$\hat R(\bk,l)$ does not vanish identically on any hyperplane
$H_\bp=\{\bk:~(\bk\cdot\bp)=0\}$.}
\end{equation}
Here $\hat R(\bk,l)=\int R(\vx,l)\exp(-i\bk\cdot\vx)d\vx$ is the power spectrum
of $H_1$.

The above assumptions are satisfied, for example, if
$H_1(\bx,k)=c_1(\bx)h(k)$, where $c_1(\vx)$ is a stationary uniformly
mixing random field with a smooth correlation function, and $h(k)$ is
a smooth deterministic function.

\subsection{Certain path-spaces}

For fixed integers $d,m\ge1$ we let ${\cal
C}^{d,m}:=C([0,+\infty);\R^{d}\times\R^m_*)$: we shall omit the
subscripts in the notation of the path space if $m=d$.
We define $(X(t),K(t)):{\cal C}^{d,m}\rightarrow
\R^{d}\times\R^m_*$ as the canonical mapping
$(X(t;\pi),K(t;\pi)):=\pi(t)$, $\pi\in{\cal C}^{d,m}$ and also let
$\theta_s(\pi)(\cdot):=\pi(\cdot+s)$ be the standard shift
transformation.

For any $u\leq v$ denote by ${\cal M}^{v}_{u}$ the $\si$-algebra of
subsets of ${\cal C}$ generated by $(X(t),K(t))$, $t\in[u,v]$. We
write ${\cal M}^{v}:={\cal M}_{0}^{v}$ and ${\cal M}$ for the $\si$
algebra of Borel subsets of ${\cal C}$. It coincides with the smallest
$\si$--algebra that contains all ${\cal M}^{t}$, $t\ge0$.

Let $\delta_*(M):=H_0\left(M^{-1}\right)/(2\tilde D(M))$.
For a given $M>0$ and $\delta\in(0,\delta_*(M)]$ we let
 \begin{equation}\label{073104}
M_\delta:=
\max\left\{H_0^{-1}(H_0(M)+
2\sqrt{\delta}\tilde D(M)),\left[H_0^{-1}\left(H_0\left(\frac1M\right)-
2\sqrt{\delta}\tilde D(M)\right)
\right]^{-1}\right\}.
\end{equation}
For a particle that is governed by the Hamiltonian flow generated by
$H_\delta(\vx,\bk)$ we
have
$
M_\delta^{-1}\le |K(t)|\le M_\delta
$
for all $t$ provided that $K(0)\in A(M)$.  Accordingly, we define ${\cal C}(T,\delta)$
as the set of paths $\pi\in{\cal C}$ so that both $(2M_\delta)^{-1}\le |K(t)|\le
2M_\delta$, and
\begin{eqnarray*}
\left|X(t)-X(u)-
\int\limits_u^t H_0'(K(s))\hat K(s)ds\right|\le
\tilde D(2M_{\delta})\sqrt{\delta}(t-u),\,\mbox{ for all }
\,0\le u<t\le T.\,
\end{eqnarray*}
In the case when $\delta=1$, or $T=+\infty$ we shall write simply
${\cal C}(T)$, or ${\cal C}(\delta)$ respectively.

\commentout{
Given $s\geq\si>0$, $\pi\in {\cal C}$ we define the linear
approximation of the trajectory
\begin{equation}\label{72802}
   \bmL(\si,s;\pi):=X(\si)+(s-\si)H_0'(K(\si))\hat K(\si)
\end{equation}
and for any $v\in[0,1]$ let
\begin{equation}\label{72803}
\bmR(v,\si,s;\pi):=(1-v)\bmL(\si,s;\pi)+vX(s).
\end{equation}
The following simple fact can be verified by a direct
calculation, see Lemma 5.4 of \cite{bakoryz}.
\begin{proposition}
\label{lm1}
Suppose that $s\geq\si\ge0$ and $\pi\in {\cal C}(\delta)$. Then,
\[
|X(s)-\bmL(\si,s;\pi)|\leq \tilde D(2M_\delta)\sqrt{\delta}(s-\si)+
\int\limits_{\si}^s|H_0'(K(\rho))\hat{K}(\rho)-H_0'(K(\si))\hat{K}(\si)|d\rho.
\]
\end{proposition}
}

\subsection{The main results}
\label{sec2.4}
Let  the function $\phi_\delta(t,\vx,\bk)$ satisfy the Liouville equation
\begin{eqnarray}\label{liouv-intro-0}
&&\pdr{\phi^{\delta}}{t}+\nabla_\vx H_{\delta}\left({\vx},\bk\right)\cdot\nabla_\bk
\phi^{\delta}
-\nabla_\bk H_{\delta}\left({\vx},\bk\right)\cdot\nabla_\vx
\phi^{\delta}=0,\\
&&\phi^{\delta}(0,\vx,\bk)=\phi_0(\delta\vx,\bk).\nonumber
\end{eqnarray}
We assume that the initial data $\phi_0(\bx,\bk)$ is a compactly supported function
four times differentiable in $\bk$, twice differentiable  in $\bx$ whose support is
contained inside a spherical shell ${\cal A}(M)=\{(\bx,\bk): M^{-1}<|\bk|< M\}$ for some
positive $M>0$.

Let us define the diffusion matrix $D_{mn}$ by
\begin{equation}\label{diff-matrix1-main}
D_{mn}(\vhatk,l)=-\frac{1}{2}
\int_{-\infty}^\infty\frac{\partial^2R(H_0'(l)s\vhatk,l)}{\partial x_n\partial x_m}
ds=-\frac{1}{2H_0'(l)}\int_{-\infty}^\infty
\frac{\partial^2R(s\vhatk,l)}{\partial x_n\partial x_m}ds
,\quad\,m,n=1,\ldots,d.
\end{equation}
Then we have the following result.
\begin{theorem}\label{main-thm}
Let $\phi^\delta$ be the solution of $(\ref{liouv-intro-0})$ and let $\bar\phi$ satisfy
\begin{eqnarray}\label{eq-mainthm-1}
&&\pdr{\bar\phi}{t}=
\sum\limits_{m,n=1}^d\pdr{}{k_m}\left(D_{mn}(\vhatk,k)\pdr{\bar\phi}{k_n}\right)+
H_0'(k)\,\hat\bk\cdot \nabla_\bx \bar\phi\\
&&\bar\phi(0,\vx,\bk)=\phi_0(\vx,\bk).\nonumber
\end{eqnarray}
Suppose that $M\ge M_0>0$ and $T\ge T_0>0$. Then, there exist two
constants $C,\,\alpha_0>0$ such that for all $T\ge T_0$
\begin{equation}\label{decorrel-main1}
\sup\limits_{(t,\bx,\bk)\in [0,T]\times K}
\left|\E \phi^{\delta}\left(\frac{t}{\delta},\frac{\vx}{\delta},\bk\right)-
 \bar \phi(t,\vx,\bk)\right|\le
CT(1+\|\phi_0\|_{1,4})\delta^{\alpha_0}
\end{equation}
for all compact sets $K\subset{\cal A}(M)$.
\end{theorem}
\begin{remark}
\label{rm1112}
{\em We shall denote by $C$, $C_1,\ldots$, $\al_0$, $\al_1,\ldots$,
$\ga_0$, $\ga_1,\ldots$ throughout this article generic positive
constants.  Unless specified otherwise the constants denoted this way
\emph{shall depend neither on $\delta$, nor on $T$.} We will also assume that
$T\ge T_0>0$ and $M\ge M_0>0$.}
\end{remark}
\begin{remark}
\label{rm14111}
{\em
Classical results of the theory of stochastic
differential equations, see e.g. Theorem 6 of Chapter 2, p. 176 and Corollary 4 of Chapter 3,
p. 303 of \cite{giksk}, imply that
there exists a unique solution to the Cauchy problem
\eqref{eq-mainthm-1} that belongs to the class
$C_b^{1,1,2}([0,+\infty)\times\R^{2d}_*)$. This solution admits a
probabilistic representation using the law of a time homogeneous
diffusion $\mathfrak Q_{\bx,\bk}$ whose Kolmogorov equation is given by
\eqref{eq-mainthm-1}, see Section \ref{sec3.6} below.}
\end{remark}
Note that
\begin{eqnarray*}
&&\sum\limits_{m=1}^dD_{nm}(\vhatk,k)\hat k_m=
-\sum\limits_{m=1}^d\frac{1}{2H_0'(k)}\int_{-\infty}^\infty
\frac{\partial^2R(s\vhatk,k)}{\partial x_n\partial x_m}\hat k_mds
\\
&&~~~~~~~~~~~~~~~~~~~~~~~=
-\sum\limits_{m=1}^d\frac{1}{2H_0'(k)}\int_{-\infty}^\infty\frac{d}{ds}
\left(\pdr{R(s\vhatk,k)}{x_n}\right)ds=0
\end{eqnarray*}
and thus the $K$-process generated by (\ref{eq-mainthm-1}) is indeed a
diffusion process on a sphere $k=\hbox{const}$, or, equivalently,
equations (\ref{eq-mainthm-1}) for different values of $k$ are
decoupled. Assumption (\ref{non-trivial}) implies the following.
\begin{proposition}\label{prop-non-deg}
The matrix $D(\vhatk,l)$ has rank $d-1$ for each $\vhatk\in{\mathbb S}^{d-1}$ and
each $l>0$.
\end{proposition}
The proof is the same as that of Proposition 4.3 in \cite{bakoryz}.
It can  be shown, using the argument given on pp. 122-123 of
ibid., that, under  assumption \eqref{non-trivial}, equation
(\ref{eq-mainthm-1}) is hypoelliptic on the manifold
$\Rm^{d}\times{\mathbb S}_k^{d-1}$ for each $k>0$.

We also show that solutions of (\ref{eq-mainthm-1}) converge in the
long time limit to the solutions of the spatial diffusion
equation. More, precisely, we have the following result. Let $\bar
\phi_\gamma(t,\vx,\bk)=\bar\phi(t/\gamma^2,\vx/\gamma,\bk)$, where
$\bar \phi$ satisfies (\ref{eq-mainthm-1}) with an initial data $\bar
\phi_\gamma(0,t,\vx,\bk)=\phi_0(\gamma\vx,\bk)$.
We also let $w(t,\vx,k)$ be the solution
of the spatial diffusion equation:
\begin{eqnarray}\label{6-diff-space-main}
&&\pdr{w}{t}=\sum\limits_{m,n=1}^d a_{mn}(k)
\frac{\partial^2 w}{\partial x_n\partial x_m},\\
&&w(0,\bx,k)=\bar\phi_0(\bx,k)\nonumber
\end{eqnarray}
with the averaged initial data
\[
\bar\phi_0(\vx,k)=\frac{1}{\Gamma_{d-1}}\int_{{\mathbb S}^{d-1}}
\phi_0(\vx,\bk)d\Omega(\vhatk).
\]
Here $d\Omega(\vhatk)$ is the surface measure on the unit sphere
${\mathbb S}^{d-1}$ and $\Gamma_n$ is the area of an $n$-dimensional
sphere.
The diffusion matrix $A:=[a_{nm}]$ in (\ref{6-diff-space-main}) is given explicitly as
\begin{equation}\label{6-wd-diffm-main}
a_{nm}(k)=\frac{1}{\Gamma_{d-1}}\int_{{\mathbb S}^{d-1}}H_0'(k)\hat k_n
\chi_m(\bk)d\Omega(\vhatk).
\end{equation}
The functions $\chi_j$ appearing above are the mean-zero solutions of
\begin{equation}\label{6-chim-eq-main}
\sum\limits_{m,n=1}^d\pdr{}{k_m}\left(D_{mn}(\vhatk,k)\pdr{\chi_j}{k_n}\right)=-H_0'(k)\hat k_j.
\end{equation}
Note that equations (\ref{6-chim-eq-main}) for $\chi_m$ are elliptic
on each sphere $\{|\bk|=k\}$. This follows from the fact that the
equations for each such sphere are all decoupled and Proposition
\ref{prop-non-deg}.  Also note that the matrix $A$ is positive
definite. Indeed, let ${\bf c}=(c_1,\ldots,c_d)\in\Rm^d$ be a fixed
vector and let $\chi_\vc:=\sum_{m=1}^dc_m\chi_m$.  Since the matrix
$D$ is non-negative we have \begin{eqnarray}
\label{exxtra2}
&&(A\vc,\vc)_{\R^d}=-\frac{1}{\Gamma_{d-1}}\sum\limits_{m,n=1}^d\int_{{\mathbb
S}^{d-1}}\chi_\vc(\vhatk,l)
\pdr{}{k_m}\left(D_{mn}(\vhatk,l)\pdr{\chi_\vc(\vhatk,l)}{k_n}\right)d\Omega(\vhatk)\\
&&~~~~~~~~~~~~~~~=-
\frac{1}{\Gamma_{d-1}}\sum\limits_{m,n=1}^d\int_{{\mathbb
R}^{d}}\chi_\vc(\vhatk,l)
\pdr{}{k_m}\left(D_{mn}(\vhatk,l)\pdr{\chi_\vc(\vhatk,l)}{k_n}\right)
\delta(k-l)\frac{d\bk}{l^{d-1}}\nonumber\\ &&~~~~~~~~~~~~~~~=
\frac{1}{\Gamma_{d-1}}\int_{{\mathbb
S}^{d-1}}(D(\vhatk,l)\nabla\chi_\vc(\vhatk,l),
\nabla\chi_\vc(\vhatk,l))_{\R^d}d\hat\Omega(\vhatk)\ge 0.\nonumber
\end{eqnarray}
The last equality holds after integration by parts
because $D(\vhatk,l)\vhatk =0$.  Moreover, the inequality
appearing in the last line of \eqref{exxtra2} is strict. This can be seen as follows. Since
the null-space of the matrix $D(\vhatk,l)$ is one-dimensional and
consists of the vectors parallel to $\vhatk $, in order for
$(A\vc,\vc)_{\R^d}$ to vanish one needs that the gradient
$\nabla\chi_\vc(\vhatk,l)$ is parallel to $\vhatk $ for all
$\vhatk\in{\mathbb S}^{d-1}$. This, however, together with
(\ref{6-chim-eq-main}) would imply that $\vhatk \cdot\vc=0$ for all
$\vhatk$, which is impossible.

The following theorem holds.
\begin{theorem}\label{thm-wave-space-main}
For every $0<T_*<T<+\infty$ the re-scaled solution $\bar
\phi_\gamma(t,\vx,\bk)=\bar\phi(t/\gamma^2,\vx/\gamma,\bk)$ of
$(\ref{eq-mainthm-1})$ converges as $\gamma\to 0$ in
$C([T_*,T];L^\infty(\Rm^{2d}))$ to $w(t,\vx,\bk)$. Moreover, there exists a constant $C>0$
so that we have
\begin{equation}\label{6-error-main}
\|w(t,\cdot)-\bar \phi_\gamma(t,\cdot)\|_{0,0}\le
C\left(\gamma T+\sqrt{\gamma}\right)\|\phi_0\|_{1,1}
\end{equation}
for all $T_*\le t\le T$.
\end{theorem}
\begin{remark}
\label{rm118}
{\em In fact, as it will become apparent in the course of the proof, we have a
stronger result, namely $T_*$ can be made to vanish as $\ga\to0$.
For instance, we can choose $T_*=\ga^{3/2}$, see
\eqref{bums4}.}
\end{remark}
The proof of Theorem \ref{thm-wave-space-main} is based on some classical
asymptotic expansions and is quite straightforward.
As an immediate corollary of Theorems \ref{main-thm} and
\ref{thm-wave-space-main} we obtain the following result, which is the main result
of this paper.
\begin{theorem}\label{thm-liouv-diff}
Let $\phi_\delta$ be solution of $(\ref{liouv-intro-0})$ with the
initial data $\phi_\delta(0,\vx,\bk)=\phi_0(\delta^{1+\alpha}\vx,\bk)$ and
let $\bar w(t,\vx)$ be the solution of the diffusion equation
$(\ref{6-diff-space-main})$ with the initial data $w(0,\vx,k)=\bar\phi_0(\vx,k)$. Then,
there exists $\alpha_0>0$ and a constant $C>0$ so that for all $0\le\alpha<\alpha_0$
and all $0<T_*\le T$ we have for all compact sets $K\subset{\cal A}(M)$:
\begin{equation}\label{6-error-twice}
\sup_{(t,\vx,\bk)\in [T_*,T]\times K}
\left|w(t,\vx,k)-{\mathbb E}\bar\phi_\delta(t,\vx,\bk)\right|
\le CT\delta^{\alpha_0-\alpha},
\end{equation}
where
$\bar\phi_\delta(t,\vx,\bk):=
\phi_\delta\left(t/\delta^{1+2\alpha},\vx/\delta^{1+\alpha},\bk\right).$
\end{theorem}
Theorem \ref{thm-liouv-diff} shows that the movement of a particle in
a weakly random quenched Hamiltonian is, indeed, approximated by a
Brownian motion in the long time-large space limit, at least for times
$T\ll \delta^{-\alpha_0}$. In fact, according to Remark \ref{rm118} we
can allow $T_*$ to vanish as $\delta\to0$ choosing
$T_*=\delta^{3\al/2}$.

In the isotropic case when $R=R(|\vx|,k)$ we may simplify the above expressions for the
diffusion matrices $D_{mn}$ and $a_{mn}$. In that case we have
\begin{eqnarray*}
&&D_{mn}(\vhatk,k)=-\frac{1}{2}\int_{-\infty}^\infty\pdrt{R(H_0'(k)s\vhatk,k
)}{x_n}{x_m}\,ds\\
&&~~~~~~~~~~~= -\int_{0}^\infty\left[\frac{k_nk_m}{k^2}R''(H_0'(k)s,k)+
\left(\delta_{nm}-\frac{k_nk_m}{k^2}\right)\frac{R'(H_0'(k)s,k)}{H_0'(k)s}\right]\,ds\\
&&~~~~~~~~~~~=
-\frac{1}{H_0'(k)}\int_{0}^\infty\frac{R'(s,k)}{s}\,ds\,
\left(\delta_{nm}-\frac{k_nk_m}{k^2}\right),
\end{eqnarray*}
so that the matrix $[D_{mn}(\vhatk,k)]$ has the form
\[
D(\vhatk,k)=D_0(k)\left(I-\vhatk\otimes\vhatk\right),~~
D_0(k)=-\frac{1}{H_0'(k)}\int_{0}^\infty\frac{R'(s,k)}{s}ds.
\]
In that case the functions $\chi_j$ are given explicitly by
\[
\chi_j(\vhatk,k)=-\frac{|H_0'(k)|^2|k|^2\hat k_j }{(d-1)\bar D_0(k)},~~
\bar D_0(k)=-\int_{0}^\infty\frac{R'(s,k)}{s}ds
\]
and
\[
a_{nm}(k)=\frac{|H_0'(k)|^3|k|^2}{\Gamma_{d-1}(d-1)\bar D_0(k)}
\int_{{\mathbb S}^{d-1}}\hat k_n
\hat k_md\Omega(\vhatk)=\frac{|H_0'(k)|^3|k|^2}{d(d-1)\bar D_0(k)}\delta_{nm}.
\]

\subsection{A formal derivation of the momentum diffusion}

We now recall how the diffusion operator in (\ref{eq-mainthm-1}) can
be derived in a quick formal way.  We represent the solution of
(\ref{liouv-intro-0}) as $\phi^\delta(t,\vx,\bk)=
\psi^\delta(\delta t,\delta\vx,\bk)$ and write an asymptotic
multiple scale expansion for $\psi^\delta$
\begin{equation}\label{asymp-bliams}
\psi^\delta(t,\vx,\bk)=\bar\phi(t,\vx,\bk)+
\sqrt{\delta}\phi_1\left(t,\vx,\frac{\vx}{\delta},\bk\right)+
{\delta}\phi_2\left(t,\vx,\frac{\vx}{\delta},\bk\right)+\dots
\end{equation}
We assume formally that the leading order term $\bar\phi$ is
deterministic and independent of the fast variable $\vz=\vx/\delta$.
We insert this expansion into (\ref{liouv-intro-0}) and obtain in the
order $O\left(\delta^{-1/2}\right)$:
\begin{equation}\label{bliam-regul}
\nabla_\vz H_1(\vz,\bk)\cdot\nabla_\bk\bar\phi-
H_0'(k)\vhatk\cdot\nabla_\vz\phi_1=0.
\end{equation}
Let $\theta\ll 1$ be a small positive regularization parameter that
will be later sent to zero, and consider a regularized version of
(\ref{bliam-regul}):
\[
\frac{1}{H_0'(k)}\nabla_\vz H_1(\vz,\bk)\cdot\nabla_\bk\bar\phi-
\vhatk\cdot\nabla_\vz\phi_1+\theta\phi_1=0,~~
\]
Its solution is
\begin{equation}\label{phi_1}
\phi_1(\vz,\bk)=-\frac{1}{H_0'(k)}\int_0^{\infty}
\sum_{m=1}^d\pdr{H_1(\vz+s\vhatk,k)}{z_m}\,\pdr{\bar\phi(t,\vx,\bk)}{k_m}\,
e^{-\theta s}ds.
\end{equation}
The next order equation becomes upon averaging
\begin{equation}\label{bliams}
\pdr{\bar\phi}{t}=
\E\left(\pdr{H_1(\vz,k)}{k}\,\vhatk\cdot\nabla_\vz\phi_1\right)
-\E\left(\nabla_\vz H_1(\vz,k)\cdot\nabla_\bk\phi_1\right)
+H_0'(k)\vhatk\cdot\nabla_\vx\bar\phi.
\end{equation}
The first two terms on the right hand side above may be computed
explicitly using expression (\ref{phi_1}) for $\phi_1$:
\begin{eqnarray*}
&&\E\left(\pdr{H_1(\vz,k)}{k}\vhatk\cdot\nabla_\vz\phi_1\right)-
\E\left(\nabla_\vz H_1(\vz,k)\cdot\nabla_\bk\phi_1\right)
\\
&&=-\E\left[\sum_{m,n=1}^d\pdr{H_1(\vz,k)}{k}\hat k_m\pdr{}{z_m}
\left(\frac{1}{H_0'(k)}\int_0^\infty
\pdr{H_1(\vz+s\vhatk,k)}{z_n}\pdr{\bar\phi(t,\vx,\bk)}{k_n}\,
e^{-\theta s}ds\right)\right]
\\
&&+\E\left[\sum_{m,n=1}^d\pdr{H_1(\vz,k)}{z_m}\pdr{}{k_m}
\left(\frac{1}{H_0'(k)}\int_{0}^{\infty}\pdr{H_1(\vz+s\vhatk,k)}{z_n}
\pdr{\bar\phi(t,\vx,\bk)}{k_n} \,e^{-\theta s}
ds\right)\right].
\end{eqnarray*}
Using spatial stationarity of $H_1(\vz,k)$ we may rewrite the above as
\begin{eqnarray*}
&&-\E\left[\sum_{m,n=1}^d\pdr{H_1(\vz,k)}{k}\hat k_m\pdr{}{z_m}
\left(\frac{1}{H_0'(k)}\int_0^{\infty}\pdr{H_1(\vz+s\vhatk,k)}{z_n}
\pdr{\bar\phi(t,\vx,\bk)}{k_n}\,
e^{-\theta s}ds\right)\right]\\
&&-\E\left[\sum_{m,n=1}^dH_1(\vz,k)\pdr{}{z_m}\pdr{}{k_m}
\left(\frac{1}{H_0'(k)}\int_0^{\infty}\pdr{H_1(\vz+s\vhatk,k)}{z_n}
\pdr{\bar\phi(t,\vx,\bk)}{k_n}\,
e^{-\theta s}ds\right)\right]\\
&&=-\sum_{m,n=1}^d\pdr{}{k_m}\left[\frac{1}{H_0'(k)}\int_0^{\infty}
\E\left(H_1(\vz,k)\frac{\partial^2H_1(\vz+s\vhatk,k)}
{\partial z_n\partial z_m}\right)
\pdr{\bar\phi(t,\vx,\bk)}{k_n}\,e^{-\theta s}ds\right]\\
&&=-\sum_{m,n=1}^d\pdr{}{k_m}\left(\frac{1}{H_0'(k)}\int_0^{\infty}
\frac{\partial^2 R(s\vhatk,k)}{\partial x_n\partial x_m}
\pdr{\bar\phi(t,\vx,\bk)}{k_n}\,e^{-\theta s}ds\right)\\
&&\to -\frac12\sum_{m,n=1}^d\pdr{}{k_m}
\left(\frac{1}{H_0'(k)}\int_{-\infty}^\infty
\frac{\partial^2 R(s\vhatk,k)}{\partial x_n\partial x_m}
\pdr{\bar\phi(t,\vx,\bk)}{k_n}ds\right)\hbox{, as $\theta\to 0^+$.}
\end{eqnarray*}
We insert the above expression into (\ref{bliams}) and obtain
\begin{equation}\label{bliams-1}
\pdr{\bar\phi}{t}=\sum_{m,n=1}^d
\pdr{}{k_n}\left(D_{nm}(\vhatk,k)\pdr{\bar\phi}{k_m}\right)+
H_0'(k)\vhatk\cdot\nabla_\vx\bar\phi
\end{equation}
with the diffusion matrix $D(\vhatk,k)$ as in
(\ref{diff-matrix1-main}). Observe that (\ref{bliams-1}) is nothing
but (\ref{eq-mainthm-1}).  However, the naive asymptotic expansion
(\ref{asymp-bliams}) may not be justified. The rigorous proof
presented in the next section is based on a quite different method.

\section{From the Liouville equation to the momentum diffusion.
Estimation of the convergence rates: proof of Theorem \ref{main-thm}}
\label{sec:liouv}


\subsection{Outline of the proof}
\label{sec2.1a}
The basic idea of the proof of Theorem \ref{main-thm} is a modification of that
of \cite{bakoryz,KP}. We consider the trajectories corresponding to
the Liouville equation (\ref{liouv-intro-0}) and introduce a stopping
time, called $\tau_\delta$, that, among others, prevents near self-intersection
of trajectories. This fact ensures that until the stopping time occurs the
particle is ``exploring a new territory'' and, thanks to the strong mixing
properties of the medium, ``memory effects" are lost. Therefore,
roughly speaking, until the stopping time the process is
approximately characterized by the Markov property. Furthermore, since the
amplitude of the random Hamiltonian is not strong enough to destroy the
continuity of its path, it becomes a diffusion in the limit, as
$\delta\to0$.  We introduce also an augmented process that follows
the trajectories of the Hamiltonian flow until the stopping time
$\tau_\delta$ and becomes a diffusion after $t=\tau_\delta$. We show
that the law of the augmented process is close to the law of a
diffusion, see Proposition \ref{prop706041}, with an explicit error
bound. We also prove that the stopping time tends to infinity as
$\delta\to 0$, once again with the error bound that is proved in
Theorem \ref{cor10211}. The combination of these two results allows us
to estimate the difference between the solutions of the Liouville and
the diffusion equations in a rather straightforward manner (see
Section \ref{sec3.9}):
they are close until the stopping time as the law of the diffusion is
always close to that of the augmented process, while the latter
coincides with the true process until $\tau_\delta$.  On the other
hand, the fact that $\tau_\delta\to\infty$ as $\delta\to 0$ shows that
with a large probability the augmented process is close to the true
process. This combination finishes the proof.

\subsection{The random characteristics corresponding to (\ref{liouv-intro-0})}

Consider the motion of a particle governed by a Hamiltonian
system of equations
\begin{equation}\label{eq1b}
\left\{
  \begin{array}{l}
\frac{d\bz^{(\delta)}(t;\bbx,\bbk)}{dt}=
(\nabla_\bbk H_\delta)\left(\frac{\bz^{(\delta)}(t;\bbx,\bbk)}{\delta},
\obm^{(\delta)}(t;\bbx,\bbk)\right)\vphantom{\int\limits_0^1}\\
\frac{d\obm^{(\delta)}(t;\bbx,\bbk)}{dt}
=-\frac{1}{\sqrt{\delta}}(\nabla_\bz H_\delta)
\left(\frac{\bz^{(\delta)}(t;\bbx,\bbk)}{\delta},
\obm^{(\delta)}(t;\bbx,\bbk)\right)\vphantom{\int\limits_0^1}\\
\bz^{(\delta)}(0;\bbx,\bbk)=\bbx,
\quad\obm^{(\delta)}(0;\bbx,\bbk)=\bbk\vphantom{\int\limits_0^1},
  \end{array}
\right.
\end{equation}
where the Hamiltonian $H_\delta(\bx,\bk):=
H_0(k)+\sqrt{\delta}H_1(\bx,k)$, $k=|\bk|$. The trajectories of
(\ref{eq1b}) are the characteristics of the Liouville equation
(\ref{liouv-intro-0}).  The hypotheses made in Section
\ref{prelim} imply that the trajectory
$(\bz^{(\delta)}(t;\bbx,\bk),\obm^{(\delta)}(t;\bbx,\bk))$
necessarily lies in ${\cal C}(T,\delta)$ for each $T>0$,
$\delta\in(0,\delta_*(M)]$, provided that the initial data
$(\bbx,\bk)\in{\cal A}(M)$.  Indeed, it follows
from the Hamiltonian structure of
\eqref{eq1b} that the Hamiltonian
$H_\delta(\vx,m)=H_0(m)+\sqrt{\delta}H_1(\bz,m)$ must be conserved
along the trajectory.  Hence, the definition (\ref{073104}) implies
that $M_\delta^{-1}\le |\obm^{(\delta)}(\cdot;\bbx,\bk)|\le M_\delta$.
We denote by $Q^\delta_{s,\bbx,\bbk}(\cdot)$ the law over ${\cal C}$
of the process corresponding to \eqref{eq1b} starting at $t=s$ from
$(\bbx,\bbk)$ (this law is actually supported in ${\cal
C}(\delta)$). We shall omit writing the subscript $s$ when it equals
to $0$.

\commentout{
We obtain from Proposition \ref{lm1} for each $s\ge\si$ an error for
the first-order approximation of the trajectory
\[
|\bz^{(\delta)}(s)-\bl^{(\delta)}(\si,s)|\leq
\tilde D(2M_\delta)\sqrt{\delta}(s-\si)+
\frac{C(s-\si)^2}{2\sqrt{\delta}},
\quad\,\delta\in(0,\delta_*(M)].
\]
Here $\bl^{(\delta)}(\si,s)
:=\bz^{(\delta)}(\si)+(s-\si)\hat{\obm}^{(\delta)}(\si)$ is the linear
approximation between the times $\sigma$ and $s$ and $$
C:=\sup_{\delta\in(0,\delta_*(M)]}(M_\delta h_0^*(M_\delta)+\tilde
h_0^*(M_\delta))\tilde D(2M_\delta).$$
}

\subsection{The stopping times}
\label{sec44}

We now define the stopping time $\tau_\delta$, described in Section
\ref{sec2.1a}, that prevents the trajectories of
(\ref{eq1b}) to have near self-intersections (recall that the intent
of the stopping time is to prevent any ``memory effects'' of the
trajectories).  As we have already mentioned, we will later show that the
probability of the event  $[\,\tau_\delta< T\,]$ for a fixed $T>0$ goes to zero, as $\delta\to 0$.

Let $0<\ep_1<\ep_2<1/2$, $\ep_3\in(0,1/2-\ep_2)$,
$\ep_4\in(1/2,1-\ep_1-\ep_2)$ be small positive constants that will
be further determined later and set
\begin{equation}\label{102302}
N=[\delta^{-\ep_1}],\quad p=[\delta^{-\ep_2}],\quad
q=p\,[\delta^{-\ep_3}],\quad N_1=Np\,[\delta^{-\ep_4}].
\end{equation}
We will specify additional restrictions on the constants $\ep_j$ as
the need for such constraints arises. However, the basic requirement
is that $\ep_i$, $i=1,2,3$ should be sufficiently small and $\ep_4$ is
bigger than $1/2$, less than one and can be made as close to one as we
would need it. It is important that $\ep_1<\ep_2$ so that $N\ll p$
when $\delta\ll 1$.  We introduce the following $({\cal
M}^{t})_{t\geq0}$--stopping times. Let $t^{(p)}_k:=kp^{-1}$ be a mesh
of times, and $\pi\in {\cal C}$ be a path.  We define the ``violent
turn'' stopping time
\begin{eqnarray}\label{Sdelta}
&&S_\delta(\pi):=\inf\left[\,t\geq0:\vphantom{\int_0^1}\mbox{ for some }k\geq 0\mbox{ we
have }t\in\left[t_k^{(p)},t_{k+1}^{(p)}\right)\right. \mbox{ and }\\ &&\left.
~~~~~~~ \hat{K}(t_{k-1}^{(p)})\cdot \hat
K(t)\le 1-\frac{1}{N},\mbox{ or
}\, \hat{K}\left(t_{k}^{(p)}-\frac{1}{N_1}\right)\cdot \hat
K(t)\le1-\frac{1}{N}\,\right],\nonumber
\end{eqnarray}
where by convention we set $\hat K(-1/p):=\hat K(0)$.  Note that with the
above choice of $\ep_4$ we have
$\hat{K}\left(t_{k}^{(p)}-1/N_1\right)\cdot \hat
K(t_{k}^{(p)})>1-1/N$, provided that $\delta\in(0,\delta_0]$ and
$\delta_0$ is sufficiently small.  We adopt in (\ref{Sdelta}) a customary convention
that the infimum of an empty set equals $+\infty$.  The stopping time $S_\delta$ is triggered when
the trajectory performs a sudden turn -- this is undesirable as the trajectory may then return back
to the region it has already visited and create correlations with the past.

For each $t\ge 0$, we denote by
 $ \mathfrak X_t(\pi):=\mathop{\bigcup\limits_{0\le s\leq
t}}X\left(s;\pi\right)$
the trace of the spatial component of the path $\pi$ up to time $t$, and by
$\mathfrak X_{t}(q;\pi):=[\bx:\mbox{dist }(\bx,
\mathfrak X_{t}(\pi))\le
1/q]$ a tubular region around the path. We
introduce the stopping time
\begin{equation}\label{Udelta}
U_\delta(\pi):=\inf\left[\,t\ge0:\,\exists\,k\ge1\, \mbox{ and }t\in[t_k^{(p)},t_{k+1}^{(p)})\mbox{ for
which }X(t)\in \mathfrak X_{t_{k-1}^{(p)}}(q)\,\right].
\end{equation}
It is associated with the return of the $X$ component of the trajectory to the tube around its past -- this is
again an undesirable way to create correlations with the past.
Finally, we set the stopping time
\begin{equation}
\label{W:delta}
 \tau_\delta(\pi):=S_\delta(\pi)\wedge U_\delta(\pi).
\end{equation}

\subsection{The cut-off functions and the corresponding dynamics}

Let $M>0$ be fixed and $p,q,N, N_1$  be the positive integers defined in Section \ref{sec44}.
We define now several auxiliary functions that will be used to
introduce the cut-offs in the dynamics. These cut-offs will ensure that the particle
moving under the modified dynamics will avoid self-intersections, will have no violent turns and
the changes of its momentum will be under control. In addition, up to the stopping time $\tau_\delta$
the motion of the particle will coincide with the motion under the original Hamiltonian flow.

Let $a_1=2$ and $a_2=3/2$.  The functions
$\psi_j:\R^d\times \mathbb S_1^{d-1}\rightarrow [0,1]$, $j=1,2$ are of $C^\infty$ class and
satisfy
\begin{equation} \label{def1}
\psi_j(\bbk,\bbl)=\left\{
\begin{array}{l}
1, \phantom{aaaaaa}\mbox{ if    }~~\hat{\bbk}\cdot \bbl\geq
1-1/N\phantom{aaaaaa}\mbox{ and
}\phantom{aaaaaa}M_\delta^{-1}\leq |\bbk|\leq M_\delta\\
0,\phantom{aaaaaa}\vphantom{\int\limits^{s^{(p_1)}_{k}}} \mbox{ if
}~~\hat{\bbk}\cdot \bbl\leq 1-a_j/N,\phantom{aa}\mbox{ or
}~~|\bbk|\leq
(2M_\delta)^{-1}\vphantom{\int\limits^{s^{(p_1)}_{k}}},\phantom{aaa}
\mbox{ or }\phantom{aaa} |\bbk|\geq 2M_\delta.
\end{array}
\right.
\end{equation}
One can construct $\psi_j$ in such a way that
for arbitrary nonnegative  integers $m,n$ it is possible to find a constant $C_{m,n}$
for which $\|\psi_j\|_{m,n}\le C_{m,n}N^{m+n}$. The cut-off
function
\begin{equation} \label{def21}
\Psi(t,\bbk;\pi):=\left\{
\begin{array}{ll}
\psi_1\left(\bbk,\hat{K}\left(t^{(p)}_{k-1}\right)\right)\psi_2\left(\bbk,
\hat{K}\left(t^{(p)}_{k}-1/N_1\right)\right)&\mbox{ for }t\in[t_k^{(p)},t_{k+1}^{(p)})
\mbox{ and }k\geq1\\
\psi_2(\bbk,\hat{K}(0))&\mbox{ for }t\in[0,t_{1}^{(p)})
\end{array}
\right.
\end{equation}
will allow us to control the direction of the particle motion over
each interval of the partition as well as not to allow the trajectory
to escape to the regions where the change of the size of the velocity
can be uncontrollable.

Let $\phi: \R^d\times\R^d\rightarrow [0,1]$ be a function of the
$C^\infty$ class that satisfies $\phi(\bby,\bbx)=1$, when
$|\bby-\bbx|\ge 3/q$ and $\phi(\bby,\bbx)=0$, when $|\bby-\bbx|\le
2/q$. Again, in this case we can construct $\phi$ in such a way that
$\|\phi\|_{m,n}\le Cq^{m+n}$ for arbitrary integers $m,n$ and a
suitably chosen constant $C$.  The function $\phi_k:
\R^d\times{\cal C}\rightarrow [0,1]$ for a fixed path $\pi$ is given
by
\begin{equation}
\label{def11} \phi_k(\bby;\pi)=
\prod\limits_{0\le l/q\le t_{k-1}^{(p)}}\phi\left(\bby,X\left(\frac{l}{q}\right)\right).
\end{equation}
We set
\begin{equation}
\label{def2} \Phi(t,\bby;\pi):=\left\{
\begin{array}{ll}
1,& \mbox{ if    }0\leq t< t^{(p)}_1\\
\phi_k(\bby;\pi),& \mbox{ if }t^{(p)}_k\leq t< t^{(p)}_{k+1}.
\end{array}
\right.
\end{equation}
The function $\Phi$ shall be used to modify the dynamics of the
particle in order to avoid a possibility of near
self-intersections of its trajectory.

For a given $t\ge0$, $(\bby,\bbk)\in\R^{2d}_*$ and $\pi\in{\cal C}$ let us denote
$
\Theta(t,\bby,\bbk;\pi):=\Psi(t,\bbk;\pi)\Phi\left(t,\bby;\pi\right).
$
The following lemma can be verified by a direct calculation.
\begin{lemma}
\label{lm3} Let $(\bt_1,\bt_2)$ be a multi-index with
nonnegative integer valued components, $m=|\bt_1|+|\bt_2|$.
There exists a constant $C$ depending
only on  $m$ and $M$  such that
$|\partial_\bby^{\beta_1}\partial_\bbk^{\beta_2}\Theta(t,\bby,\bbk;\pi)|
\leq CT^{|\bt_1|}q^{2|\bt_1|}N^{|\bt_2|} $ for all
$t\in[0,T],\,(\bby,\bbk)\in{\cal A}(2M),\,\pi\in{\cal C}$.
\end{lemma}

Finally, let us set
\begin{equation}
\label{70613} F_\delta(t,\bby,\bbl;\pi,\om)
=\Theta(t,\delta\bby,\bbl;\pi) \nabla_{\bby}
H_1\left(\bby,|\bbl|;\om\right).
\end{equation}
For a fixed $(\bbx,\bbk)\in\R^{2d}_*$, $\delta>0$ and $\om\in\Om$ we consider
the modified  particle dynamics with the cut-off that
 is described by the stochastic process
 $(\by^{(\delta)}(t;\bbx,\bbk,\om),\bl^{(\delta)}(t;\bbx,\bbk,\om))_{t\ge0}$
whose paths are
 the
 solutions of the following equation
\begin{equation}\label{eq2}
\left\{
  \begin{array}{l}
\frac{d\by^{(\delta)}(t;\bbx,\bbk)}{dt}=\left[H_0'(|\bl^{(\delta)}(t;\bbx,\bbk)|)+\sqrt{\delta}\,
\partial_l H_1\left(\frac{\by^{(\delta)}(t;\bbx,\bbk)}{\delta},|\bl^{(\delta)}(t;\bbx,\bbk)|\right)
\right]\hat{\bl}^{(\delta)}(t;\bbx,\bbk,)\\
\frac{d\bl^{(\delta)}(t;\bbx,\bbk)}{dt}=-\frac{1}{\sqrt{\delta}}\,F_\delta\left(t,
\frac{\by^{(\delta)}(t;\bbx,\bbk)}{\delta},\bl^{(\delta)}(t;\bbx,\bbk);
\by^{(\delta)}(\cdot;\bbx,\bbk),\bl^{(\delta)}(\cdot;\bbx,\bbk)\right)
\vphantom{\int\limits_{\frac{1}{1}}^{\frac{1}{1}}}\\
\by^{(\delta)}(0;\bbx,\bbk)=\bbx,\quad\bl^{(\delta)}(0;\bbx,\bbk)=\bbk.
  \end{array}
\right.
\end{equation}
We will denote by $\tilde
Q^{(\delta)}_{\bbx,\bbk}$ the law of the modified process
$(\by^{(\delta)}(\cdot;\bbx,\bbk),\bl^{(\delta)}(\cdot;\bbx,\bbk))$
over ${\cal C}$ for a given $\delta>0$ and by $\tilde
E^{(\delta)}_{\bbx,\bbk}$ the corresponding expectation.  We assume
that the initial momentum $\bbk\in A(M)$.  From the construction of
the cut-offs we immediately conclude that
\begin{equation}\label{80402}
    \hat{\bl}^{(\delta)}(t)\cdot \hat{\bl}^{(\delta)}(t_{k-1}^{(p)})\ge 1-\frac{2}{N},\quad
    t\in[t_{k-1}^{(p)},t_{k+1}^{(p)}),\quad\forall\,k\ge0.
\end{equation}
\subsection{Some consequences of the mixing assumption}
\label{secmix}

For any $t\geq0$ we denote by ${\cal F}_t$ the
$\si$-algebra generated by
$(\by^{(\delta)}(s),\bl^{(\delta)}(s))$, $s\leq t$. Here we suppress, for the sake of abbreviation,
writing the initial data in the notation of the trajectory.
In this section we assume that $M>0$ is fixed, $X_1,X_2:(\R\times\R^d\times
\R^{d^2})^2\rightarrow\R$ are certain continuous functions, $Z$ is a
random variable and $g_1,g_2$ are $\R^d\times[M^{-1},M]$-valued random vectors. We
suppose further that $Z,g_1,g_2$, are ${\cal F}_t$-measurable,
while $\tilde X_1,\tilde X_2$ are random fields of the form
\[
\tilde X_i(\bbx,k)=X_i\left(\left(\partial_k^j H_1(\bbx,k),
\nabla_\bbx\partial_k^j H_1(\bbx,k),\nabla_\bbx^2
\partial_k^j H_1(\bbx,k)\right)_{j=0,1}\right) .
\]
For $i=1,2$ we denote $g_i:=(g_i^{(1)},g_i^{(2)})$  where $g_i^{(1)}\in\R^d$ and
$g_i^{(2)}\in[M^{-1},M]$. We also let
\begin{equation}\label{80101}
U(\theta_1,\theta_2):= \bbE\left[\tilde X_1(\theta_1)\tilde X_2(\theta_2)\right]
,\quad \theta_1,\theta_2\in\R^d\times [M^{-1},M].
\end{equation}
The following mixing lemma is useful in formalizing the ``memory loss effect'' and
can be proved in the same way as Lemmas 5.2 and 5.3 of \cite{bakoryz}.
\begin{lemma}\label{mix1}
(i) Assume that $r,t\geq0$ and
\begin{equation} \label{70202}
 \inf\limits_{u\leq t}\left|g_i^{(1)}-\frac{\by^{(\delta)}(u)}{\delta}\right|\geq
\frac{r}{\delta},
\end{equation}
$\bbP$--a.s. on the set $Z\not=0$ for  $i=1,2$. Then, we have
\begin{equation}
\label{70201} \left|\bbE\left[\tilde X_1(g_1)\tilde X_2(g_2)Z\right]-\bbE\left[
U(g_1,g_2) Z\right]\right| \leq
2\phi\left(\frac{r}{2\delta}\right)\|X_1\|_{L^\infty}\|X_2\|_{L^\infty}\|Z\|_{L^1(\Om)}.
\end{equation}
\item(ii) Let
$\bbE X_1(\bze,k)=0$ for all $k\in[M^{-1},M]$. Furthermore, we assume that $g_2$ satisfies
$(\ref{70202})$,
\begin{equation} \label{70202b}
 \inf\limits_{u\leq t}\left|g_1^{(1)}-\frac{\by^{(\delta)}(u)}{\delta}\right|\geq
\frac{r+r_1}{\delta}
\end{equation}
and
$|g_1^{(1)}-g_2^{(1)}|\geq r_1\delta^{-1}$
for some $r_1\geq0$, $\bbP$-a.s. on the event  $Z\not=0$.
 Then, we have
\begin{equation}
\label{70201b}
\left|\bbE\left[\tilde X_1(g_1)\tilde X_2(g_2)\,Z\right]-\bbE\left[ U(g_1,g_2)
Z\right]\right| \leq C\phi^{1/2}\left(\frac{r}{2\delta}\right)
\phi^{1/2}\left(\frac{r_1}{2\delta}\right)\|X_1\|_{L^\infty}\|X_2\|_{L^\infty}\|Z\|_{L^1(\Om)}
\end{equation}
for some absolute constant $C>0$.
Here the function $U$ is given by $(\ref{80101})$.
\end{lemma}

\subsection{The momentum diffusion}
\label{sec3.6}
Let $\bbk(t)$ be a diffusion, starting at $\bk\in\R^d_*$ at $t=0$, with the generator of the form
\begin{eqnarray}
\label{61102}
&&{\cal L}F(\bbk)= \sum\limits_{m,n=1}^d
D_{mn}(\hat{\bbk},|\bk|)\partial_{k_m,k_n}^2F(\bbk) +
\sum\limits_{m=1}^d E_{m}(\hat{\bbk},|\bk|)\partial_{k_m}F(\bbk)\\
&&~~~~~~~~~=  \sum\limits_{m,n=1}^d\partial_{k_m}\left(
D_{m,n}(\hat{\bbk},|\bk|)\partial_{k_n}F(\bbk)\right),
\quad F\in C^\infty_0(\R^d_*).
\nonumber
\end{eqnarray}
Here the diffusion matrix is given by (\ref{diff-matrix1-main})
\commentout{

:
\begin{equation}\label{diff-matrix1}
D_{mn}(\vhatk,l)=-\frac{1}{2H_0'(l)}\int_{-\infty}^\infty\frac{\partial^2
R(s\vhatk,l)}{\partial x_n\partial x_m}ds,\quad\forall\,m,n=1,\ldots,d
\end{equation}

}
and the drift vector is
\[
E_{m}(\hat{\bbk},l)= -\frac{1}{H_0'(l)l}\sum\limits_{n=1}^d
\int_{0}^{+\infty}s\frac{\partial^3R(s\hat{\bbk},l)}{\partial x_m\partial x_n^2}\,
ds,\quad\,m=1,\ldots,d.
\]
Employing exactly the same argument as the one used in Section 4 of
\cite{bakoryz} it can be easily seen that this diffusion is supported
on ${\mathbb S}^{d-1}_k$, where $k=|\bk|$.  Moreover, it is
non-degenerate on the sphere, for instance, under the assumption
\eqref{non-trivial}, cf. Proposition 4.3 of ibid.

Let $\mathfrak Q_{\bbx,\bbk}$ be the law of the process
$(\bbx(t),\bbk(t))$ that starts at $t=0$ from $(\bbx,\bbk)$ given by
$\bbx(t)=\vx+\int_0^tH_0'(|\bbk(s)|)\hat\bbk(s)ds$, where
$\bbk(t)$ is the diffusion described by (\ref{61102}). This
process is a degenerate diffusion whose generator is given by
\begin{equation}
\label{61102b} \tilde{\cal L}F(\bbx,\bbk)={\cal L}_\bk F(\bbx,\bbk)+
H_0'(|\bbk|)\,\hat\bbk\cdot \nabla_\bx F(\bbx,\bbk),\quad F\in C_0^\infty(\R^{2d}_*).
\end{equation}
Here the notation ${\cal L}_\bk$ stresses that the operator ${\cal L}$
defined in \eqref{61102} acts on the respective function in the $\bk$
variable.  We denote by $\mathfrak M_{\bx,\bk}$ the expectation
corresponding to the path measure $\mathfrak Q_{\bbx,\bbk}$.

\subsection{The augmented process}

The following construction of the augmentation of path measures  has been
carried out in Section 6.1 of \cite{stroock-varadhan}.
Let $s\ge0$ be fixed and $\pi\in {\cal C}$. Then, according to Lemma 6.1.1 of ibid.
there exists a unique probability measure, that is denoted by
 $\delta_\pi\otimes_s \mathfrak Q_{X(s),K(s)}$, such that
 for any pair of events $A \in{\cal M}^{s}$, $B\in{\cal M}$  we have
 $\delta_\pi\otimes_s \mathfrak Q_{X(s),K(s)}[A]=\bone_A(\pi)$ and
 $\delta_\pi\otimes_s \mathfrak Q_{X(s),K(s)}[\theta_s(B)]=\mathfrak Q_{X(s),K(s)}[B]$.
The following result is a direct consequence of Theorem 6.2.1 of \cite{stroock-varadhan}.
\begin{proposition}
\label{prop70501}
There exists a unique probability measure $R^{(\delta)}_{\bx,\bk}$ on
${\cal C}$ such that $
R^{(\delta)}_{\bx,\bk}[A]:=Q^{(\delta)}_{\bx,\bk}[A] $ for all
$A\in{\cal M}^{\tau_\delta}$ and the regular conditional probability
distribution  of $R^{(\delta)}_{\bx,\bk}[\,\cdot\,|{\cal
M}^{\tau_\delta}]$ is given by $\delta_\pi\otimes_{\tau_\delta(\pi)}
\mathfrak Q_{X(\tau_\delta(\pi)),K(\tau_\delta(\pi))}$, $\pi\in {\cal
C}$.  This measure shall be also denoted by
$Q^{(\delta)}_{\bx,\bk}\otimes_{\tau_\delta} \mathfrak
Q_{X(\tau_\delta),K(\tau_\delta)}$.
\end{proposition}
Note that for any $(\bbx,\bbk)\in{\cal A}(M)$ and $A\in {\cal M}^{\tau_\delta}$
we have
\begin{equation}\label{72801}
   R^{(\delta)}_{\bbx,\bbk}[A]= Q^{(\delta)}_{\bbx,\bbk}[A]=\tilde Q^{(\delta)}_{\bbx,\bbk}[A],
\end{equation}
that is, the law of the augmented process coincides with that of the
true process, and of the modified process with the cut-offs until the
stopping time $\tau_\delta$.  Hence, according to the uniqueness part
of Proposition \ref{prop70501}, in such a case
$Q^{(\delta)}_{\bx,\bk}\otimes_{\tau_\delta} \mathfrak
Q_{X(\tau_\delta),K(\tau_\delta)}=\tilde
Q^{(\delta)}_{\bx,\bk}\otimes_{\tau_\delta} \mathfrak
Q_{X(\tau_\delta),K(\tau_\delta)}$.  We denote by
$E^{(\delta)}_{\bx,\bk}$ the expectation with respect to the augmented
measure described by the above proposition. Let also
$R^{(\delta)}_{\bx,\bk,\pi}$, $E^{(\delta)}_{\bx,\bk,\pi}$ denote the
respective conditional law and expectation obtained by conditioning
$R^{(\delta)}_{\bx,\bk}$ on ${\cal M}^{\tau_\delta}$.

The following proposition is of crucial importance for us, as it shows
that the law of the augmented process is close to that of the
momentum diffusion as $\delta\to 0$.  To abbreviate the notation we let
\[
N_t(G):=G(t,X(t),K(t))- G(0,X(0),K(0))-
\int\limits_0^t(\partial_\varrho  +\tilde{\cal
L})G(\varrho,X(\varrho),K(\varrho)))\,d\varrho
\]
for any $G\in C^{1,1,3}_b([0,+\infty)\times\R^{2d}_*)$ and $t\ge0$.
\begin{proposition}
\label{prop706041}
Suppose that $(\bx,\bk)\in{\cal A}(M)$ and $\zeta\in
C_b((\R^{2d}_*)^{n})$ is nonnegative.  Let $\ga_0\in(0,1/2)$ and let
$0\leq t_1<\cdots< t_n\leq T_*\le t<v\le T$. We assume further that $v-t\ge
\delta^{\ga_0}$.  Then, there exist constants $\ga_1$, $C$ such
that for any function $G\in C^{1,1,3}([T_*,T]\times\R^{2d}_*)$
we have
\begin{equation}
\label{73101}
\left|E^{(\delta)}_{\bx,\bk}\left\{\left[N_v(G)-N_t(G)\right] \tilde\zeta\right\}\right|
\leq
C\delta^{\ga_1}(v-t)\|G\|_{1,1,3}^{[T_*,T]}T^2
E^{(\delta)}_{\bx,\bk}\tilde\zeta.
\end{equation}
Here
$\tilde\zeta(\pi):=\zeta(X(t_1),K(t_1),\ldots,
X(t_n),K(t_n))$, $\pi\in{\cal C}(T,\delta)$.
 The choice of the constants $\gamma_1,\,C$ does not depend on
$(\bx,\bk)$, $\delta\in(0,1]$, $\zeta$, times $t_1,\ldots, t_n,T_*, T, v,t$, or the function $G$.
\end{proposition}
{\bf Proof.}  Let $0=s_0\le s_1\le \ldots \le s_n\le t$ and
$B_1,\ldots,B_n\in {\cal B}(\R^{2d}_*)$ be Borel sets.  We denote $A_0:={\cal
C}$ and for any $k\in\{1,\ldots,n\}$, $s\le s_k$ we define the events
\[
A_k:=[\pi:\,(X(s_1),K(s_1))\in B_1,\ldots,(X(s_k),K(s_k))\in B_k]
\]
and their shifted counterparts
\[
A_{k}^{(s)}:=[\pi:\,(X(s_k-s),K(s_k-s))\in
B_k,\ldots,(X(s_n-s),K(s_n-s))\in B_n].
\]
For
$(\bbx,\bbk)\in\R^{2d}_*$, $\pi\in{\cal C}$ and $G\in
C^{1,1,2}([0,+\infty)\times\R^{2d}_*)$ we let
\begin{eqnarray*}
&&\widehat{\cal L}_tG(t,\bx,\bbk;\pi) :=
H_0'(|\bk|)\,\hat\bbk\cdot \nabla_\bx G(t,\bbx,\bbk)+\Theta^2(t,X(t),K(t);\pi){\cal
L}_\bk G(t,\bx,\bbk)\\
&&~~~~~~~~~~~~~~~~~~~~
-\Theta(t,X(t),K(t);\pi)\sum\limits_{m,n=1}^d\partial_{K_m}
\Theta(t,X(t),K(t);\pi)D_{m,n}(\hat\bbk,|\bk|)\partial_{k_n}G(t,\bx,\bbk)
\end{eqnarray*}
and
\[
\widehat{ N}_t(G):=G(t,X(t),K(t))-
G(0,X(0),K(0))-
\int\limits_0^t(\partial_\varrho  +\widehat{\cal L}_\varrho)
G(\varrho,X(\varrho),K(\varrho);\pi)\,d\varrho.
\]
It follows from the definition of the stopping time $\tau_\delta(\pi)$ and the cut-off function $\Theta$
that
\[
\nabla_{K}
\Theta(t,X(t),K(t);\pi)=\bze,~~~t\in[0, \tau_\delta(\pi)],
\]
hence
\[
\widehat{\cal L}_tG(t,X(t),K(t);\pi) =\tilde{\cal
L}G(t,X(t),K(t);\pi),~~~ t\in[0, \tau_\delta(\pi)].
\]
We need the following result.
\begin{lemma}
\label{lmA1}
Suppose that   $(\bx,\bk)\in{\cal A}(M)$
and $\zeta\in C_b((\R^{2d}_*)^{n})$ is nonnegative.
Let $\ga_0'\in(0,1)$,
 $0\leq t_1<\cdots< t_n\le T_*\leq  t<v\le T$ and $t-T_*\ge\delta^{\ga_0'}$.
Then, there exist   constants
 $\ga_1',$ $C'>0$ 
  such that for any
 function $G\in C^{1,1,3}([T_*,T]\times\R^{2d}_*)$
 we have
\begin{equation}
\label{73101b}
\left|\tilde E^{(\delta)}_{\bx,\bk}
\left\{[\widehat{ N}_v(G)-\widehat{ N}_t(G)]\tilde\zeta\right\}\right|
\leq
C'\delta^{\ga_1'}(v-t)\|G\|_{1,1,3}^{[T_*,T]}T^2
\tilde E^{(\delta)}_{\bx,\bk}\tilde\zeta.
\end{equation}
The choice of the constants $\gamma_1',\,C'$ does not depend on
$(\bx,\bk)$, $\delta\in(0,1]$, times  $t_1,\ldots, t_n,T_*$, $T, v,t$, or  function $G$.
\end{lemma}
The proof of this lemma follows very closely the argument presented in
Section 5.3 of \cite{bakoryz} and we postpone it until the Appendix.
In the meantime we apply this result to conclude the proof of Proposition \ref{prop706041}.
We write
\begin{eqnarray}
&&\!\!\!\!\!\!\!\!\!\!
E^{(\delta)}_{\bx,\bbk,\pi}[N_v(G)-N_{v\wedge \tau_\delta(\pi)}(G),A_n]\!=\!
\sum\limits_{p=0}^{n-1}\!\bone_{[s_p,s_{p+1})}(\tau_\delta(\pi))\bone_{A_p}(\pi)
\mathfrak M_{X(\tau_{\delta}(\pi)),K(\tau_{\delta}(\pi))}
[N_{v-\tau_\delta(\pi)}(G),A^{(\tau_\delta(\pi))}_{p+1}]\nonumber\\
&&~~~~~~~~~~~~~~~~~~~~~~~~~~~~~~~~~~~~~~
+\bone_{[s_n,v)}(\tau_\delta(\pi))\bone_{A_n}(\pi)
\mathfrak M_{X(\tau_{\delta}(\pi)),K(\tau_{\delta}(\pi))}[N_{v-\tau_\delta(\pi)}(G)].\label{80104}
\end{eqnarray}
When $\tau_\delta(\pi)\in [s_p,s_{p+1})$ we obviously have
\[
\mathfrak M_{X(\tau_{\delta}(\pi)),K(\tau_{\delta}(\pi))}
[N_{v-\tau_\delta(\pi)}(G),A^{(\tau_\delta(\pi))}_{p+1}]
=\mathfrak M_{X(\tau_{\delta}(\pi)),K(\tau_{\delta}(\pi))}
[N_{t-\tau_\delta(\pi)}(G),A^{(\tau_\delta(\pi))}_{p+1}]
\]
and $\mathfrak
M_{X(\tau_{\delta}(\pi)),K(\tau_{\delta}(\pi))}[N_{v-\tau_\delta(\pi)}(G)]=0$. Hence
the left hand side of \eqref{80104} equals
\begin{equation}\label{80104b}
\sum\limits_{p=0}^{n-1}\bone_{[s_p,s_{p+1})}(\tau_\delta(\pi))\bone_{A_p}(\pi)
\mathfrak M_{X(\tau_{\delta}(\pi)),K(\tau_{\delta}(\pi))}
[N_{t-\tau_\delta(\pi)}(G),A^{(\tau_\delta(\pi))}_{p+1}]
\end{equation}
$$
=
E^{(\delta)}_{\bx,\bbk,\pi}[N_{t}(G)-N_{t\wedge \tau_\delta(\pi)}(G),A_n].
$$
\commentout{
Note that, if in addition we have  $t'=t-1/L$,  $\tau_\delta(\pi)\in (t',t]$, $u- \tau_\delta(\pi)<1/L$,
 then we obtain the following estimate
$$
\mathfrak M_{X(\tau_{\delta}(\pi)),K(\tau_{\delta}(\pi))}[N_{u-\tau_\delta(\pi)}(G)]=
\mathfrak M_{X(\tau_{\delta}(\pi)),K(\tau_{\delta}(\pi))}[\mathfrak M_{X(t'),K(t')}N_{u-t'}(G),
N_{t'-\tau_\delta(\pi)}(G)]
$$
}
We conclude from \eqref{80104}, \eqref{80104b} that
\begin{eqnarray}\label{80204}
&&E^{(\delta)}_{\bx,\bbk,\pi}[N_v(G),A_n]=
E^{(\delta)}_{\bx,\bbk,\pi}[N_{v\wedge \tau_\delta(\pi)}(G)+N_t(G)-
N_{t\wedge \tau_\delta(\pi)}(G),A_n]\\
&&~~~~~~~~~~~~~~~~~~~~~~~=
E^{(\delta)}_{\bx,\bbk,\pi}[N_{(v\wedge \tau_\delta(\pi))\vee t}(G),A_n]\nonumber
\end{eqnarray}
and therefore
\begin{eqnarray}\label{80304}
&&\!\!\!\!E^{(\delta)}_{\bx,\bbk}[N_v(G),A_n]=E^{(\delta)}_{\bx,\bbk}\left[
E^{(\delta)}_{\bx,\bbk,\pi}[N_{(v\wedge \tau_\delta(\pi))\vee t}(G),A_n]\right]
\\
&&\!\!\!\!=E^{(\delta)}_{\bx,\bbk}\left[
E^{(\delta)}_{\bx,\bbk,\pi}
\left[N_{(v\wedge \tau_\delta(\pi))\vee t}(G),A_n\,\right],
\tau_\delta(\pi)\le t\right]
+
E^{(\delta)}_{\bx,\bbk}\left[E^{(\delta)}_{\bx,\bbk,\pi}
\left[N_{(v\wedge \tau_\delta(\pi))\vee t}(G),A_n\,\right],\,
\tau_\delta(\pi)> t\right].\nonumber
\end{eqnarray}
The first term on the utmost right hand side of \eqref{80304} equals
$E^{(\delta)}_{\bx,\bbk}\left[N_{t}(G),A_n,\,
\tau_\delta\le t\right]$, while the second one equals
$
\tilde E^{(\delta)}_{\bx,\bbk}\left[N_{(v\wedge \tau_\delta)\vee t}(G),B\right].
$ Here $B:=A_n\cap[\tau_\delta> t]$ is an ${\cal M}_t$--measurable
event.  Suppose that $\ga_0'\in(\ga_0+1/2,1)$ and let
$L:=[\delta^{-\ga_0'}]$ be yet another mesh size parameter.
\commentout{

DO NOT REMOVE; THE EXPLANATION TO THE ABOVE!!!

Note  that when $ \tau_\delta(\pi)>t$
we have
$$
E^{(\delta)}_{\bx,\bbk,\pi}\left[N_{(v\wedge \tau_\delta(\pi))\vee t}(G),A_n\,\right]=
E^{(\delta)}_{\bx,\bbk,\pi}\left[N_{v\wedge \tau_\delta(\pi)}(G),A_n\,\right].
$$
Hence the second term on the utmost right hand side of \eqref{80304} equals
$$
\tilde E^{(\delta)}_{\bx,\bbk}\left[N_{v\wedge \tau_\delta}(G),A_n,\,\tau_\delta> t\right]=
\tilde E^{(\delta)}_{\bx,\bbk}\left[N_{(v\wedge \tau_\delta)\vee t}(G),A_n,\,\tau_\delta> t\right].
$$

}
We define
\[
\sigma:=L^{-1}[([L(v\wedge \tau_\delta)]+2)\vee ([Lt]+2)]
\]
and note that
\begin{equation}\label{80201}
\tilde E^{(\delta)}_{\bx,\bbk}\left[N_{\sigma}(G),\,B\right]=
\sum\limits_{p=[Lt]+2}^{[Lv]+2}\tilde E^{(\delta)}_{\bx,\bbk}\left[N_{p/L}(G),\,B,\,
\sigma=\frac{p}{L}\right]
\end{equation}
Representing the event $[\sigma=p/L]$ as the difference of $[\sigma\ge
p/L]$ and $[\sigma\ge(p+1)/L]$ (note that
$[\sigma\ge([Lv]+3)/L]=\emptyset$) and grouping the terms of the sum
that correspond to the same index $p$ we obtain that the right hand
side of
\eqref{80201} equals
\begin{equation}\label{80201b}
\tilde E^{(\delta)}_{\bx,\bbk}\left[N_{([Lt]+2)/L}(G),\,B\,\right]+
\sum\limits_{p=[Lt]+2}^{[Lv]+2}\tilde E^{(\delta)}_{\bx,\bbk}\left[N_{p+1/L}(G)-N_{p/L}(G),\,B,\,
\sigma\ge\frac{p+1}{L}\right].
\end{equation}
Since the event $B\cap[\sigma\ge (p+1)/L]$ is ${\cal
M}^{(p-1)/L}$-measurable, from Lemma \ref{lmA1} we conclude that the
absolute value of each term appearing under the summation sign in
\eqref{80201b} can be estimated by $C'\|G\|_{1,1,3}
\delta^{\gamma_1'}L^{-1}\tilde Q^{(\delta)}_{\bx,\bbk}[B]$ which implies
\[
\left|\tilde E^{(\delta)}_{\bx,\bbk}\left[N_{\sigma}(G),\,B\right]-
\tilde E^{(\delta)}_{\bx,\bbk}\left[N_{([Lt]+2)L^{-1}}(G),\,B\right]\right|\le
C'
\delta^{\gamma_1'}\|G\|_{1,1,3}^{[T_*,T]}T^2\,\tilde Q^{(\delta)}_{\bx,\bbk}[B]\,\frac{[Lv]+1-[Lt]}{L}.
\]
A direct calculation using formulas \eqref{eq1b} allows us to conclude
also that both $|N_{\sigma}(G)-N_{(v\wedge \tau_\delta)\vee t}(G)|$
and $|N_{([Lt]+2)L^{-1}}(G)-N_t(G)|$ are estimated by $C\|G\|_{1,1,3}^{[T_*,T]}\delta^{\ga_0'-1/2}$.
Hence, (since $\ga_0'>1/2+\ga_0$)
\begin{eqnarray}
&&\left|\tilde E^{(\delta)}_{\bx,\bbk}\left[N_{(v\wedge \tau_\delta)\vee t}(G),\,B\right]-
\tilde E^{(\delta)}_{\bx,\bbk}\left[N_{t}(G),\,B\right]\right|\le
 \left|\tilde E^{(\delta)}_{\bx,\bbk}\left[
N_{\sigma}(G)-N_{(v\wedge \tau_\delta)\vee t}(G),\,B\right]\right|
\nonumber\\
&&+
\left|\tilde E^{(\delta)}_{\bx,\bbk}\left[N_{\sigma}(G),\,B\right]-
\tilde E^{(\delta)}_{\bx,\bbk}\left[N_{([Lt]+2)L^{-1}}(G),\,B\right]\right|+
\left|\tilde E^{(\delta)}_{\bx,\bbk}\left[N_{([Lt]+2)L^{-1}}(G)-N_t,\,B\right]\right|\nonumber\\
&&\le C\delta^{\gamma_1}\|G\|_{1,1,3}^{[T_*,T]}T^2\tilde Q^{(\delta)}_{\bx,\bbk}[B]
\left(v-t\right)\vee \delta^{\gamma_0}\label{bliams-dec7}
\end{eqnarray}
for a certain constant $C>0$ and $\ga_1:=\min[\ga_0'-\ga_0-1/2,\ga_1']$.
From \eqref{80304}, \eqref{bliams-dec7} and the observation just below (\ref{80304}), we obtain
\[
\left|E^{(\delta)}_{\bx,\bbk}[N_v(G)-N_t(G),A_n]\right|\le C\delta^{\gamma_1}
\|G\|_{1,1,3}^{[T_*,T]}T^2R^{(\delta)}_{\bx,\bbk}[A_n]
\left(v-t\right)\vee \delta^{\gamma_0}
\]
for a certain constant $C>0$ and the conclusion of Proposition \ref{prop706041} follows. $\Box$

\subsection{An estimate of the stopping time}

The purpose of this section is to prove the following estimate for
$R^{(\delta)}_{\bbx,\bbk}\left[\tau_{\delta}< T\right]$.
\begin{theorem}
Assume that the dimension $d\ge3$.
Then, one can choose $\ep_1,\ep_2,\ep_3,\ep_4$ in such a way that
there exist  constants $C,\ga>0$ for which
\label{cor10211}
\begin{equation}
\label{92601b}
R^{(\delta)}_{\bx,\bk}\left[  \,\tau_{\delta}< T\right]\le
C\delta^{\gamma}T,\quad\forall\,\delta\in(0,1],\,T\ge1,\,
(\bx,\bk)\in{\cal A}(M).
\end{equation}
\end{theorem}
{\bf Proof.}  We obviously have
\begin{equation}\label{70701}
\left[  \,\tau_{\delta}< T\right]=\left[  \,U_{\delta}\le \tau_{\delta},\,U_{\delta}< T\right]
\cup\left[  \,S_{\delta}\le \tau_{\delta},\,S_{\delta}< T\right]
\end{equation}
with the stopping times $S_\delta$ and $U_\delta$ defined in (\ref{Sdelta}) and
(\ref{Udelta}).
Let us denote the first and  second event appearing on the right hand side of
\eqref{70701} by $A(\delta)$ and $B(\delta)$ respectively.
To show that \eqref{70701} holds we prove that the
$R^{(\delta)}_{\bx,\bk}$ probabilities of both events can be estimated
by $C\delta^{\gamma}T$ for some $C,\ga>0$: see (\ref{Adelta-estimate}), (\ref{bliams-dec71})
and (\ref{72501}).

\subsubsection{An estimate of $R^{(\delta)}_{\bx,\bk}[A(\delta)]$}

The first step towards obtaining the desired estimate will be to
replace the event $A(\delta)$ whose definition involves a stopping
time by an event $C(\delta)$ whose definition depends only on
deterministic times, see \eqref{91804} below. Next we use the estimate
\eqref{73101} of Proposition \ref{prop706041} for an appropriately
chosen function $G$ to reduce the question of bounding the
$R^{(\delta)}_{\bx,\bk}$ probability of $\tilde A(\delta)$ by an
easier problem of estimating its $\mathfrak Q_{\bx,\bk}$ probability
($\mathfrak Q_{\bx,\bk}$ corresponds to a degenerate diffusion
determined by \eqref{61102b}). The latter is achieved by using bounds
on heat kernels corresponding to hypoelliptic diffusions due to
Kusuoka and Stroock.

We assume in this section to simplify the notation and without any loss of generality that $h^*(M)=1$.
Note that then
\begin{equation}\label{91804}
A(\delta)\subset \tilde A(\delta):=\left[\left|X\left(\frac{j}{q}\right)-X\left(\frac{i}{q}\right)\right|
\le \frac{3}{q}:
\,1\le i\le j\le[Tq],\quad|i-j|\ge \frac{q}{p}\right]
\end{equation}
and thus
\begin{equation}\label{71203}
R^{(\delta)}_{\bx,\bk}[A(\delta)]\le
[Tq]^2
\max \left\{R^{(\delta)}_{\bx,\bk}\left[\left|X\left(\frac{j}{q}\right)-X\left(\frac{i}{q}\right)\right|
\le \frac{3}{q}\right]:
\,1\le i\le j\le[Tq],~|i-j|\ge \frac{q}{p}\right\}.
\end{equation}
Suppose that $f^{(\delta)}:\R^d\to[0,1]$ is a $C^\infty$--regular
function that satisfies $f(\bx)=1$, if $|\bx|\le 4h/q$ and
$f^{(\delta)}(\bx)=0$, if $|\bx|\ge 5/q$. We assume furthermore
that $i,j$ are positive integers such that $(j-i)/q\in[0,1]$ and
$\|f^{(\delta)}\|_{3} \le 2q^3.  $ For any $\bx_0\in\bbR$ and $i/q\le
t\le j/q$ define
\[
G_j(t,\bx,\bk;\bx_0):=\mathfrak
M_{\bx,\bk}f^{(\delta)}\left(X\left(\frac{j}{q}-t\right)-\bx_0\right).
\]
\commentout{

$$
=\mathop{\int\!\int}\limits_{\R^{d}\times {\mathbb S}^{d-1}_k}f^{(\delta)}\left(\bby-\bx_0\right)
p\left(\frac{j}{q}-t,\bby-\bx,\bk,\bbl\right)\,d\by S_{d-1}(d\bbl),
$$
where $S_{d-1}(d\bbl)$ is the surface element on the sphere ${\mathbb S}^{d-1}_k$
and $(t,\bx,\bk,\bby,\bbl)\mapsto p^{(\delta)}
\left(t,\bby-\bx,\bk,\bbl\right)$ is the transition probability density with respect to $m_d\times S_{d-1}$
of the path measure $\mathfrak Q_{\bx,\bbk}$.
Here $m_d$ is the $d$--dimensional Lebesgue measure.
The existence of the transition of probability density in question follows from the fact that
corresponding
diffusion has a hypoelliptic generator $\tilde{\cal L}$, see pp 122 and 123 of \cite{bakoryz}.
We conclude therefore that $p
\left(\cdot,\cdot,\cdot,\cdot\right)\in C^\infty((0,+\infty)\times \R^d\times \mathbb S^{d-1}\times \mathbb S^{d-1})$.
Note that $....$

}
Obviously, we have
\[
\partial_t G_j(t,\bx,\bk;\bx_0)+\tilde {\cal L}G_j(t,\bx,\bk;\bx_0)=0.
\]
Hence, using Proposition \ref{prop706041} with $v=j/q$ and $t=i/q$
(note that $v-t\ge 1/p\ge\delta^{\ep_2}$ and $\ep_2\in(0,1/2)$), we
obtain that there exists $\ga_1>0$ such that
\begin{eqnarray}\label{10703}
&&\left|E^{(\delta)}_{\bx,\bk}\left[ f^{(\delta)}\left(X\left(\frac{j}{q}\right)-\bx_0\right)-
G_j\left(\frac{i}{q},X\left(\frac{i}{q}\right),K\left(\frac{i}{q}\right);\bx_0\right)
 \left|\vphantom{\int_0^1}\right.{\cal M}^{i/q}\right]\right|\\
&&\leq
C\,\frac{j-i}{q}\,\|G_j(\cdot,\cdot,\cdot;\bx_0)\|_{1,1,3}^{[i/q,j/q]}T^2\delta^{\ga_1}
,\quad\forall\,\delta\in(0,1].\nonumber
\end{eqnarray}
According to \cite{stroock} Theorem 2.58, p. 53 we have
\begin{equation}\label{100702}
\|G_j(\cdot,\cdot,\cdot;\bx_0)\|_{1,1,3}^{[i/q,j/q]}\le C\|f^{(\delta)}\|_{3}\le
Cq^3\le C\delta^{-3(\ep_2+\ep_3)},\quad\,j\in\{0,\ldots,[qT]\}.
\end{equation}
Hence combining \eqref{10703} and \eqref{100702} we obtain that the left hand side of
\eqref{10703} is less than, or equal to
$C\,\delta^{\ga_1-3(\ep_2+\ep_3)}$ for all $\delta\in(0,1].$ Let now
$i_0=j-\dfrac{q}{p}$ so that $1\le i\le i_0\le j\le[Tq]$. We have
\begin{eqnarray}\label{71201}
&&R^{(\delta)}_{\bx,\bk}\left[\left|X\left(\frac{j}{q}\right)-X\left(\frac{i}{q}\right)\right|\le
\frac{3}{q}\right]\le
E^{(\delta)}_{\bx,\bk}
\left[ f^{(\delta)}\left(X\left(\frac{j}{q}\right)-X\left(\frac{i}{q}\right)\right)\right]
\\
&&~~~~~
=E^{(\delta)}_{\bx,\bk}\left[E^{(\delta)}_{\bx,\bk}\left[ f^{(\delta)}\left(X\left(\frac{j}{q}\right)-
\bby\right)
\left|\vphantom{\int_0^1}\right.{\cal M}^{i_0/q}\right]\mathop{\vphantom{\int}}\limits_{\left|
\vphantom{\int_{\int}^{\int}}\right.\bby=X\left(i/q\right)}
\right].\nonumber
\end{eqnarray}
According to \eqref{10703} and \eqref{100702} we can estimate the
utmost right hand side of \eqref{71201} by
\begin{eqnarray}
&&\sup_{\bx,\bby,\bk}\,\left\{\mathfrak M_{\bx,\bk}f^{(\delta)}
\left(X\left(\frac{1}{p}\right)-\bby\right):
\bx,\bby\in\R^d, \,\bk\in A(2M) \right\}
\label{71203bb}
+C\,\delta^{\ga_1-3(\ep_2+\ep_3)}T^2.
\end{eqnarray}
To estimate the first term in \eqref{71203bb} we use
the following.
\begin{lemma}
\label{lm70901}
Let $p,q$ be as in \eqref{102302}. Then, there exist positive constants $C_1$, $C_2$ and $C_3$
such that for all $\bx,\bby\in\R^d$,
$\bk\in A(2M)$, $j\in\{1,\ldots,[p T]\}$, $\delta\in(0,1]$ we have
\begin{equation}\label{70901}
\mathfrak Q_{\bx,\bk}\left[\left|X\left(\frac{j}{p}\right)-
\bby\right|\le \frac{5}{q}\right]\le C_1\left(\frac{p^{\,C_2}}{q^d}+e^{-C_3p}\right).
\end{equation}
\end{lemma}
We postpone the proof of the lemma for a moment in order to finish the
estimate of $R^{(\delta)}_{\bx,\bk}[A(\delta)]$.  Using \eqref{70901}
we obtain that the expression in \eqref{71203bb} can be estimated by
\[
C_1\left(\frac{p^{\,C_2}}{q^d}+e^{-C_3p}\right)+C\,\delta^{\ga_1-3(\ep_2+\ep_3)}T^2\le
C_1\delta^{(d-C_2)\ep_2+d\ep_3}+\exp\left\{-C_3\delta^{-\ep_2}\right\}+C\,\delta^{\ga_1-3(\ep_2+\ep_3)}T^2.
\]
Hence, from \eqref{71203}, we obtain that
\begin{eqnarray}\label{Adelta-estimate}
&&R^{(\delta)}_{\bx,\bk}[A(\delta)]\le
[Tq]^2\left(C_1\delta^{(d-C_2)\ep_2+d\ep_3}+\exp\left\{-C_3\delta^{-\ep_2}\right\}+
C\,\delta^{\ga_1-3(\ep_2+\ep_3)}T^2\right)\\
&&
\le CT^2\left(\delta^{(d-2-C_1)\ep_2+(d-2)\ep_3}+\delta^{-2(\ep_2+\ep_3)}\exp\left\{-C_3\delta^{-\ep_2}\right\}+\delta^{\ga_1-5(\ep_2+\ep_3)}T^2\right)
\le C\delta^{\ga_{2}}T^4\nonumber
\end{eqnarray}
for  $\ga_{2}:=\min[(d-2-C_1)\ep_2+(d-2)\ep_3,\ga_1-5(\ep_2+\ep_3)]>0$, provided that
$\ep_2+\ep_3<\ga_1/5$ and $\ep_2\in(0,(d-2)\ep_3/(C_1+2-d))$. Here with no loss
of generality we have assumed that $C_1+2>d$. Recall also that $d\ge3$.
Now suppose that $\ga_3\in(0,\ga_2)$. Consider two cases: $T^3<\delta^{-\ga_3}$
and $T^3\ge\delta^{-\ga_3}$. In the first one, the utmost right hand side of \eqref{Adelta-estimate}
can be bound from above by $C\delta^{\ga_{2}-\ga_3}T$. In the second we have
a trivial bound of the left side by $\delta^{\ga_3/3}T$. We have proved therefore that
\begin{equation}\label{bliams-dec71}
R^{(\delta)}_{\bx,\bk}[A(\delta)]\le C\delta^{\ga}T
\end{equation}
for some $C,\ga>0$
independent of $\delta$ and $T$.

{\bf The proof of Lemma $\ref{lm70901}$.} We prove this lemma by
induction on $j$. First, we verify it for $j=1$.  Without any loss of
generality we may suppose that $\bk=(k_1,\ldots,k_d)$ and
$k_d>(4dM_\delta)^{-1}$.  Let $\tilde{D}_{mn}:\R^{d-1}\rightarrow\R$,
$m,n=1,\dots,d-1$, $\tilde{E}_{m}:\R^{d-1}\rightarrow\R$,
$m=1,\dots,d$ be given by
\[
\tilde{D}_{pq}(\bbl):=D_{pq}(k^{-1}\bbl,k^{-1}\sqrt{k^2-l^2},k),\quad
\tilde{E}_p(\bbl):=E_p(k^{-1}\bbl,k^{-1}\sqrt{k^2-l^2},k),
\]
when $\bbl\in Z:=[\bbl\in\mathbb
B^{d-1}_k:\,k^{-1}\sqrt{k^2-l^2}>(4dM_\delta)^{-1}]$, $l=|\bbl|$.
These functions are $C^\infty$ smooth and bounded together with all
their derivatives.  Note also that the matrix ${\bf
\tilde{D}}=[\tilde{D}_{mn}]$ is symmetric and ${\bf
\tilde{D}}\xi\cdot\xi\ge \la_0|\xi|^2$ for all $\xi\in\R^{d-1}$ and a
certain $\la_0>0$. The projection
$K(t)=\left(K_1(t),\ldots,K_d(t)\right)$ of the canonical path process
$(X(t;\pi),K(t;\pi))$ considered over the probability space $({\cal
C},{\cal M},\mathfrak Q_{\bx,\bk_0})$, where
$\bk_0:=(\bbl,\sqrt{k^2-l^2})$, with $\bbl\in Z$, is a diffusion whose
generator equals ${\cal L}$, see \eqref{61102}.  It can be easily seen
that $\left(K_1(t),\ldots,K_{d-1}(t)\right)_{t\ge0}$, is then a
diffusion starting at $\bbl$, whose generator ${\cal N}$ is of the
form
\begin{equation}\label{061130}
{\cal N}
F(\bbl,\bbx):=
\sum\limits_{p=1}^{d-1}X_p^2F(\bbl)+
\sum\limits_{q=1}^{d-1}a_q(\bbl)\partial_{l_q}F(\bbl),\quad F\in
C^\infty_0(\R^{d-1}),
\end{equation}
where $a_q(\bbl)$, $q=1,\ldots,d-1$ are certain $C^\infty$-functions and
\[
X_p(\bbl):=\sum\limits_{q=1}^{d-1}\tilde
D_{pq}^{1/2}(\bbl)\partial_{l_q},\quad p=1,\ldots,d-1.
\]
The $(d-1)\times(d-1)$ matrix $[\tilde D_{pq}^{1/2}(\bbl)]$ is non-degenerate  when $\bbl\in Z$.
Let
\[
\tilde{\cal N}
F(\bbl,\bbx):=
\sum\limits_{p=1}^{d-1}\tilde X_p^2F(\bbl,\bbx)+\tilde X_0F(\bbl,\bbx),\quad F\in
C^\infty_0(\R^{d-1}\times\R^d),
\]
where $\tilde X_0$ is a $C^\infty$--smooth extension of the field
\begin{equation*}
X_0(\bbl):=\frac{H_0'(k)}{k}\,\sum\limits_{q=1}^{d-1}l_q\partial_{x_q}+
\frac{H_0'(k)}{k}\,\sqrt{k^2-l^2}\,\partial_{x_d}+
\sum\limits_{q=1}^{d-1}a_q(\bbl)\partial_{l_q},\quad\bbl\in Z.
\end{equation*}
It can  be shown, by the same type of argument as that given on pp. 122-123 of
\cite{bakoryz}, that for each $(\bx,\bbl)$, with  $\bbl\in Z$, the linear space
spanned at that point  by the fields
belonging to the Lie algebra generated by $X_0,\ldots,X_{d-1}$ is of dimension $2d-1$.
One can also guarantee that the extensions $\tilde X_0,\ldots,\tilde X_{d-1}$ satisfy the same condition.
We shall  denote  the respective extension  of ${\cal N}$ by the same symbol.

Set $\bbl_0:=(k_1,\ldots,k_{d-1})$.  Let ${\cal R}_{\bbl_0}$,
$\tilde{\cal R}_{\bbx,\bbl_0}$ be the path measures supported on
${\cal C}^{d-1}$ and ${\cal C}^{d,d-1}$ respectively that solve the
martingale problems corresponding to the generators $\cal N$ and
$\tilde{\cal N}$ with the respective initial conditions at $t=0$ given
by $\bbl_0$ and $(\bbx,\bbl_0)$.  Let $r(t,\bx-\bby,\bbl_1,\bbl_2)$,
$t\in(0,+\infty)$, $\bx,\bby\in \R^d$, $\bbl_1,\bbl_2\in\R^{d-1}$ be
the transition of probability density that corresponds to $\tilde{\cal
R}_{\bbx,\bbl_0}$.  Using Corollary 3.25 p. 22 of \cite{kustr} we have
that for some constants $C,m>0$
\begin{equation}\label{80701}
r\left(t,\bby,\bk,\bbl\right)\le C t^{-m},\quad\forall\,\bby\in\R^d,\,\bk,\,
\bbl\in\mathbb R^{d-1},\,t\in(0,1].
\end{equation}
Denote by $\tau_Z(\pi)$ the exit time
of a path  $\pi\in {\cal C}^{d-1}$  from the set $Z$.
For any $\pi\in {\cal C}^{d,d-1}$ we set also $\tilde{\tau}_Z(\pi)=\tau_Z(K(\cdot;\pi))$.
Let $S:\mathbb B_k^{d-1}\rightarrow \mathbb S_k^{d-1}$ be given by
\[
S(\bbl):=(l_1,\ldots,l_{d-1},\sqrt{k^2-l^2}),\quad \bbl=(l_1,\ldots,l_{d-1})\in
\mathbb B_k^{d-1},\,l:=|\bbl|
\]
and let $\tilde S:{\cal C}^{d,d-1}\to {\cal C}$ be given by $\tilde
S(\pi)(t):=(X(t;\pi),S\circ K(t;\pi))$, $t\ge0$.  For any $A\in{\cal
M}^{\tilde{\tau}_Z}$ we have $\tilde{\cal R}_{\bbx,\bbl_0}[\tilde
S^{-1}(A)]={\mathfrak Q}_{\bbx,S(\bbl_0)}[A]$.  Since the event
$[\left|X\left(1/p\right)-\bby\right|\le 5/q]\cap [\tilde{\tau}_Z\ge
1/p]$ is ${\cal M}^{\tilde{\tau}_Z}$--measurable we have
\begin{eqnarray}\label{100701}
&&\mathfrak Q_{\bx,\bk}\left[\left|X\left(\frac{1}{p}\right)-\bby\right|\le \frac{5}{q}\right]\le
\tilde{\cal R}_{\bx,\bbl_0}
\left[\left|X\left(\frac{1}{p}\right)-\bby\right|\le \frac{5}{q},\tilde{\tau}_Z\ge \frac{1}{p}\right]+
{\cal R}_{\bbl_0}\left[\tau_Z< \frac{1}{p}\right]\\
&&~~~~~~~~~~~~~~~~~~~~~~~~~~~~~~~~~~
\le C\bar{\om}_d p^{m}\left(\frac{4}{q}\right)^d+Ce^{-C_3p}.\nonumber
\end{eqnarray}
Here $\bar{\om}_d$ denotes the volume of $\mathbb B^d$.  To obtain the
last inequality we have used \eqref{80701} and an estimate for
non-degenerate diffusions stating that ${\cal R}_{\bbl_0}\left[\tau_Z<
1/p\right]<Ce^{-C_3p}$ for some constants $C,C_3>0$ depending only on
$d$ and $\la_0$, see e.g. (2.1) p. 87 of \cite{stroock-varadhan}.
Inequality \eqref{100701} implies easily \eqref{70901} for $j=1$ with
$C_1=m$. To finish the induction argument assume that
\eqref{70901} holds for a certain $j$. We
show that it holds for $j+1$ with the same constants $C_1,C_2$ and $C_3>0$.
The latter follows easily from the Chapman-Kolmogorov equation, since
\begin{eqnarray*}
&&\mathfrak Q_{\bx,\bk}\left[\left|X\left(\frac{j+1}{p}\right)-\bby\right|\le \frac{5}{q}\right]=
\mathop{\int\!\int}\limits_{\R^d\times\mathbb S^{d-1}_k}
\mathfrak Q_{\bby,\bbl}\left[\left|X\left(\frac{j}{p}\right)-\bby\right|\le \frac{5}{q}\right]
Q\left(\frac{1}{p},\bx,\bk,d\bby,d\bbl\right)\\
&&\stackrel{\mbox{\tiny induction assumpt.}}{\le} C_1 \left[\frac{p^{C_2}}{q^d}+e^{-C_3p}\right]
\mathop{\int\!\int}Q\left(\frac{1}{p},\bx,\bk,d\bby,d\bbl\right)=C_1 \left[\frac{p^{C_2}}{q^d}+
e^{-C_3p}\right]
\end{eqnarray*}
and the formula \eqref{70901} for $j+1$ follows. Here $Q(t,\bx,\bk,\cdot,\cdot)$ is
the transition of probability  corresponding to the path measure $\mathfrak Q_{\bx,\bk}$. $\Box$

\subsubsection{An estimate of $R^{(\delta)}_{\bx,\bk}[B(\delta)]$}

We start with a simple observation concerning the H\"older regularity
of the $K$ component of any path $\pi\in B(\delta)$.  Let us denote
$\rho:=2M_\delta^{-1} N^{-1/2}$ and
\[
D:=\left[\,\pi\in {\cal C}(T,\delta):
|K(t)-K(s)|\ge \rho
\mbox{  for some $k$ s.t. $t_{k}^{(p)}\le T$}\right.
\left.\vphantom{\int_0^1}\mbox{ and  }t\in [t_{k}^{(p)},t_{k+1}^{(p)}],\,s\in
 [t_{k-1}^{(p)},t_{k}^{(p)}]\,\right],
\]
where $M_\delta$ has been defined in \eqref{073104} and $N$ in (\ref{102302}).
Suppose that $\pi\in B(\delta)$, then
we can find $t\in [t_{k}^{(p)},t_{k+1}^{(p)}]$, $s\in [t_{k-1}^{(p)},t_{k}^{(p)}]$ for which
$\hat K(t)\cdot\hat K(s)\le1-1/N$. This, however, implies that
\[
|K(t)-K(s)|^2\ge \frac{1}{M^2_\delta}\,|\hat K(t)-\hat K(s)|^2\ge\frac{2}{M^2_\delta N},
\]
thus $\pi\in D$.  Hence the desired estimate of
$R^{(\delta)}_{\bx,\bk}[B(\delta)]$ follows from the following lemma.
\begin{lemma}
\label{lm806041}
Under the assumptions of Theorem $\ref{cor10211}$
 there exist $C,\ga>0$  such that
\begin{equation}\label{72501}
R^{(\delta)}_{\bx,\bk}[D]\le CT\delta^{\ga},\quad\forall\,
\delta\in(0,1],\,T\ge1,\,(\bx,\bk)\in{\cal A}(M).
\end{equation}
\end{lemma}
{\bf Proof.} We define the following events:
\begin{eqnarray*}
&&F_1:=\left[\,
|K(t)-K(s)|\ge \rho\quad
\mbox{  for some  }s,t\in [0,T],\,0<t-s<\frac{2}{p},\,t\le \tau_\delta\,\right],
\\
&&
F_2:=\left[\,
|K(t)-K(s)|\ge  \rho\quad
\mbox{  for some  }s,t\in [0,T],\,0<t-s<\frac{2}{p},\,s\ge \tau_\delta\,\right],
\\
&&
F_3:=\left[\,
|K(\tau_\delta)-K(s)|\ge  \frac{\rho}{2}\quad
\mbox{  for some  }s\in [0,T],\,0<\tau_\delta-s<\frac{2}{p},\, \tau_\delta\le T\,\right],
\\
&&
F_4:=\left[\,
|K(\tau_\delta)-K(t)|\ge  \frac{\rho}{2}\quad
\mbox{  for some  }t\in [0,T],\,0<t-\tau_\delta<\frac{2}{p}\,\right].
\end{eqnarray*}
Observe that $D\subset\bigcup\limits_{i=1}^4 F_i$.
Note that $F_1,F_3$ are ${\cal M}^{\tau_\delta}$--measurable, hence
\begin{equation}\label{073103}
R^{(\delta)}_{\bx,\bk}[F_i]=\tilde
Q^{(\delta)}_{\bx,\bk}[F_i],\quad i=1,3.
\end{equation}
On the other hand for $i=2,4$ we have
\[
R^{(\delta)}_{\bx,\bk}[F_i]=
\int\mathfrak Q_{X(\tau_\delta(\pi)),K(\tau_\delta(\pi))}[F_{i,\pi}] \tilde Q^{(\delta)}_{\bx,\bk}(d\pi),
\]
where for a given $\pi\in{\cal C}$
\begin{eqnarray*}
&&F_{2,\pi}:=\left[\,
|K(t)-K(s)|\ge  \rho\quad
\mbox{  for some  }s,t\in [0,(T-\tau_\delta(\pi))\wedge0],\,0<t-s<\frac{2}{p}\,\right],
\\
&&
F_{4,\pi}:=\left[\,
|K(0)-K(t)|\ge  \frac{\rho}{2}\quad
\mbox{  for some  }t\in [0,(T-\tau_\delta(\pi))\wedge0],\,0<t<\frac{2}{p}\,\right].
\end{eqnarray*}
Since all $F_i$, $i=1,3$ and $F_{i,\pi}$, $i=2,4$, $\pi\in{\cal C}$ are contained in the event
\[
F:=\left[\,
|K(t)-K(s)|\ge   \frac{\rho}{2}\quad
\mbox{  for some  }s,t\in [0,T],\,0<t-s<\frac{2}{p}\,\right],
\]
\eqref{72501} would follow if we show that
there exist $C>0$ and $\ga>0$ for which
\begin{equation}\label{72501b}
    \tilde Q^{(\delta)}_{\bx,\bk}[F]\le CT\delta^{\ga}
\mbox{  for all $(\bx,\bbk)\in {\cal A}(M)$}
\end{equation}
and
\begin{equation}\label{72501c}
    \mathfrak Q_{\bx,\bk}[F]\le CT\delta^{\gamma}
\mbox{  for all $(\bx,\bbk)\in {\cal A}(M_\delta)$}.
\end{equation}
The estimate \eqref{72501c} follows from elementary properties of
diffusions, see e.g. (2.46) p. 47 of \cite{stroock}.  We carry on with
the proof of \eqref{72501b}. The argument is analogous to  the proof of
Theorem 1.4.6 of \cite{stroock-varadhan}.  Let $L$ be a multiple of
$p$ such that $L:=[\delta^{-\ga_0'}]$, where $\ga_0'\in(1/2,1)$ is to be
specified even further later on. Let also $s_k^{(L)}:=k/L$,
$k=0,1,\ldots$.  We now define the stopping times $\tau_k(\pi)$ that
determine the times at which the $K$ component of the path $\pi$
performs $k$--th oscillation of size $\rho/8$.  Let $\tau_0(\pi):=0$
and for any $k\ge0$
\[
\tau_{k+1}(\pi):=\inf\left[s_k^{(L)}\ge \tau_k(\pi):\,
|K(s_k^{(L)})-K(\tau_k(\pi))|\ge\frac{\rho}{8}\right],
\]
with the convention that $\tau_{n+1}=+\infty$ when $\tau_n=+\infty$, or
when the respective event is impossible.  Let
$N_\#:=\min[n:\,\tau_{n+1}>T]$ and
$\delta^*:=\min[\tau_{n}-\tau_{n-1}:n=1,\ldots,N_\#]$.  Then, for a
sufficiently small $\delta_0$ and $\delta\in(0,\delta_0)$ we have
$F\subset [\delta^*\le 1/p]$ so we only need to estimate $\tilde
Q^{(\delta)}_{\bx,\bk}$ probability of the latter event.

\commentout{

       Indeed, this can be seen as follows. Suppose that  $\pi\in F$ and $\delta^*>
       1/p$. Let $0<t-s\le 1/p$ be such that $|K(t)-K(s)|\ge \rho/2$.
       Then  for some $n$ we have either $t,s\in [\tau_n,\tau_{n+1})$, or $s\in [\tau_n,\tau_{n+1})$
       and $t\in [\tau_{n+1},\tau_{n+2})$. In the first case  for all times of
       the form $p/L\in[\tau_n,\tau_{n+1})$ we have $|K(p/L)-K(\tau_n)|\le \rho/8$ and for some $p_1$, $p_2$
       we have $|K(t)-K(p_1/L)|\le \delta^{-1/2}D_1M_\delta/L$, $|K(s)-K(p_2/L)|\le \delta^{-1/2}D_1M_\delta/L$
       hence
       $$
       |K(t)-K(s)|\le |K(t)-K(p_1/L)|+|K(s)-K(p_2/L)|+|K(\tau_n)-K(p_1/L)|+
       |K(\tau_n)-K(p_2/L)|
       $$
       $$
       \le 2\delta^{-1/2}D_1M_\delta/L+2\rho/4<\rho/2.
       $$
In the other case we have
$$
       |K(t)-K(s)|\le |K(t)-K(p_1/L)|+|K(s)-K(p_2/L)|+|K(\tau_{n+1})-K(p_1/L)|+
       |K(\tau_{n})-K(p_2/L)|
       $$
       $$
       \le 2\delta^{-1/2}D_1M_\delta/L+2\rho/4<\rho/2.
       $$

\begin{equation}\label{72503}
      \tilde Q^{(\delta)}_{\bx,\bk}[F]\le \tilde Q^{(\delta)}_{\bx,\bk}\left[\delta^*\left(\frac{\rho}{2}\right)\le
       \frac{1}{p}\right].
\end{equation}

}
Let $f:\R^d\to[0,1]$ be a function
of $C^\infty_0(\R^d)$ class such that $f(\bze)\equiv1$, when $|\bk|\le
\rho/16$ and $f(\bk)\equiv0$, when $|\bk|\ge \rho/8$.  Let also
$f_\bbl(\cdot):=f(\cdot-\bbl)$ for any $\bbl\in\R^d$.  Note that
according to Lemma \ref{lmA1} we can choose constants
$A_\rho,\,C>0$, where $C$ is independent of $\rho$, in such
a way that $A_\rho<CT^2\rho^{-3}$ and the random sequence
\begin{equation}\label{72507}
S_N^\bbl:=\tilde E^{(\delta)}_{\bx,\bk}\left[f_\bbl\left(K\left(\frac{N+1}{L}\right)\right)
\left|\vphantom{\int_0^1}\right.\,{\cal M}^{N/L}\,\right]+A_\rho \frac{N}{L}, \quad N\ge0
\end{equation}
 is a $\tilde Q^{(\delta)}_{\bx,\bk}$--submartingale with respect to
 the filtration $\left({\cal M}^{N/L}\right)_{N\ge0}$ for all $\bbl$
 with $|\bbl|\in ((3M_\delta)^{-1},3M_\delta)$ provided that $\delta$
 is sufficiently small.
\commentout{

 Note that
from \eqref{eq1b} and the definition of the path measure $\tilde Q^{(\delta)}_{\bx,\bk}$
 then with $\tilde Q^{(\delta)}_{\bx,\bk}$--almost surely $|K(t)|\in ((3M)^{-1},3M)$, $t\ge0$,
 provided that $\bk\in A(M)$.

 }
We can decompose
\begin{eqnarray}\label{72504}
&&\tilde Q^{(\delta)}_{\bx,\bk}\left[\delta^*\le \frac{2}{p}\right]\le
\tilde Q^{(\delta)}_{\bx,\bk}\left[\delta^*\le  \frac{2}{p},\,N_\#\le [\delta^{-\al}]\,\right]+
\tilde Q^{(\delta)}_{\bx,\bk}\left[\delta^*\le \frac{2}{p}, \,N_\#> [\delta^{-\al}]\,\right]
\\
&&~~~~~~~~~~~~~~~~~~~\le \sum\limits_{i=1}^{[\delta^{-\al}]}\tilde Q^{(\delta)}_{\bx,\bk}
\left[\tau_{i}-\tau_{i-1}\le \frac{2}{p}\right]+
\tilde Q^{(\delta)}_{\bx,\bk}[N_\#> [\delta^{-\al}]],\nonumber
\end{eqnarray}
where $\al>0$ is to be determined later.  We will show that
\begin{equation}\label{bliams-dec72}
\tilde Q^{(\delta)}_{\bx,\bk}[N_\#> [\delta^{-\al}]]\le
C e^T\left(1-\frac{\delta^{1/2(\ep_1+\ep_2)}}{2}\right)^{\delta^{-\al}}
\end{equation}
and
\begin{equation}\label{bliams-dec73}
\tilde Q^{(\delta)}_{\bx,\bk}\left[\tau_{n+1}-\tau_{n}\le
[L\delta^\ep_2]/L\left|\vphantom{\int_0^1}\right.  {\cal
M}^{\tau_{n}}\right]\le C\delta^{\ga}T^2,
\end{equation}
for $0<\ga<\min[\ep_2-3\ep_1/2,\ga_0'-(1+\ep_1)/2]$.
From \eqref{72507}, \eqref{72504}
\eqref{bliams-dec72} and \eqref{bliams-dec73} we further conclude that
\begin{equation}\label{xv2-0}
\tilde Q^{(\delta)}_{\bx,\bk}\left[\delta^*\le \frac{1}{p}\right]\le
CT^2\delta^{\ga-\al}+
Ce^T\left(1-\frac{\delta^{1/2(\ep_1+\ep_2)}}{2}\right)^{\delta^{-\al}}
\end{equation}
for some $C>0$, independent of $\delta\in(0,1]$ and $T\ge T_0$,
provided that we choose $\al\in(1/2(\ep_1+\ep_2),\ga)$.  This is
possible if $\min[\ep_2-3\ep_1/2,\ga_0'-(1+\ep_1)/2]>(\ep_1+\ep_2)/2$,
which is true if we assume $\ep_2>10\ep_1>0$ and
$1>\ga_0'>(1+\ep_2)/2+\ep_1$.  Now, by the argument made after
\eqref{Adelta-estimate} we can always replace the first term on the
right side of (\ref{xv2-0}) by $CT\delta^{\ga_1}$. We can also assume
that the second term on the right hand side of \eqref{xv2-0} is less
than or equal to $CT\delta^{\ga_1}$.  This can be seen as follows. Let
$\bt:=\al-1/2(\ep_1+\ep_2)$. The term in question is bounded by
$C\exp\left\{T-C_1\delta^{-\bt}\right\}$ with $
C_1:=\inf_{\rho\in(0,1]}\rho^{-1}\log\left(1-\rho/2\right)^{-1}$.  For
$\delta^{-\bt}\ge 2 T/C_1$ we get that
$\exp\left\{T-C_1\delta^{-\bt}\right\}$ is less than or equal to
$\exp\left\{-C_1\delta^{-\bt}/2\right\}$, while for $\delta^{-\bt}< 2
T/C_1$ the left side of (\ref{xv2-0}) is obviously less than $2
T\delta^{\bt}/C_1$.  In both cases we can find a bound as claimed.
This proves (\ref{72501b}) and hence the proof of Lemma \ref{lm806041}
will be complete if we prove (\ref{bliams-dec72}) and
\eqref{bliams-dec73}.

To this end, let $\tilde
Q^{(\delta)}_{\bx,\bk,\pi}$, $\pi\in\cal C$ denote the family of the
regular conditional probability distributions that corresponds to
$\tilde Q^{(\delta)}_{\bx,\bk}\left[\,\cdot\,\left|\right.{\cal
M}^{\tau_n}\right]$.  Then, there exists a ${\cal M}^{\tau_n}$
measurable, null $\tilde Q^{(\delta)}_{\bx,\bk}$ probability event $Z$
such that for each $\pi\not \in Z$ and each $\bbl\in \R^d_*$ the
random sequence \[
S_{N,\pi}^\bbl:=S_{N}^\bbl\bone_{[0,N/L]}(\tau_n(\pi)),\quad N\ge0 \]
is an $\left({\cal M}^{N/L}\right)_{N\ge0}$ submartingale under
$\tilde Q^{(\delta)}_{\bx,\bk,\pi}$.  Let $T_{n,\pi}:=\tau_{n+1}\wedge
(\tau_n(\pi)+2[L\delta^{\ep}]/L)$, where $\ep\in(0,1)$ is a constant
to be chosen later on.  We can choose the event $Z$ in such a way that
\begin{equation}\label{30801}
\tilde Q^{(\delta)}_{\bx,\bk,\pi}[T_{n,\pi}\ge \tau_n(\pi)]=1,\quad \forall\,\pi\not \in Z.
\end{equation}
Let $\tilde S_{N,\pi}:=S_{N,\pi}^{K(\tau_n(\pi))}$, then
the submartingale property of $\left(\tilde S_{N,\pi}\right)_{N\ge0}$
and \eqref{30801}  imply that
\begin{equation}\label{dec74}
\tilde E^{(\delta)}_{\bx,\bk,\pi} \tilde S_{LT_{n,\pi},\pi}\ge
\tilde E^{(\delta)}_{\bx,\bk,\pi} \tilde
S_{L\tau_{n}(\pi),\pi}=1+A_\rho \tau_n(\pi),
\end{equation}
provided that $\gamma_0\ge (1+\ep_1)/2$. The latter condition assures that
$\rho\ge C/(L\sqrt{\delta})$ so that $K$ does not change by more than
$\rho$ during the time $1/L$.  In consequence of (\ref{dec74}) we have
\begin{equation}\label{80302}
\tilde E^{(\delta)}_{\bx,\bk,\pi}\left[f_{K(\tau_n(\pi))}\left(K\left(T_{n,\pi}+\frac{1}{L}\right)\right)
\,\right]+2A_\rho\delta^\ep\ge 1,
\end{equation}
as $T_{n,\pi}-\tau_n(\pi)\le 2[L\delta^\eps]/L$.  Since
\[
\left|f_{K(\tau_n(\pi))}\left(K\left(T_{n,\pi}+\frac{1}{L}\right)\right)-
f_{K(\tau_n(\pi))}\left(K\left(T_{n,\pi}\right)\right)\right|\le \frac{C}{L\rho\delta^{1/2}}
\]
we obtain from \eqref{80302}
\[
2A_\rho\delta^\ep\ge \tilde E^{(\delta)}_{\bx,\bk,\pi}
\left[1-f_{K(\tau_n(\pi))}\left(K\left(T_{n,\pi}\right)\right)
\right]-\frac{C}{ L\rho\delta^{1/2}}
\]
so in particular
\begin{eqnarray}\label{80303}
&&2A_\rho\delta^\ep+\frac{C}{ L\rho\delta^{1/2}}\ge
\tilde E^{(\delta)}_{\bx,\bk,\pi}\left[1-f_{K(\tau_n(\pi))}\left(K\left(\tau_{n+1}\right)\right)
,\,\tau_{n+1}\le \tau_n(\pi)+\frac{[L\delta^\ep]}{L}\right]
\\
&&~~~~~~~~~~~~~~~~~~~~~
=\tilde Q^{(\delta)}_{\bx,\bk,\pi}\left[\tau_{n+1}\le \tau_n(\pi)+\frac{[L\delta^\ep]}{L}\right].
\nonumber
\end{eqnarray}
We have  shown, therefore, that
\begin{equation}\label{72505}
\tilde Q^{(\delta)}_{\bx,\bk}\left[\tau_{n+1}-\tau_{n}\le \frac{[L\delta^\ep]}{L}
\left|\vphantom{\int_0^1}\right.
{\cal M}^{\tau_{n}}\right]\le
\frac{CT^2\delta^\ep}{\rho^3 }+\frac{C}{ L\rho\delta^{1/2}}\le C
(\delta^{\ep-3\ep_1/2}T^2+\delta^{\ga_0'-(1+\ep_1)/2})\le C\delta^{\ga_1}T^2
\end{equation}
for $\ga_1<\min[\ep-3\ep_1/2,\ga_0'-(1+\ep_1)/2]$ and some constant
$C>0$. We can always assume that $T^2\delta^{\ga_1/2}\le 1$.  If
otherwise, we can always write $\tilde Q^{(\delta)}_{\bx,\bk}[F]\le
T\delta^{\ga/4}$ and
\eqref{72501b} follows.
In particular, selecting $\ep:=(\ep_1+\ep_2)/2$, one concludes from \eqref{72505}  that
\begin{eqnarray}
&&\tilde E^{(\delta)}_{\bx,\bk}[\exp\{-(\tau_{n+1}-\tau_{n})\}|{\cal
M}^{\tau_{n}}]\le e^{-\delta^{(\ep_1+\ep_2)/2}}\tilde
Q^{(\delta)}_{\bx,\bk}\left[\tau_{n+1}-\tau_{n}\ge
\frac{[L\delta^{(\ep_1+\ep_2)/2}]}{L}\left|\vphantom{\int_0^1}\right.
{\cal M}^{\tau_{n}}\right]\nonumber\\
&&
+\tilde Q^{(\delta)}_{\bx,\bk}\left[\tau_{n+1}-\tau_{n}\le
\frac{[L\delta^{(\ep_1+\ep_2)/2}]}{L}\left|\vphantom{\int_0^1}\right.
{\cal M}^{\tau_{n}}\right]\stackrel{\eqref{72505}}{\le} e^{-\delta^{(\ep_1+\ep_2)/2}}+
C\left(1-e^{-\delta^{(\ep_1+\ep_2)/2}}\right)\delta^{\ga/2}\nonumber\\
&&~~~~~~~~~~~~~~~~~~~~~~~~~~~~~~~~~~~~~~~~~~~~~~~~~~~~~~~~~
<1-\frac{\delta^{(\ep_1+\ep_2)/2}}{2}\label{72508}
\end{eqnarray}
provided that $\delta$ is sufficiently small.  From \eqref{72508} one
concludes easily, see e.g. Lemma 1.4.5 p. 38 of
\cite{stroock-varadhan}, that (\ref{bliams-dec72}) holds.

On the other hand, taking $\ep=\ep_2$ in \eqref{72505} we obtain
\eqref{bliams-dec73} with
$0<\ga<\min[\ep_2-3\ep_1/2,\ga_0'-(1+\ep_1)/2]$.
Hence the proof of Lemma \ref{lm806041} is now complete.
$\Box$
\commentout{

$\le C_{85}T\delta^{\ga}$.

The last inequality of \eqref{xv2}
can be proved using the same argument as the one used after \eqref{Adelta-estimate}.

Note that we can always assume that the second term on the right hand side of
\eqref{xv2} is less than or equal to $C_{85}T\delta^{\ga}$ for all $\delta\in(0,1]$, $T\ge1$.
This can be seen as follows. Let $\bt:=\al-1/2(\ep_1+\ep_2)$. The term in question
is bound by $\exp\left\{T-C\delta^{-\bt}\right\}$ with
$
C:=\sup_{\rho\in(0,1]}\rho^{-1}\log\left(1-\rho/2\right)^{-1}.
$
For $\delta^{-\bt}\ge 2 T/C$ we get that $\exp\left\{T-C\delta^{-\bt}\right\}$ is less than or equal to
$\exp\left\{-C\delta^{-\bt}/2\right\}$, while for $\delta^{-\bt}< 2 T/C$ it is obviously less than $2 T\delta^{\bt}/C$.
In both cases we can find a bound as claimed.

Let $\bt:=\al-1/2(\ep_1+\ep_2)$.
Note that
\begin{equation}\label{xv1}
e^T\left(1-\frac{\delta^{1/2(\ep_1+\ep_2)}}{2}\right)^{\delta^{-\al}}=
\exp\left\{T-\delta^{-\bt}\delta^{-1/2(\ep_1+\ep_2)}\log\left(1-\delta^{1/2(\ep_1+\ep_2)}/2\right)^{-1}\right\}
\end{equation}
$$
\le \exp\left\{T-C\delta^{-\bt}\right\},
$$
where
$$
C:=\sup\limits_{\delta\in(0,1]}\delta^{-1/2(\ep_1+\ep_2)}\log\left(1-\delta^{1/2(\ep_1+\ep_2)}/2\right)^{-1}(=8).
$$
For $\delta^{-\bt}\ge 2 T/C$ we get that the left hand side of \eqref{xv1} is less than or equal to
$\exp\left\{-C\delta^{-\bt}/2\right\}$, while for $\delta^{-\bt}< 2 T/C$ it is obviously less than $2 T\delta^{\bt}/C$.

}

\subsection{The estimation of the convergence rate. The proof of Theorem \ref{main-thm}.}
\label{sec3.9}

Recall that $\phi_{\delta}$, $\bar \phi$ satisfy
\eqref{liouv-intro-0},
\eqref{eq-mainthm-1}, respectively, with the initial condition $\phi_0$.
We start with the following lemma.
\begin{lemma}
\label{lm-estimate}
Assume that $\phi_0$ satisfies the hypotheses formulated in Section $\ref{sec2.4}$. Then,
\begin{equation}\label{013011}
   \|\bar\phi\|_{0,0,0}^{[0,T]}\le \|\phi_0\|_{0,0},\quad
   \sum\limits_{i=1}^d\|\partial_{x_i}\bar\phi\|_{0,0,0}^{[0,T]}\le \|\phi_0\|_{1,0}.
\end{equation}
Furthermore, there exists a constant $C>0$ such that for all $T\ge1$
\begin{equation}\label{013011c}
   \|\partial_t\bar\phi\|_{0,0,0}^{[0,T]}\le C\|\phi_0\|_{1,2}.
\end{equation}
In addition, for any nonnegative integer valued multi-index $\ga=(\al_1,\al_2,\al_3)$ satisfying $|\ga|\le 3$ we have
\begin{equation}\label{013011b}
    \sum\limits_{i_1,i_2,i_3=1}^d\|\partial^{\ga}_{k_{i_1},k_{i_2},k_{i_3}}\bar\phi\|_{0,0,0}^{[0,T]}\le CT^{|\ga|}
    \|\phi_0\|_{1,4},
\end{equation}
\end{lemma}
{\bf Proof.}
The estimates \eqref{013011} follow directly from differentiating
(\ref{eq-mainthm-1}) with respect to $\vx$.  To obtain the estimates
\eqref{013011c} and \eqref{013011b} we note first that the application
of the operator $\tilde{\cal L}$ to both sides of \eqref{eq-mainthm-1}
and the maximum principle leads to the estimate $\|\tilde{\cal
L}\bar\phi(t,\bx,\cdot)\|_{L^\infty(A(M))}\le \|\tilde{\cal
L}\phi_0\|_{L^\infty({\cal A}(M))}$ for all $t\ge 0$, hence we
conclude bound \eqref{013011c}.

In fact, thanks to already proven estimate \eqref{013011} we conclude
that $\|{\cal L}\bar\phi(t,\bx,\cdot)\|_{L^\infty(A(M))}\le
C\|\phi_0\|_{1,2}$ for some constant $C>0$ and all
$(t,\bx)\in[0,+\infty)\times\R^d$. Let $Z$ be as in the proof of Lemma
\ref{lm70901}.  Define $S:Z\times[M^{-1}, M]\to A(M)$ as
$S(\bbl,k):=(\bbl,\sqrt{k^2-l^2})$, where $l=|\bbl|$.  Let also
$\psi(\bbl,k)=\bar\phi\circ S(\bbl,k)$. We have $({\cal
L}_\bk\bar\phi)\circ S(\bbl,k)={\cal N}\psi(\bbl,k)$, see
\eqref{061130}. The $L^p$ estimates for elliptic partial differential
equations, see e.g. Theorem 9.13 p. 239 of
\cite{gilbarg} allow us to estimate
$$
\|\psi\|_{W^{2,p}(Z)}\le C(\|\psi\|_{L^p(Z)}+\|{\cal N}\psi\|_{L^p(Z)})\le
C\|\phi_0\|_{1,2}.  $$
Choosing $p$ sufficiently large we obtain that
$\sum_{i}\|\partial_{l_i}\psi\|_{L^\infty(Z)}\le C\|\phi_0\|_{1,2}$,
which in fact implies that $\|{\bf
D}(\cdot)\nabla_\bk\bar\phi(t,\cdot)\|_{L^\infty(S(Z))}\le
C\|\phi_0\|_{1,2}$. Obviously, one can find a covering of $A(M)$ with
charts corresponding to different choices of the components of $\bk$
being projected onto the hyperplane $\mathbb R^{d-1}$ and we obtain in
that way that $\|{\bf
D}(\cdot)\nabla_\bk\bar\phi(t,\cdot)\|_{L^\infty({\cal A}(M))}\le
C\|\phi_0\|_{1,2}$ for all $t\ge0$.  Since the rank of the matrix
${\bf D}(\hat\bk,k)$ equals $d-1$, with the kernel spanned by the
vector $\bk$, we obtain in that way the $L^\infty$ estimates of
directional derivatives in any direction perpendicular to $\bk$.  We
still need to obtain the $L^\infty$ bound on the
derivative in the direction $\bk$, denoted by
$\partial_n:=k_1\partial_{k_1}+\ldots+k_d\partial_{k_d}$. To that
purpose we apply $\partial_n$ to both sides of \eqref{eq-mainthm-1}
and after a straightforward calculation we get
$
\partial_t\partial_n\bar\phi=\tilde{\cal L}\partial_n\bar\phi-
2{\cal L}_{\bk}\bar\phi+{\cal L}_1\bar\phi+H_0'\!'(k)\hat
\bk\cdot\nabla_\bx\bar\phi, $ where
$$ {\cal
L}_1\bar\phi:=\sum\limits_{m,n=1}^d\pdr{}{k_m}
\left(\partial_kD_{mn}(\vhatk,k)\pdr{\bar\phi}{k_n}\right).
$$
Note that ${\bf D}(\vhatk,k)\vhatk=\bze$ implies that
$\partial_k{\bf D}(\vhatk,k)\vhatk=\bze$ hence $\|{\cal
L}_1\bar\phi(t,\cdot)\|_{L^\infty({\cal A}(M))}\le
C\|\phi_0\|_{1,2}$. We already know that ${\cal L}_{\bk}\bar\phi$ and
$\|\nabla_\bx\bar\phi\|_{L^\infty({\cal A}(M))}$ are bounded,
hence $\|\partial_n\bar\phi(t,\cdot)\|_{L^\infty({\cal
A}(M))}\le C\|\phi_0\|_{1,2}T$ for $t\in[0,T]$. We have shown
therefore that $\|\bar\phi(t,\cdot)\|_{1,1}\le C\|\phi_0\|_{1,2}T$ for
$t\in[0,T]$. The above procedure can be iterated in order to obtain
the estimates of the suprema of derivatives of the higher order.
$\Box$

\commentout{

We have the following formulas
$$
\partial_n\nabla\cdot {\bf F}+\nabla\cdot {\bf F}=\nabla\cdot \partial_n{\bf F},
$$
$$
\partial_n\nabla\phi+\nabla\phi=\nabla\partial_n\phi.
$$
Hence
$$
\partial_n\nabla\cdot {\bf D}\nabla \phi=\nabla\cdot \partial_n({\bf D}\nabla \phi)-
\nabla\cdot {\bf D}\nabla \phi
$$
$$
=\nabla\cdot \partial_k{\bf D}\nabla \phi+\nabla\cdot {\bf D}\partial_n\nabla \phi-
\nabla\cdot {\bf D}\nabla \phi
$$
$$
=\nabla\cdot \partial_k{\bf D}\nabla \phi+\nabla\cdot {\bf D}\nabla \partial_n\phi-
2\nabla\cdot {\bf D}\nabla \phi
$$

}
{\bf Proof of Theorem \ref{main-thm}.}  Let $u\in[\delta^{\ga_0'},T]$,
where we assume that $\ga_0'$ (as in the statement of Lemma
\ref{lmA1}) belongs to the interval $(1/2,1)$.  Substituting for
$G(t,\bx,\bk):=\bar \phi(u-t,\bx,\bk)$, $\zeta\equiv1$ into
(\ref{73101}) we obtain (taking $v=u$, $t=\delta^{\ga_0'}$)
\begin{eqnarray}
&&\left|\tilde E^{(\delta)}_{\bx,\bk}\vphantom{\int\limits_t^u}
\left[\phi_0(X(u),K(u))-\bar \phi(u-\delta^{\ga_0'},X(\delta^{\ga_0'}),K(\delta^{\ga_0'}))
-\int\limits_{\delta^{\ga_0'}}^u
(\partial_\varrho  +\widehat{\cal L}_\varrho)G(\varrho,X(\varrho),K(\varrho))\,d\varrho\right]
\right|\nonumber
\\
&&
\leq
C\|G\|_{1,1,3}^{[0,T]}\delta^{\ga_1}T^2,\quad\forall\,\delta\in(0,1].\label{73101bb}
\end{eqnarray}
Using the fact that $|X(\delta^{\ga_0'})-\bx|\le C\delta^{\ga_0'}$,
$|K(\delta^{\ga_0'})-\bk|\le C\delta^{\ga_0'-1/2}$, $\tilde Q_{\bx,\bk}^{(\delta)}$--a.s.
for some deterministic constant $C>0$, cf. \eqref{eq2}, and Lemma \ref{lm-estimate}
we obtain that there exist constants $C,\ga>0$ such that
\begin{eqnarray}
&&\left|\tilde E^{(\delta)}_{\bx,\bk}\vphantom{\int\limits_t^u}
\left[\phi_0(X(u),K(u))-\bar \phi(u,\bx,\bk)
-\int\limits_{0}^u
(\partial_\varrho  +\widehat{\cal L}_\varrho)G(\varrho,X(\varrho),K(\varrho))\,d\varrho\right]
\right|\nonumber\\
&&
\leq
C\|G\|_{1,1,3}^{[0,T]}\delta^{\ga}T^2,\quad\delta\in(0,1],\,T\ge 1,\,u\in[0,T].\label{73101bbb}
\end{eqnarray}
We have however
\begin{eqnarray}
&&\left| E^{(\delta)}_{\bx,\bk}\vphantom{\int\limits_t^u}
\left[\phi_0(X(u),K(u))-\bar \phi(u,\bx,\bk),\, \tau_{\delta}\ge T\right]\right|
=\left|\tilde E^{(\delta)}_{\bx,\bk}\vphantom{\int\limits_t^u}
\left[\phi_0(X(u),K(u))-\bar \phi(u,\bx,\bk),\, \tau_{\delta}\ge T\right]\right|
\nonumber\\
&&\stackrel{\eqref{73101bbb}}{\leq}
C\|G\|_{1,1,3}^{[0,T]}\delta^{\ga}T^2
+\left(2\|\phi_0\|_{0,0}+T\|G\|_{1,1,2}^{[0,T]}\right)\tilde Q_{\bx,\bk}^{(\delta)}[\tau_{\delta}<T].
\label{051130}
\end{eqnarray}
Using ${\cal M}^{\tau_\delta}$ measurability of the event $[\tau_\delta<T]$ we obtain that
$\tilde Q^{(\delta)}_{\bx,\bk}\left[
\tau_\delta< T\right]=R^{(\delta)}_{\bx,\bk}\left[
\,\tau_\delta< T\right]$
and by virtue of Theorem \ref{cor10211} we can estimate the right hand side of \eqref{051130} by
$$
 C\|G\|_{1,1,3}^{[0,T]}\delta^{\ga}T^2
+C\delta^{\gamma}T\left(2\|\phi_0\|_{0,0}+T\|G\|_{1,1,2}^{[0,T]}\right)
\stackrel{\mbox{\tiny Lemma \ref{lm-estimate}}}{\le} C\delta^{\gamma}T^5.
$$

On the other hand, the expression under the absolute value on
the utmost left hand side of \eqref{051130}
equals
$$
 E^{(\delta)}_{\bx,\bk}\vphantom{\int\limits_t^u}
\left[\phi_0(X(u),K(u))-\bar \phi(u,\bx,\bk)\,\right]-
E^{(\delta)}_{\bx,\bk}\vphantom{\int\limits_t^u}
\left[\phi_0(X(u),K(u))-\bar \phi(u,\bx,\bk),\, \tau_{\delta}< T\right].
$$
The second term
 can be estimated by
\[
2\|\phi_0\|_{0,0}R_{\bx,\bbk}^{(\delta)}\left[\tau_\delta< T\right]\stackrel{\eqref{92601b}}{\le}
C\delta^{\ga}\|\phi_0\|_{0,0}T,
\]
by virtue of Theorem \ref{cor10211}.
Since
$$
\E \phi_{\delta}\left(\frac{u}{\delta},\frac{\vx}{\delta},\bk\right)=
\E\phi_0(\bz^{(\delta)}(u;\bx,\bk),\obm^{(\delta)}(u;\bx,\bk))=
 E^{(\delta)}_{\bx,\bk}\phi_0(X(u),K(u))
$$
we conclude from the above
 that the left hand side of \eqref{decorrel-main1} can be estimated by
$C\delta^{\gamma}\|\phi_0\|_{1,4}T^5$ for some constants
$C,\gamma>0$ independent of $\delta>0$, $T\ge1$.
The bound appearing on the right hand side of \eqref{decorrel-main1} can be now
concluded by the same argument as the one used after \eqref{Adelta-estimate}.
$\Box$

\section{Momentum diffusion to spatial diffusion:
proof of Theorem \ref{thm-wave-space-main}}\label{sec:wd-sd}

We show in this section that solutions of the momentum diffusion
equation (\ref{eq-mainthm-1}) in the long-time, large space limit
converge to the solutions of the spatial diffusion equation
(\ref{6-diff-space-main}). We first recall the setup of Theorem
\ref{thm-wave-space-main}.  Let $\bar \phi_\gamma(t,\vx,\bk)=\bar
\phi(t/\gamma^2,\vx/\gamma,\bk)$, where $\bar \phi$ satisfies
(\ref{eq-mainthm-1})
and let $w(t,\vx,k)$ be the solution
of the spatial diffusion equation (\ref{6-diff-space-main}).
In order to prove Theorem \ref{thm-wave-space-main} we need to show that
the re-scaled solution $\phi_\gamma(t,\vx,\bk)$ converges as $\gamma\to
0$ in the space $C([0,T];L^\infty({\cal A}(M)))$ to $w(t,\vx,\bk)$, so that
\begin{equation}\label{6-error}
\|w(t)-\bar \phi_\gamma(t)\|_{L^\infty({\cal A}(M))}\le
C\left(\gamma T+\gamma^{1/2}\right)\|\phi_0\|_{2,0},\quad0\le t\le T.
\end{equation}


{\bf Proof of Theorem \ref{thm-wave-space-main}.}  The proof is quite
standard. We present it for the sake of completeness and convenience
to the reader.  The function $\bar \phi_\gamma$ is the unique
$C^{1,1,2}_b([0,+\infty),\R^{2d}_*)$-solution to
\begin{eqnarray}\label{wd-diff}
&&\gamma^2\pdr{\bar \phi_\gamma}{t}=
\sum\limits_{m,n=1}^d\pdr{}{k_m}\left(D_{mn}(\vhatk,k)\pdr{\bar
\phi_\gamma}{k_n}\right)+\gamma H_0'(k)\vhatk\cdot\nabla_\vx\bar
\phi_\gamma.\\
&&\bar\phi_\gamma(0,\vx,k)=\phi_0(\vx,\bk),\nonumber
\end{eqnarray}
see Remark \ref{rm14111}.
We represent $\bar \phi_\gamma$ as
\begin{equation}\label{6-expand}
\bar \phi_\gamma=w+\gamma w_1+\gamma^2 w_2+R.
\end{equation}
Here $w$ is the solution of the diffusion equation (\ref{6-diff-space-main}),
the correctors $w_1$ and $w_2$ will be constructed explicitly, and the remainder $R$
will be shown to be small.
The first corrector $w_1$ is the unique  solution of zero mean over each sphere $\mathbb S^{d-1}_k$ of
the equation
\begin{equation}\label{xtra1}
\sum_{m,n=1}^d\pdr{}{k_m}\left(D_{mn}(\vhatk,k)\pdr{w_1}{k_n}\right)=
-H_0'(k)\vhatk\cdot\nabla_\vx w.
\end{equation}
It has an explicit form
\begin{equation}\label{wd-w1}
w_1(t,\vx,\bk)=\sum_{j=1}^d\chi_j(\bk)\pdr{w(t,\vx,k)}{x_j}
\end{equation}
with the functions $\chi_j$ defined in (\ref{6-chim-eq-main}).  The
second order corrector $w_2$ is the unique zero mean over each sphere
$\mathbb S^{d-1}_k$ solution of the equation
\begin{equation}\label{wd-w2eq}
\sum_{m,n=1}^d\pdr{}{k_m}\left(D_{mn}(\vhatk,k)\pdr{w_2}{k_n}\right)=
\pdr{w}{t}-H_0'(k)\vhatk\cdot\nabla_\vx w_1.
\end{equation}
Note that the expression
on the right hand side of \eqref{wd-w2eq} is of zero mean since thanks
 to (\ref{6-diff-space-main}) and equality (\ref{6-wd-diffm-main}) we have
\[
\pdr{w}{t}=\frac{1}{\Gamma_{d-1}}\int_{{\mathbb S}^{d-1}}
H_0'(k)\vhatk\cdot\nabla_\vx
w_1 d\Omega(\vhatk).
\]
Equations (\ref{xtra1}) and (\ref{wd-w2eq}) for various values of $k=|\bk|$ are
decoupled. As a consequence of this fact and the  regularity properties
for solutions of elliptic equations on a sphere we have that
$w_1$, $w_2$  belong to $C([0,T];L^\infty({\cal A}(M)))$. More explicitly, we may
represent the function $w_2$ as
\[
w_2(t,\vx,\bk)=\sum\limits_{j,l=1}^d\psi_{jl}(\bk)\pdrt{w(t,\vx,\bk)}{x_j}{x_l}.
\]
The functions $\psi_{jm}(\bk)$ satisfy
\begin{equation}\label{wd-psieq}
\sum_{m,n=1}^d\pdr{}{k_m}\left(D_{mn}(\vhatk,k)\pdr{\psi_{jl}}{k_n}\right)
=-H_0'(k)\hat k_j\chi_l(\bk)+a_{jl}(k).
\end{equation}
A unique mean-zero,  bounded solution of (\ref{wd-psieq})
exists according to the Fredholm alternative combined the  regularity properties
for solutions of (\ref{wd-psieq}) on each sphere  $\mathbb S^{d-1}_k$.
With the above definitions of $w$, $w_1$, $w_2$, equation
(\ref{wd-diff}) for $\bar \phi_\gamma$ implies that the remainder $R$ in
(\ref{6-expand}) satisfies
\[
\gamma^2\pdr{R}{t} +\gamma^3\pdr{w_1}{t}+\gamma^4\pdr{w_2}{t}
-\gamma H_0'(k)\vhatk\cdot\nabla_\vx R- \gamma^3 H_0'(k)\vhatk\cdot\nabla_\vx w_2
=\sum_{m,n=1}^d\pdr{}{k_m}\left(D_{mn}(\vhatk,k)\pdr{R}{k_n}\right).
\]
We re-write this equation in the form
\begin{eqnarray}
&&\pdr{R}{t} -\frac{1}{\gamma}H_0'(k)\vhatk\cdot\nabla_\vx
R-\frac{1}{\gamma^2}\sum_{m,n=1}^d\pdr{}{k_m}\left(D_{mn}(\vhatk,k)\pdr{R}{k_n}\right)=f
\label{011411}\\
&&R(0,\vx,\bk)=\phi_0(\vx,\bk)-\bar\phi_0(\vx,\bk)-\gamma w_1(0,\vx,\bk)-
\gamma^2w_2(0,\vx,\bk),\nonumber
\end{eqnarray}
where $f:=-\gamma\partial_t w_1-\gamma^2\partial_t w_2 -
\gamma H_0'(k)\vhatk\cdot\nabla_\vx w_2$.
Here, as before, $R$ is understood as the unique solution
to \eqref{011411} that belongs to $C^{1,1,2}_b([0,+\infty),\R^{2d}_*)$.
We may split $R=R_1+R_2$ according to the initial data and forcing in the
equation: $R_1$ satisfies
\begin{eqnarray}
&&\!\!\!\!\!\pdr{R_1}{t}- \frac{1}{\gamma}H_0'(k)\vhatk\cdot\nabla_\vx
R_1-\frac{1}{\gamma^2}\!\sum_{m,n=1}^d\!
\pdr{}{k_m}\left(D_{mn}(\vhatk,k)\pdr{R_1}{k_n}\right)\!
=f,\nonumber\\
&&R(0,\vx,\bk)=-\gamma w_1(0,\vx,\bk)-
\gamma^2w_2(0,\vx,\bk)\label{bliams-force}
\end{eqnarray}
and the initial time boundary layer corrector $R_2$ satisfies
\begin{eqnarray}\label{init-bliams}
&&\pdr{R_2}{t} -\frac{1}{\gamma}H_0'(k)\vhatk\cdot\nabla_\vx
R_2-\frac{1}{\gamma^2}\sum_{m,n=1}^d
\pdr{}{k_m}\left(D_{mn}(\vhatk,k)\pdr{R_2}{k_n}\right)
=0\\
&&R_2(0,\vx,\bk)=\phi_0(\vx,\bk)-\bar\phi_0(\vx,\bk).\nonumber
\end{eqnarray}
Using the probabilistic representation for the solution of
(\ref{bliams-force}) as well as the regularity of $w_1$ and $w_2$
we obtain that
\begin{equation}\label{bums3}
\|R_1(t)\|_{L^\infty({\cal A}(M))}\le C\gamma T,~~0\le t\le T.
\end{equation}
To obtain the bound for $R_2$
we consider $R_2^\gamma(t,\vx,\bk):=R_2(\gamma^{3/2}t,\vx,\bk)$. This function
satisfies
\begin{eqnarray*}
&&\pdr{R_2^\gamma}{t} -{\gamma}^{1/2}H_0'(k)\vhatk\cdot\nabla_\vx
R_2^\gamma-
\frac{1}{\gamma^{1/2}}
\sum_{m,n=1}^d\pdr{}{k_m}\left(D_{mn}(\vhatk,k)\pdr{R_2^\gamma}{k_n}\right)
=0\\
&&R_2^\gamma(0,\vx,\bk)=\phi_0(\vx,\bk)-\bar\phi_0(\vx,\bk).
\end{eqnarray*}
We also define $\tilde R_2^\gamma$, the solution of
\begin{eqnarray}\label{bums1}
&&\pdr{\tilde R_2^\gamma}{t}
-\frac{1}{\gamma^{1/2}}\sum_{m,n=1}^d\pdr{}{k_m}\left(D_{mn}(\vhatk,k)
\pdr{\tilde R_2^\gamma}{k_n}\right)
=0\\
&&\tilde R_2^\gamma(0,\vx,\bk)=\phi_0(\vx,\bk)-\bar\phi_0(\vx,\bk).\nonumber
\end{eqnarray}
The uniform ellipticity of the right hand side of (\ref{bums1}) on
each sphere $\mathbb S^{d-1}_k$
implies, see e.g. Proposition 13.3, p. 55 of \cite{Taylor} that the
function $\tilde R_2^\gamma$ satisfies the decay estimate on each
sphere
\begin{equation}\label{nash0}
\|\tilde R_2^\gamma(t)\|_{L^\infty({\mathbb S}^{d-1})}\le
\frac{C\gamma^{(d-1)/4}}{t^{(d-1)/2}}\|\phi_0\|_{L^1({\mathbb S}^{d-1}_k)}
\le\frac{C\gamma^{(d-1)/4}}{t^{(d-1)/2}}\|\phi_0\|_{L^\infty({\mathbb S}^{d-1}_k)}
\end{equation}
for $t\in[0,T]$ and, similarly,
\[
\|\nabla_\vx\tilde R_2^\gamma(t)\|_{L^\infty({\mathbb S}^{d-1}_k)}\le
\frac{C\gamma^{(d-1)/4}}{t^{(d-1)/2}}
\|\phi_0\|_{1,0}.
\]
Furthermore, the difference $q^\gamma=R_2^\gamma-\tilde R_2^\gamma$ satisfies
\begin{eqnarray}
\label{051711}
&&\hspace*{-.3in}\pdr{q^\gamma}{t} -{\gamma}^{1/2}H_0'(k)\vhatk\cdot\nabla_\vx
q^\gamma-\frac{1}{\gamma^{1/2}}\sum\limits_{m,n=1}^d\pdr{}{k_m}
\left(D_{mn}(\vhatk,k)\pdr{q^\gamma}{k_n}\right)
={\gamma}^{1/2}H_0'(k)\vhatk\cdot\nabla_\vx \tilde R_2^\gamma\\
&&\hspace*{-.3in}q^\gamma(0,\vx,\bk)=0\nonumber.
\end{eqnarray}
We conclude, using the probabilistic representation of the solution of
\eqref{051711}, that
\[
\|q^\gamma(t)\|_{L^\infty({\cal A}(M))}\le
C\gamma^{1/2}t\|\phi_0\|_{1,0}
\]
and thus
\begin{eqnarray*}
&&\|R_2(\gamma^{3/2})\|_{L^\infty({\cal A}(M))}\le\|R_2^\gamma(1)\|_{L^\infty({\cal A}(M))}+
 \|q^\gamma(1)\|_{L^\infty({\cal A}(M))}\\
&&\le
C\left(\gamma^{(d-1)/4}\|\phi_0\|_{0,0}+
\gamma^{2}\|\phi_0\|_{1,0}\right).
\end{eqnarray*}
The maximum principle for (\ref{init-bliams}) implies that we have the
above estimate for all $t\ge\gamma^{3/2}$:
\begin{equation}\label{bums2}
\|R_2(t)\|_{L^\infty({\cal A}(M))}\le
C\left(\gamma^{(d-1)/4}\|\phi_0\|_{0,0}+
\gamma^{1/2}\|\nabla_\vx\phi_0\|_{1,0}\right),~~t\ge\gamma^{3/2}.
\end{equation}
Combining (\ref{6-expand}), (\ref{bums3}) and (\ref{bums2}) we conclude that
\begin{equation}\label{bums4}
\|w(t)-\bar\phi_\gamma(t)\|_{L^\infty({\cal A}(M))}\le
C\left(\gamma T+\gamma^{(d-1)/4}+\gamma^{1/2}\right)
\|\phi_0\|_{1,0},~~\ga^{3/2}\le t\le T,
\end{equation}
and thus (\ref{6-error}) follows, as $d\ge 3$. This finishes the proof of Theorem
\ref{thm-wave-space-main}. $\Box$

\section{The spatial diffusion of wave energy}\label{sec:waves}

In this section we consider an application of the previous results to
the random geometrical optics regime of propagation of acoustic waves.
We show that when the wave length is much shorter than the correlation
length of the random medium, there exist temporal and spatial scales where
the energy density of the wave undergoes the spatial diffusion.  We
start with the wave equation in dimension $d\ge 3$
\begin{equation}\label{eq-wave}
\frac{1}{c^2(\vx)}\frac{\partial^2\phi}{\partial t^2}-\Delta\phi=0
\end{equation}
and assume that the wave speed has the form
$c(\vx)=c_0+\sqrt{\delta}c_1\left(\vx\right)$.
Here $c_0>0$ is the constant
sound speed of the uniform background medium, while the small
parameter $\delta\ll 1$ measures the strength of the mean zero
random perturbation $c_1$. Rescaling the spatial and temporal
variables $\vx=\vx'/\delta$ and $t=t'/\delta$ we obtain (after
dropping the primes) equation (\ref{eq-wave}) with a rapidly
fluctuating wave speed
\begin{equation}\label{c-fluct-delta}
c_\delta(\vx)=c_0+\sqrt{\delta}c_1\left(\frac{\vx}{\delta}\right).
\end{equation}
It is convenient to rewrite (\ref{eq-wave}) as the system of
acoustic equations for the ``pressure'' $p=\phi_t/c$ and
``acoustic velocity'' $\vu=-\nabla \phi$:
\begin{eqnarray}\label{acoust-eq-symm}
&&\pdr{\vu}{t}+\nabla\left(c_\delta(\vx)p\right)=0\\
&&\pdr{p}{t}+ c_\delta(\vx)\nabla\cdot\vu=0.\nonumber
\end{eqnarray}
We will denote for brevity
$\bv=(\bu,p)\in{\mathbb R}^{d+1}$ and write (\ref{acoust-eq-symm})
in the more general form of a first order linear symmetric
hyperbolic system.
To do so we introduce
symmetric matrices $A_\delta$ and $D^j$
 defined by
\begin{equation}
  \label{eq:symmatADj}
  A_\delta(\bx)=\hbox{diag}(1,1,1,c_\delta(\vx)), \qquad \mbox{ and } \qquad
  D^j={\bf e}_j\otimes{\bf e}_{d+1}+{\bf e}_{d+1}\otimes{\bf e}_j,\,\,
  j=1,\dots,d.
\end{equation}
Here ${\bf e}_m\in{\mathbb R}^{d+1}$ is the standard orthonormal
basis: $({\bf e}_m)_k=\delta_{mk}$.

We consider the initial data for (\ref{acoust-eq-symm}) as a mixture
of states.  Let $S$ be a measure space equipped with a non-negative
finite measure $\mu$.  A typical example is that the initial data is
random, $S$ is the state space and $\mu$ is the corresponding
probability measure. We assume that for each parameter $\zeta\in S$
and $\eps,\delta>0$ the initial data is given by
$\bv_\eps^{\delta}(0,\vx;\zeta):=
(-\eps\nabla\phi_0^\eps(\bx),1/c_\delta(\bx)\dot\phi_0^\eps(\bx))$ and
$\bv_\eps^{\delta}(t,\vx;\zeta)$ solves the system of equations
\begin{equation}\label{symm-version-intro}
\pdr{\bv_{\eps}^{\delta}}{t}+\sum\limits_{j=1}^dA_\delta(\vx)D^j\pdr{}{x^j}
\left(A_\delta(\vx)\bv_{\eps}^{\delta}(\vx)\right)=0.
\end{equation}
The set of initial data is assumed to form an $\eps$-oscillatory and
compact at infinity family \cite{GMMP} as $\eps\to 0$. By the above we mean that
for any continuous, compactly supported function $\varphi:\R^d\to \R$
we have
\[
\lim_{R\to+\infty}\limsup_{\eps\to0^+}\int_{|\bk|\ge R/\eps}|\widehat{\vphi
\bv_\eps^{\delta}}|^2d\bk\to0
\hbox{  and }
\lim_{R\to +\infty}\limsup_{\eps\to0^+}\int_{|\bx|\ge
R}| \bv_\eps^{\delta}|^2d\bx\to0
\]
for a fixed realization $\zeta\in S $ of the initial data and each
$\delta>0$.  In the regime of geometric acoustics the scale $\eps$ of
oscillations of the initial data is much smaller than the correlation
length $\delta$ of the medium: $\eps\ll \delta\ll 1$.

The dispersion matrix for (\ref{symm-version-intro}) is
\begin{equation}\label{P0-intro}
P_0^\delta(\vx,\bk)=
i\sum_{j=1}^dA_\delta(\vx)k_jD^jA_\delta(\vx)=i\sum_{j=1}^dc_\delta(\vx)k_jD^j
=ic_\delta(\vx)\left(\tilde\bk\otimes{\bf e}_{d+1}+{\bf
e}_{d+1}\otimes\tilde\bk\right),
\end{equation}
where $\tilde\bk=\sum_{j=1}^dk_j{\bf e}_j$.  The self-adjoint matrix
$(-iP_0^\delta)$ has an eigenvalue $H_0=0$ of the multiplicity $d-1$, and
two simple eigenvalues
\begin{equation}\label{ac-ham}
H_{\pm}^\delta(\vx,\bk)=\pm c_\delta(\vx)|\bk|.
\end{equation}
Its eigenvectors are
\begin{equation}
\label{eq:eigbs}
\vb_{m}^0=\left(\bk_m^\perp,0\right),~m=1,\dots,d-1;~~
\vb_\pm=\frac{1}{\sqrt{2}}\left(\frac{\tilde\bk}{|\bk|}\pm{\bf
e}_{d+1}\right),
\end{equation}
where $\bk_m^\perp\in{\mathbb R}^d$ is the orthonormal basis of
vectors orthogonal to $\bk$.

The $(d+1)\times(d+1)$ Wigner matrix of a mixture of solutions of
(\ref{symm-version-intro}) is defined by
\begin{equation}\label{wigner}
W_\eps^{\delta}(t,\vx,\bk)=
\frac{1}{(2\pi)^d}\int\limits_{{\mathbb R}^d}\int\limits_S
e^{i\bk\cdot\vy} \bv_\eps^{\delta}(t,\vx-\frac{\eps\vy}{2};\zeta)
\bv_\eps^{\delta*}(t,\vx+\frac{\eps\vy}{2};\zeta)d\vy\mu(d\zeta) .
\end{equation}
It is well-known, see \cite{GMMP,LP,RPK-WM}, that for each fixed $\delta>0$
(and even without introduction of a mixture of states) when
$W_\eps^{\delta}(t=0)$ converges weakly in ${\cal
S}'({\mathbb R}^d\times{\mathbb R}^d)$, as $\eps\to 0$, to
\begin{equation}\label{no-offdiagonal}
W_0(\vx,\bk)=u_+^0(\vx,\bk)\vb_+(\bk)\otimes\vb_+(\bk)
+u_-^0(\vx,\bk)\vb_-(\bk)\otimes\vb_-(\bk).
\end{equation}
then $W_\eps^{\delta}(t)$  converges weakly in ${\cal S}'({\mathbb
 R}^d\times{\mathbb R}^d)$ to
\[
U^{\delta}(t,\vx,\bk)=u_+^{\delta}(t,\vx,\bk)\vb_+(\bk)\otimes\vb_+(\bk)
+u_-^{\delta}(t,\vx,\bk)\vb_-(\bk)\otimes\vb_-(\bk).
\]
The scalar amplitudes $u_\pm^{(\delta)}$ satisfy the Liouville
equations:
\begin{equation}\label{liouv-intro}
\left\{\begin{array}{l}
\partial_t u_\pm^{\delta}+\nabla_\bk H_\pm^\delta\cdot\nabla_\vx
u_\pm^{\delta}-\nabla_\vx H_\pm^{\delta}\cdot\nabla_\bk
u_\pm^{\delta}=0,\vphantom{\int_0^1}\\
\\
u_\pm^{\delta}(0,\vx,\bk)=u_\pm^0(\vx,\bk)\vphantom{\int_0^1}.
\end{array}
\right.
\end{equation}
These equations are of the form (\ref{liouv-intro-0}), written in the
macroscopic variables, with the Hamiltonian given by (\ref{ac-ham}).

One may obtain an $L^2$-error estimate for this convergence when a
mixture of states is introduced, as in (\ref{wigner}). In order to make the
scale separation $\eps\ll\delta\ll 1$ precise we define the set
\[
{\cal K}_{\mu}:=\left\{(\eps,\delta):~~|\ln\eps|^{-2/3+\mu}\le\delta\le1
\right\}.
\]
The parameter $\mu$ is a fixed number in the interval $(0,2/3)$.  The
following proposition has been proved in Theorem 3.2 of
\cite{bakoryz}, using straightforward if tedious asymptotic
expansions.
\begin{proposition}\label{prop-bakoryz}
Let the acoustic speed $c_\delta(\vx)$ be of the form
$(\ref{c-fluct-delta})$ and such that the Hamiltonian $H_\delta(\vx)$ given by
$(\ref{ac-ham})$ satisfies assumptions $(\ref{d-i})$. We assume
that the Wigner transform $W_\eps^\delta$ satisfies
\begin{equation}\label{assump-conv-intro}
\hbox{$W_\eps^\delta(0,\vx,\bk)\to W_0(\vx,\bk)$ strongly in
$L^2({\mathbb R}^d\times{\mathbb R}^d)$ as
${\cal K}_\mu\ni(\eps,\delta)\to0$.}
\end{equation}
We also assume that the limit $W_0\in C_c^2({\mathbb
R}^d\times{\mathbb R}^d)$ with a support that satisfies
\begin{equation}\label{assump-conv-2-intro}
\mbox{supp}~ W_0(\vx,\bk)\subseteq {\cal A}(M)
\end{equation}
for some $M>0$.
Moreover, we assume that the initial limit Wigner transform $W_0$ is
of the form
\begin{equation}\label{no-offdiagonal-1}
W_0(\vx,\bk)=\sum_{q=\pm} u_q^0(\vx,\bk)\Pi_q(\bk),~~~\Pi_q(\bk)=\vb_q(\bk)\otimes\vb_q(\bk).
\end{equation}
Let $U^\delta(t,\vx,\bk)=\disp\sum_{p=\pm} u_p^\delta(t,\vx,\bk)\Pi_p(\bk)$,
where the functions $u_p^\delta$ satisfy the Liouville equations
$(\ref{liouv-intro})$.
Then there exists a constant $C_1>0$ that is independent of $\delta$ so that
\begin{equation}\label{eq-prop1}
\|W_\eps^\delta(t,\vx,\bk)-U^\delta(t,\vx,\bk)\|_2\le
C(\delta)
\left( \eps \|W_0\|_{H^2}e^{C_1t/\delta^{3/2}}+
\eps^2\|W_0\|_{H^3}e^{C_1t/\delta^{3/2}}\right)+\|W_\eps^\delta(0)-W_0\|_2,
\end{equation}
where $C(\delta)$ is a rational function of $\delta$  with
deterministic coefficients
that may depend on the constant $M>0$ in the bound
$(\ref{assump-conv-2-intro})$ on the support of $W_0$.
\end{proposition}
The Liouville equations (\ref{liouv-intro}) are of the form (\ref{liouv-intro-0}).
Therefore, one may pass to
the limit $\delta\to 0$ in (\ref{liouv-intro}) using Theorem
\ref{main-thm} and conclude that $\E u_\pm^{\delta}$ converge to the
respective solutions of
\begin{equation}\label{diff-intro}
\pdr{\bar u_\pm}{t}=\sum\limits_{m,n=1}^d\pdr{}{k_m}\left(|\bk|^2D_{mn}(\vhatk)\pdr{\bar
u_\pm}{k_n}\right)\pm c_0\vhatk\cdot\nabla_\vx\bar
u_\pm
\end{equation}
with the  initial conditions as in \eqref{liouv-intro}.
Here  the diffusion matrix $D(\vhatk)=[D_{mn}(\vhatk)]$ is given by
\begin{equation}\label{diff-matrix}
D_{mn}(\vhatk)=-\frac{1}{2}\int_{-\infty}^\infty
\frac{\partial^2R(c_0s\vhatk)}{\partial x_n\partial x_m}ds,
\end{equation}
where $R(\vx)$ is the correlation function of the random field $c_1(\vx)$:
$
\E\left[c_1(\vz)c_1(\vx+\vz)\right]=R(\vx).
$
Furthermore, it follows from Theorem \ref{thm-liouv-diff} that there
exists $\alpha_0>0$ so that solutions of (\ref{liouv-intro}) with the
initial data of the form $u_\pm^{\delta}(0,\vx,\bk)=
u_\pm^0(\delta^{\alpha}\vx,\bk)$ with $0<\alpha<\alpha_0$, converge in
the long time limit to the solutions of the spatial diffusion equation. More
precisely, in that case the function $\bar
u^\delta(t,\vx,\bk)=u_+^{\delta}(t/\delta^{2\alpha},\vx/\delta^\alpha,\bk)$
(and similarly for $u_-^{\delta}$) converges as $\delta\to 0$ to $w(t,\vx,k)$ -- the
solution of the spatial diffusion equation
\begin{eqnarray}\label{diff-space}
&&\pdr{w}{t}=\sum\limits_{m,n=1}^d a_{mn}(k)
\frac{\partial^2 w}{\partial x_n\partial x_m},\\
&&w(0,\bx,k)=\bar u_+^0(\bx;k):=
\frac{1}{\Gamma_{d-1}}\int_{{\mathbb S}^{d-1}}u_+^0(\vx,\bk)d\Omega(\vhatk).\nonumber
\end{eqnarray} with
 the diffusion matrix $a_{mn}$ given by:
\begin{equation}\label{6-wd-diffm-main-again}
a_{nm}(k)=\frac{c_0}{\Gamma_{d-1}}\int_{{\mathbb S}^{d-1}}\hat k_n
\chi_m(\bk)d\Omega(\vhatk),
\end{equation}
and
the functions $\chi_j$ above are the mean-zero solutions of
\begin{equation}\label{6-chim-eq-main-again}
\sum_{m,n=1}^d\pdr{}{k_m}\left(k^2D_{mn}(\vhatk)\pdr{\chi_j}{k_n}\right)=-c_0\hat k_j.
\end{equation}
Theorems \ref{main-thm}, \ref{thm-wave-space-main} and
\ref{thm-liouv-diff} allow us to make the passage to the limit
$\eps,\delta,\gamma\to 0$ rigorous.  The assumption that
$\eps\ll\delta\ll\gamma$ is formalized as follows. We let
\[
{\cal K}_{\mu,\rho}:=\left\{(\eps,\delta,\gamma):~~\delta\ge
  |\ln\eps|^{-2/3+\mu}\mbox{ and }\gamma\ge \delta^{\rho}\right\},
\]
with $0<\mu<2/3$, $\rho\in(0,1)$.  Suppose also that $u_0^\pm\in
C^3_c(\R^{2d}_*)$ and supp $u_0^\pm\subseteq{\cal A}(M)$.
Let
\begin{equation}\label{bums-10}
W^{0}(\bx,\bk):=u_+^0(\bx,\bk)\vb_+(\bk)\otimes \vb_+(\bk)+
u_-^0(\bx,\bk)\vb_-(\bk)\otimes \vb_-(\bk),
\end{equation}
and
\begin{equation}\label{bums-11}
W(t,\bx,\bk):=
w_+(t,\bx;\bk)\vb_+(\bk)\otimes \vb_+(\bk)+w_-(t,\bx;\bk)\vb_-(\bk)\otimes \vb_-(\bk).
\end{equation}
Our main result regarding the diffusion of wave energy can be stated
as follows.
\begin{theorem}
\label{main}
Assume that the dimension $d\ge3$ and $M\ge1$ are fixed.  Suppose for
some $0<\mu<2/3$, $\rho\in(0,1)$ we have, with $W^0$ as in
$(\ref{bums-10})$ and $W_\eps^\delta$ defined by $(\ref{wigner})$
\[
\int\limits_{\R^{2d}}\left|
\E W_\eps^{\delta}\left(0,\frac{\bx}{\gamma},\bk\right)-
W^{0}(\bx,\bk)\right|^2d\bx d\bk\to0,
\mbox{ as }(\eps,\delta,\gamma)\to0\mbox{ and }
(\eps,\delta,\gamma)\in {\cal K}_{\mu,\rho}.
\]
Then, there exists $\rho_1\in(0,\rho]$ such that for any $T>T_*>0$ we have
\[
\sup\limits_{t\in[T_*,T]}
\int\left|\E W_\eps^{\delta}\left(\frac{t}{\gamma^2},\frac{\bx}{\gamma},\bk\right)-
W(t,\bx,\bk)\right|^2d\bx d\bk\to0,
\mbox{ as }(\eps,\delta,\gamma)\to0\mbox{ and }
(\eps,\delta,\gamma)\in {\cal K}_{\mu,\rho_1}.
\]
Here $W(t,\vx,\bk)$ is of the form $(\ref{bums-11})$ with the functions
$w_\pm$ that satisfy \eqref{diff-space} with the initial data
$w_\pm(0,\vx,\bk)=u_\pm^0(\vx,\bk)$.
\end{theorem}
The proof follows immediately from Theorems \ref{main-thm},
\ref{thm-wave-space-main} and
\ref{thm-liouv-diff} as well as Proposition
\ref{prop-bakoryz}.

\begin{appendix}

\section{The proof of Lemma \ref{lmA1}.}
\label{sec:append}

Given $s\geq\si>0$, $\pi\in {\cal C}$ we define the linear
approximation of the trajectory
\begin{equation}\label{72802}
   \bmL(\si,s;\pi):=X(\si)+(s-\si)H_0'(K(\si))\hat K(\si)
\end{equation}
and for any $v\in[0,1]$ let
\begin{equation}\label{72803}
\bmR(v,\si,s;\pi):=(1-v)\bmL(\si,s;\pi)+vX(s).
\end{equation}
The following simple fact can be verified by a direct
calculation, see Lemma 5.4 of \cite{bakoryz}.
\begin{proposition}
\label{lm1}
Suppose that $s\geq\si\ge0$ and $\pi\in {\cal C}(\delta)$. Then,
\[
|X(s)-\bmL(\si,s;\pi)|\leq \tilde D(2M_\delta)\sqrt{\delta}(s-\si)+
\int\limits_{\si}^s|H_0'(K(\rho))\hat{K}(\rho)-H_0'(K(\si))\hat{K}(\si)|d\rho.
\]
\end{proposition}
We obtain from Proposition \ref{lm1} for each $s\ge\si$ an error for
the first-order approximation of the trajectory
\[
|\bz^{(\delta)}(s)-\bl^{(\delta)}(\si,s)|\leq
\tilde D(2M_\delta)\sqrt{\delta}(s-\si)+
\frac{C(s-\si)^2}{2\sqrt{\delta}},
\quad\,\delta\in(0,\delta_*(M)].
\]
Here $\bl^{(\delta)}(\si,s)
:=\bz^{(\delta)}(\si)+(s-\si)\hat{\obm}^{(\delta)}(\si)$ is the linear
approximation between the times $\sigma$ and $s$ and $$
C:=\sup_{\delta\in(0,\delta_*(M)]}(M_\delta h_0^*(M_\delta)+\tilde
h_0^*(M_\delta))\tilde D(2M_\delta).$$
With no loss of generality we may assume that $\bbx=0$ and that there exists $k$ such
that $t,u\in [t_k^{(p)},t_{k+1}^{(p)})$. We shall omit the initial
condition in the notation of the solution to \eqref{eq2}.  Throughout
this argument we use Proposition \ref{lm1} with
\begin{equation}\label{80703}
  \si(s):=s-\delta^{1-\ga_A}\mbox{ for some }\ga_A\in(0,1/16\wedge(1-\ep_4)),\quad s\in[t,u].
\end{equation}
The aforementioned proposition proves that for this choice of $\si$ we have
\begin{equation}\label{80704}
|\bmL^{(\delta)}(\si,s)-\by^{(\delta)}(s)|\le C_A\delta^{3/2-2\ga_A},\quad\forall\,\delta\in(0,1].
\end{equation}
Throughout this section we denote
$\tilde \zeta=\zeta(\by^{(\delta)}(t_1),\bl^{(\delta)}(t_1),\ldots$
$,\by^{(\delta)}(t_n),\bl^{(\delta)}(t_n))$.
We assume first that $G\in C^2(\R^d_*)$ and $\|G\|_2<+\infty$.
Note that
\begin{equation}
\label{51402}
G(\bl^{(\delta)}(u))-
G(\bl^{(\delta)}(t))
=-\frac{1}{\sqrt{\delta}}\sum\limits_{j=1}^d
\int\limits_t^u\partial_{j}G(\bl^{(\delta)}(s))
F_{j,\delta}\left(s,\frac{\by^{(\delta)}(s)}{\delta},\bl^{(\delta)}(s)\right)ds.
\end{equation}
We can rewrite then (\ref{51402}) in the form
$ I^{(1)} + I^{(2)} + I^{(3)},
$
where
\begin{eqnarray*}
&&I^{(1)}:=-\frac{1}{\sqrt{\delta}}
\sum\limits_{j=1}^d\int\limits_t^u\partial_jG(\bl^{(\delta)}(\sigma))
F_{j,\delta}\left(s,\frac{\by^{(\delta)}(s)}{\delta},\bl^{(\delta)}(\sigma)\right)ds,
\\
&&
I^{(2)}:=\frac{1}{\delta}\sum\limits_{i,j=1}^d
\int\limits_t^u\int\limits_\si^s
\partial_jG(\bl^{(\delta)}(\rho))
\partial_{\ell_i}F_{j,\delta}
\left(s,\frac{\by^{(\delta)}(s)}{\delta},\bl^{(\delta)}(\rho)\right)
F_{i,\delta}\left(\rho,\frac{\by^{(\delta)}(\rho)}{\delta},
\bl^{(\delta)}(\rho)\right)ds\,d\rho,\\
&&
I^{(3)}:=\frac{1}{\delta}\sum\limits_{i,j=1}^d
\int\limits_t^u\int\limits_\si^s
\partial^2_{i,j}G(\bl^{(\delta)}(\rho))
F_{j,\delta}\left(s,\frac{\by^{(\delta)}(s)}{\delta},\bl^{(\delta)}(\rho)\right)
F_{i,\delta}\left(\rho,\frac{\by^{(\delta)}(\rho)}{\delta},\bl^{(\delta)}(\rho)
\right)ds\,d\rho
\end{eqnarray*}
and $\sigma$ is given by \eqref{80703}. Each of these terms will be
estimated separately below.

\subsection{The term  $\bE[I^{(1)}\tilde\zeta]$}\label{seca41}

The term $I^{(1)}$ can be rewritten in the form
$
J^{(1)}+J^{(2)},
$
where
\[
J^{(1)}:=-\frac{1}{\sqrt{\delta}}\sum\limits_{j=1}^d
\int\limits_t^u\partial_jG(\bl^{(\delta)}(\sigma))
F_{j,\delta}\left(s,\frac{\bmL^{(\delta)}(\si,s)}{\delta},\bl^{(\delta)}(\sigma)
\right)ds,
\]
and
\begin{equation}
J^{(2)}:=-\frac{1}{\delta^{3/2}}\sum\limits_{i,j=1}^d
\int\limits_t^u\int\limits_0^1\partial_jG(\bl^{(\delta)}(\sigma))
\partial_{y_i}F_{j,\delta}
\left(s,\frac{\bmR^{(\delta)}(v,\si,s)}{\delta},
\bl^{(\delta)}(\si)\right)
(y_{i}^{(\delta)}(s)-L_{i}^{(\delta)}(\si,s))\,ds\,dv,\label{53105}
\end{equation}
where, see \eqref{72802} and \eqref{72803},
 $\bmL^{(\delta)}(\si,s)=\bmL(\si,s;\by^{(\delta)}(\cdot),\bl^{(\delta)}(\cdot))$,
 $\bmR^{(\delta)}(\si,s)=\bmR(\si,s;\by^{(\delta)}(\cdot),\bl^{(\delta)}(\cdot))$.
We use part (i) of Lemma \ref{mix1} to
handle the term $\bE[J^{(1)}\tilde\zeta]$. Let
$\tilde X_1(\bbx,k)=-\partial_{x_i}H_1(\bbx,k)$, $\tilde X_2(\bbx,k)\equiv1$,
\[
Z=\Theta\left(t^{(p)}_{k},\frac{\bmL^{(\delta)}(\si,s)}{\delta},\bl^{(\delta)}(\si)
\right)
\partial_jG(\bl^{(\delta)}(\si))\tilde\zeta
\]
and $g_1=(\bmL^{(\delta)}(\si,s)\delta^{-1},|\bl^{(\delta)}(\si)|)$.
Note that $g_1$ and $Z$ are both $\cal F_\si$ measurable. We need to
verify (\ref{70202}). Suppose therefore that $Z\not=0$.  For
$\rho\in[0, t^{(p)}_{k-1}]$ we have
$|\bmL^{(\delta)}(\si,s)-\by^{(\delta)}(\rho)|\geq (2q)^{-1}$,
provided that $C_A\delta^{3/2-2\ga_A}<1/(2q)$, which holds for
sufficiently small $\delta$, since our assumptions on the exponents
$\ep_2,\ep_3,\ga_A$ (namely that $\ep_2,\ep_3<1/8$, $\ga_A<1/8$)
guarantee that $\ep_2+\ep_3<3/4-\ga_A/2$. For
$\rho\in[t^{(p)}_{k-1},\si]$ we have
\begin{eqnarray}
\label{80705}
&&(\bmL^{(\delta)}(\si,s)-\by^{(\delta)}(\rho))\cdot
\hat{\bl}^{(\delta)}\left(t^{(p)}_{k-1}\right)\geq
(s-\si)H_0'(|\bl^{(\delta)}(\si)|)\,\hat{\bl}^{(\delta)}\left(\si\right)\cdot
\hat{\bl}^{(\delta)}\left(t^{(p)}_{k-1}\right)
\\
&&+
\int\limits_\rho^\si\left[H_0'(|\bl^{(\delta)}(\rho_1)|)+\sqrt{\delta}\,
\partial_l H_1\left(\frac{\by^{(\delta)}(\rho_1)}{\delta},|\bl^{(\delta)}(\rho_1)|
\right)\right]
\hat{\bl}^{(\delta)}\left(\rho_1\right)\cdot
\hat{\bl}^{(\delta)}\left(t^{(p)}_{k-1}\right)d\rho_1\nonumber\\
&&
\stackrel{\eqref{80402}}{\geq}(s-\si)h_*(2M_\delta)\,\left(1-\frac{2}{N}\right)+\left[h_*(2M_\delta)-
\sqrt{\delta}
\tilde D(2M_\delta)\right](s-\rho)\left(1-\frac{2}{N}\right)
\nonumber\\
&&
\geq(s-\si)h_*(2M_\delta)\,
\left(1-\frac{2}{N}\right),\nonumber
\end{eqnarray}
provided that $\delta\in(0,\delta_0]$ and $\delta_0$ is sufficiently
small. We see from (\ref{80705}) that
(\ref{70202}) is satisfied with $r=\left(1-2/N\right)h_*(2M_\delta)\delta^{1-\ga_A}$.
Using Lemma \ref{mix1} we estimate
\begin{eqnarray}
\label{70502}
&&\left|\bbE[J^{(1)}\tilde\zeta]\right|
\leq \frac{\tilde D(2M_\delta)}{\sqrt{\delta}}
\|G\|_{1}
\bbE[\tilde\zeta]\int\limits_t^u
\,\phi\left(C_{A}^{(1)}\frac{s-\si}{\delta}\right)ds
\\
&&~~~\leq C_{A}^{(2)}\| G\|_{1}
\bbE[\tilde\zeta]\delta^{-1/2}\phi\left(C_{A}^{(1)}\delta^{-\ga_A}\right)
(u-t)\leq C_{A}^{(3)}\| G\|_{1}
\bbE[\tilde\zeta]\delta(u-t)\nonumber
\end{eqnarray}
 and $C_{A}^{(3)}$
exists by virtue of assumption
\eqref{DR}.
On the other hand, the term $J^{(2)}$ defined by (\ref{53105}) may be
written as
$J^{(2)}=J^{(2)}_1+J^{(2)}_2,
$
where
$$
J^{(2)}_1:=-\frac{1}{\delta^{3/2}}\sum\limits_{i,j=1}^d
\int\limits_t^u \partial_jG(\bl^{(\delta)}(\si))
\partial_{y_i}F_{j,\delta}
\left(s,\frac{\bmL^{(\delta)}(\si,s)}{\delta},
\bl^{(\delta)}(\si)\right)(y_{i}^{(\delta)}(s)-L_{i}^{(\delta)}(\si,s))\,ds
$$
and
\begin{eqnarray}
\label{53106}
&&J^{(2)}_2:=-\frac{1}{\delta^{5/2}}
\sum\limits_{i,j,k=1}^d
\int\limits_t^u\int\limits_0^1\int\limits_0^1
\partial^2_{y_i,y_k}F_{j,\delta}\left(s,\frac{\bmR^{(\delta)}(\theta
v,\si,s)}{\delta},\bl^{(\delta)}(\si)\right)\,v\\
&&~~~~~~~~\times \partial_{j}G(\bl^{(\delta)}(\si))
(y_{i}^{(\delta)}(s)-L_{i}^{(\delta)}(\si,s))
(y_{k}^{(\delta)}(s)-L_{k}^{(\delta)}(\si,s))\,ds\,dv\,d\theta.\nonumber
\end{eqnarray}
The term involving $J^{(2)}_2$ may be handled easily with the help of
\eqref{80704} and Lemma \ref{lm3}. We have then
\begin{equation}
\label{53107} |\bbE[J^{(2)}_2\tilde\zeta]|\leq C_{A}^{(4)}\tilde
D(2M_\delta)\bbE[\tilde\zeta] \|
G\|_{1}(u-t)\delta^{-5/2}\delta^{3-4\ga_A-4(\ep_2+\ep_3)}T^2
\end{equation}
$$
\leq
C_{A}^{(5)}\delta^{1/2-4\ga_A-4(\ep_2+\ep_3)}T^2(u-t)\bbE[\tilde\zeta]
\| G\|_{1}.
$$
In order to estimate the term corresponding to $J^{(2)}_1$ we write
$
J^{(2)}_1=J^{(2)}_{1,1}+J^{(2)}_{1,2},
$
where
\begin{eqnarray}
&&J^{(2)}_{1,1}:=-\frac{1}{\delta^{3/2}}
\sum\limits_{i,j=1}^d\int\limits_t^u\int\limits_\si^s
 \partial_jG(\bl^{(\delta)}(\si))
\partial_{y_i}
F_{j,\delta}\left(s,\frac{\bmL^{(\delta)}(\si,s)}{\delta},\bl^{(\delta)}(\si)\right)\,
\label{J_11^2}\\
&&~~~~~~~~~~~
\times(s-\rho_1)\frac{d}{d\rho_1}\left[H_0'(|\bl^{(\delta)}(\rho_1)|)\,
\hat{l}_{i}^{(\delta)}(\rho_1)\right]\,ds\,d\rho_1\nonumber
\end{eqnarray}
and
\[
J^{(2)}_{1,2}:=\!-\frac{1}{\delta}\sum\limits_{i,j=1}^d
\int\limits_t^u\int\limits_\si^s\!
\partial_jG(\bl^{(\delta)}(\si))
\partial_{y_i}F_{j,\delta}
\left(s,\frac{\bmL^{(\delta)}(\si,s)}{\delta},\bl^{(\delta)}(\si)\right)
\partial_l H_1\!\left(\frac{\by^{(\delta)}(\rho)}{\delta},|\bl^{(\delta)}(\rho)|
\right)
\hat{l}_{i}^{(\delta)}(\rho)\,ds\,d\rho,
\]
with
\begin{eqnarray}
\label{x1}
&&\frac{d}{d\rho_1}\left[H_0'(|\bl^{(\delta)}(\rho_1)|)\,\hat{l}_{i}^{(\delta)}
(\rho_1)
\right]=
H_0'\!'(|\bl^{(\delta)}(\rho_1)|)\,(\hat{\bl}^{(\delta)}(\rho_1),
\frac{d}{d\rho_1}\,\bl^{(\delta)}
(\rho_1))_{\R^d}\,\hat{l}_{i}^{(\delta)}(\rho_1)
\\
&&+H_0'(|\bl^{(\delta)}(\rho_1)|)|\bl^{(\delta)}(\rho_1)|^{-1}
\left[\frac{d}{d\rho_1}\,l_{i}^{(\delta)}(\rho_1)-
(\hat{\bl}^{(\delta)}(\rho_1),\frac{d}{d\rho_1}\,\bl^{(\delta)}
(\rho_1))_{\R^d}\,\hat{l}_{i}^{(\delta)}(\rho_1)\right].\nonumber
\end{eqnarray}
We deal with $J^{(2)}_{1,2}$ first. It may be split as
$J^{(2)}_{1,2}=J^{(2)}_{1,2,1}+J^{(2)}_{1,2,2}+J^{(2)}_{1,2,3}$, where
\begin{eqnarray}
\label{70612}
&&J^{(2)}_{1,2,1}:=-\frac{1}{\delta}\sum\limits_{i,j=1}^d\int\limits_t^u
(s-\si)\partial_jG(\bl^{(\delta)}(\si))
\partial_{y_i}F_{j,\delta}
\left(s,\frac{\bmL^{(\delta)}(\si,s)}{\delta},\bl^{(\delta)}(\si)\right)
\\
&&~~~~~~~~~~
\times
\partial_l H_1\left(\frac{\bmL^{(\delta)}(\si,\rho)}{\delta},|\bl^{(\delta)}
(\si)|\right)
\hat{l}_{i}^{(\delta)}(\si)\,ds\,\nonumber
\end{eqnarray}
\begin{eqnarray*}
&&J^{(2)}_{1,2,2}:=
-\frac{1}{\delta^2}\sum\limits_{i,j=1}^d
\int\limits_t^u\int\limits_\si^s\int\limits_0^1
\partial_jG(\bl^{(\delta)}(\si))
\partial_{y_i}F_{j,\delta}
\left(s,\frac{\bmL^{(\delta)}(\si,s)}{\delta},\bl^{(\delta)}(\si)\right)\\
&&~~~~~~~~~~\times
(\partial_{y_i}\partial_l H_1)
\left(\frac{\bmR^{(\delta)}(v,\si,\rho)}{\delta},|\bl^{(\delta)}(\rho)|\right)
(y_{i}^{(\delta)}(\rho)-L_{i}^{(\delta)}(\si,\rho))\,
\hat{l}_{i}^{(\delta)}(\rho)\,ds\,d\rho\,dv
\end{eqnarray*}
and
\begin{eqnarray*}
&&J^{(2)}_{1,2,3}:=-\frac{1}{\delta}\sum\limits_{i,j=1}^d
\int\limits_t^u\int\limits_\si^s\int\limits_\si^{\rho}
\partial_jG(\bl^{(\delta)}(\si))
\partial_{y_i}F_{j,\delta}
\left(s,\frac{\bmL^{(\delta)}(\si,s)}{\delta},\bl^{(\delta)}(\si)\right)\\
&&~~~~~~~~~~\times
\frac{d}{d\rho_1}\left[\partial_l
H_1\left(\frac{\bmL^{(\delta)}(\si,s)}{\delta},|\bl^{(\delta)}(\rho_1)|\right)
\,\hat{l}_{i}^{(\delta)}(\rho_1)\right]\,ds\,d\rho\,d\rho_1.
\end{eqnarray*}
By virtue of (\ref{80704}), definition
(\ref{70613}) and Lemma \ref{lm3} we obtain easily that
\begin{equation}\label{80301}
 |\bbE[J^{(2)}_{1,2,2}\tilde\zeta]|\le C_{A}^{(6)}\delta^{1/2-(3\ga_A+2\ep_2+2\ep_3)}\|
G\|_{1}T(u-t)\bE\tilde\zeta.
\end{equation}
The same argument and equality \eqref{x1} also allow us to estimate
$|\bbE[J^{(2)}_{1,2,3}\zeta]|$ by the right hand side of
(\ref{80301}).

Using Lemma \ref{lm3} and the definition (\ref{70613}) we conclude
that there exists a constant $C_{A}^{(7)}>0$ independent of $\delta$ such that
\begin{eqnarray*}
&&\left|\partial_{y_i}F_{j,\delta}\left(s,\frac{\bmL^{(\delta)}(\si,s)}{\delta},
\bl^{(\delta)}(\si)\right)-\Theta\left(t^{(p)}_{k},\bmL^{(\delta)}(\si,s),
\bl^{(\delta)}(\si)\right)
\partial^2_{y_i,y_j}H_1\left(
\frac{\bmL^{(\delta)}(\si,s)}{\delta},|\bl^{(\delta)}(\si)|\right)
\right|\\
&&\le C_{A}^{(7)}\delta^{1-2(\ep_2+\ep_3)}T,\quad\,i,j=1,\ldots,d.
\end{eqnarray*}
Therefore, we can write
\begin{eqnarray}
\label{60103a}
&&\left|\bE[J^{(2)}_{1,2,1}\tilde\zeta]+\frac{1}{\delta}\sum\limits_{i,j=1}^d\int\limits_t^u\int\limits_\si^s
\bE\left[\partial_jG(\bl^{(\delta)}(\si))
\Theta\left(t^{(p)}_{k},\bmL^{(\delta)}(\si,s),
\bl^{(\delta)}(\si)\right)\right.\right.\\
&&\times
\left.\left.\partial^2_{y_i,y_j}H_1\left(\frac{\bmL^{(\delta)}(\si,s)}{\delta},|\bl^{(\delta)}(\si)|\right)
\partial_l H_1\left(\frac{\bmL^{(\delta)}(\si,\rho)}{\delta},|\bl^{(\delta)}(\si)|\right)\hat{l}_{i}^{(\delta)}(\si)\,\tilde\zeta\right]\,ds\,d\rho\right|\nonumber
\\
&&\le
C_{A}^{(8)}\delta^{1-\ga_A-2(\ep_2+\ep_3)}(u-t)\|
G\|_{1}T\bE\tilde\zeta.\nonumber
\end{eqnarray}
With our choice of the exponents we have
$\delta<(2q)^{[1-\ga_A-2(\ep_2+\ep_3)]^{-1}}$ for all
$\delta\in[0,\delta_0)$ where $\delta_0>0$ is sufficiently small. We
apply now part (ii) of Lemma
\ref{mix1} with
\begin{eqnarray*}
&&Z=\partial_jG(\bl^{(\delta)}(\si))
\Theta\left(t^{(p)}_{k},\bmL^{(\delta)}(\si,s),
\bl^{(\delta)}(\si)\right)
\hat{l}_{i}^{(\delta)}(\si)\,\tilde\zeta,\\
&& \tilde X_1(\bx,k):=\partial^2_{x_i,x_j}H_1(\bbx,k),\quad  \tilde X_2(\bx):=\partial_k H_1(\bbx,k),\\
&&
 g_1:=\left(\frac{\bmL^{(\delta)}(\si,s)}{\delta},|\bl^{(\delta)}(\si)|\right),\quad
 g_2:=\left(\frac{\bmL^{(\delta)}(\si,\rho)}{\delta},|\bl^{(\delta)}(\si)|\right),\\
 &&
 r=C_{A}^{(9)}(\rho-\si),\quad\,r_1=C_{A}^{(9)}(s-\rho).
\end{eqnarray*}
  We conclude that
\begin{eqnarray}
\label{60103}
&&\left|\bbE\left[J^{(2)}_{1,2,1}\zeta\right]+\frac{1}{\delta}
\sum\limits_{i,j=1}^d\int\limits_t^u\int\limits_\si^s\bbE\left[
\partial_jG(\bl^{(\delta)}(\si))
\Theta\left(t^{(p)}_{k},\bmL^{(\delta)}(\si,s),
\bl^{(\delta)}(\si)\right)\right.\right.\\
&&
\left.\left.\times\partial^2_{y_i,y_j} R_1\left(\frac{\bmL^{(\delta)}(\si,s)-\bmL^{(\delta)}(\si,\rho)}{\delta},\,
 |\bl^{(\delta)}(\si)|\right)\hat{l}_{i}^{(\delta)}(\si)
\tilde\zeta\right]
\,ds\,d\rho
\right|\nonumber\\
&&
\leq C_A^{(8)}\delta^{1-\ga_A-2(\ep_2+\ep_3)}(u-t)\|
G\|_{1}T\bE\tilde\zeta \nonumber\\
&&+\frac{C_{A}^{(10)}}{\delta}\|G\|_{1}\bE[\tilde\zeta]
\int\limits_t^u\int\limits_\si^s
\phi^{1/2}\left(\,\frac{C_{A}^{(9)}(\rho-\si)}{2\delta}\right)
\phi^{1/2}\left(\,\frac{C_{A}^{(9)}(s-\rho)}{2\delta}\right)\,ds\,d\rho,\nonumber
\end{eqnarray}
where
\begin{equation}\label{011210}
R_1(\bby,k):=\bbE[H_1(\bby,k)\partial_kH_1(\bze,k)],\, \quad(\bby,k)\in\R^d\times[0,+\infty).
\end{equation}
We can use assumption \eqref{DR}
to estimate the second term on the  right hand side of (\ref{60103}) e.g. by
$C_{A}^{(11)}\delta(u-t)\|
G\|_{1}\bE\tilde\zeta$. The second term appearing on the left
hand side of (\ref{60103}) equals to
\begin{equation}
\label{70206}
\!\!\!\!\!\!\!\!\!\!\!\!\!\!\!\!\!\!\!\!\!\!\!\!\!\!\!\!\!\!\!\!\!\!\!\!\!\!\!\!\!\!\!\!\!\!\!\!
\sum\limits_{j=1}^d\int\limits_t^u\bbE\left\{
\partial_jG(\bl^{(\delta)}(\si))
\Theta\left(t^{(p)}_{k},\bmL^{(\delta)}(\si,s),
\bl^{(\delta)}(\si)\right)\right.~~~~~~~~~~~~~~~~~~
\end{equation}
$$
\left.\times\frac{1}{H_0'(|\bl^{(\delta)}(\si)|)}\,\left[-\int\limits_\si^s\frac{d}{d\rho}
\partial_{y_j}R_1\left(\frac{s-\rho}{\delta}\,
H_0'(|\bl^{(\delta)}(\si)|)\,\hat{\bl}^{(\delta)}(\si),|\bl^{(\delta)}(\si)|\right)\,
d\rho\right]\tilde\zeta\right\}\,ds
$$
and integrating over $d\rho$ we obtain
that it equals
\begin{equation}
\label{70206b}
-\sum\limits_{j=1}^d\int\limits_t^u\bbE\left\{\frac{\partial_jG(\bl^{(\delta)}(\si))}{H_0'(|\bl^{(\delta)}(\si)|)}
\,\Theta\left(t^{(p)}_{k},\bmL^{(\delta)}(\si,s),
\bl^{(\delta)}(\si)\right)
\partial_{y_j}R_1\left(\bze,|\bl^{(\delta)}(\si)|\right)\,
\tilde\zeta\right\}\,ds
\end{equation}
$$
+
\sum\limits_{j=1}^d\int\limits_t^u\bbE\left\{
\frac{\partial_jG(\bl^{(\delta)}(\si))}{H_0'(|\bl^{(\delta)}(\si)|)}
\,
\Theta\left(t^{(p)}_{k},\bmL^{(\delta)}(\si,s),
\bl^{(\delta)}(\si)\right)
\partial_{y_j}R_1\left(\delta^{-\ga_A}\,\hat{\bl}^{(\delta)}(\si),
|\bl^{(\delta)}(\si)|\right)\,\tilde\zeta\right\}\,ds.
$$
By virtue of \eqref{53102-intro}
the second term appearing
in (\ref{70206b})
is bounded e.g. by $C_{A}^{(12)}\delta(u-t)\|
G\|_{1}\bE\tilde\zeta$ for some constant $C_{A}^{(12)}>0$,
thus we have shown that
\begin{equation}\label{021104}
\left|\bbE[J^{(2)}_{1,2,1}\zeta]-\sum\limits_{j=1}^d\int\limits_t^u\bbE\left\{
\frac{\partial_jG(\bl^{(\delta)}(\si))}{H_0'(|\bl^{(\delta)}(\si)|)}
\,
\Theta\left(t^{(p)}_{k},\bmL^{(\delta)}(\si,s),
\bl^{(\delta)}(\si)\right)
\partial_{y_j}R_1\left(\bze,|\bl^{(\delta)}(\si)|\right)\,
\tilde\zeta\right\}\,ds\right|
\end{equation}
$$
\le C_{A}^{(13)}\delta^{1-\ga_A-2(\ep_2+\ep_3)}(u-t)\|
G\|_{1}T\bE\tilde\zeta.
$$

Let us consider the term corresponding to $J_{1,1}^{(2)}$, cf. (\ref{J_11^2}). Note that
according to (\ref{x1}) and (\ref{eq2}) we have
$
J^{(2)}_{1,1}=J^{(2)}_{1,1,1}+J^{(2)}_{1,1,2},
$
where
\begin{eqnarray*}
&&J^{(2)}_{1,1,1}:= -\frac{1}{\delta^{2}}\sum\limits_{i,j=1}^d
\int\limits_t^u\int\limits_\si^s
\partial_{j}G(\bl^{(\delta)}(\si))
\partial_{y_i}F_{j,\delta}
\left(s,\frac{\bmL^{(\delta)}(\si,s)}{\delta},\bl^{(\delta)}(\si)\right)\\
&&~~~~~~~~\times
(s-\rho_1)\Gamma_{i}\left(\rho_1,
\frac{\by^{(\delta)}(\rho_1)}{\delta},\bl^{(\delta)}(\si)\right)
\,ds\,d\rho_1,
\end{eqnarray*}
with
\[
\Gamma_i\left(\rho,\bby,\bbl\right):=|\bbl|^{-1}H_0'(|\bbl|)\left[
\left(\hat{\bbl},F_{\delta}\left(\rho,\bby,\bbl\right)\right)_{\R^d}
l_i-F_{i,\delta}\left(\rho,\bby,\bbl\right)\right]-
H_0'\!'(|\bbl|)\left(\hat{\bbl},F_{\delta}\left(\rho,\bby,\bbl\right)\right)_{\R^d}
\hat l_i,
\]
while
\begin{eqnarray}
&& J^{(2)}_{1,1,2}:= -\frac{1}{\delta^{2}}\sum\limits_{i,j=1}^d
\int\limits_t^u\int\limits_\si^s
\int\limits_\si^{\rho_1}
\partial_{j}G(\bl^{(\delta)}(\si))
\partial_{y_i}F_{j,\delta}\left(s,\frac{\bmL^{(\delta)}(\si,s)}{\delta},\bl^{(\delta)}(\si)\right)\nonumber\\
&&~~~~~~~~\times(s-\rho_1)\frac{d}{d\rho_2}\Gamma_{i}
\left(\rho_1,\frac{\by^{(\delta)}(\rho_1)}{\delta},
\bl^{(\delta)}(\rho_2)\right)\,
ds\,d\rho_1\,d\rho_2.\label{80801}
\end{eqnarray}
Note that $|\frac{d}{d\rho_2}\Gamma_{i}|\le C_{A}^{(14)} \delta^{-1/2}$
for some constant $C_{A}^{(14)}>0$. A straightforward computation, using \eqref{80703} and Lemma \ref{lm3}, shows that
$
|\bbE[J^{(2)}_{1,1,2}\zeta]|\leq C_{A}^{(15)}\delta^{1/2-(3\ga_A+2\ep_2+2\ep_3)}(u-t)\|G\|_{1}T\bbE[\tilde\zeta].
$
An application of \eqref{80704}, in the same fashion as it was
done in the calculations concerning the terms $\bE[J^{(2)}_{1,2,2}\zeta]$ and
$\bE[J^{(2)}_{1,2,3}\zeta]$,
yields that
\begin{eqnarray}
\label{52701} &&\left|\bbE[J^{(2)}_{1,1,1}\zeta]+ \frac{1}{\delta^{2}}
\sum\limits_{i,j=1}^d\int\limits_t^u\int\limits_\si^s (s-\rho_1)
\bbE\left[\partial_{j}G(\bl^{(\delta)}(\si))
\partial_{y_i}F_{j,\delta}\left(s,\frac{\bmL^{(\delta)}(\si,s)}{\delta},\bl^{(\delta)}(\si)\right) \right.\right.\\
&&\times\left.\left.\Gamma_{i}\left(\rho_1,
\frac{\bmL^{(\delta)}(\si,\rho_1)}{\delta},\bl^{(\delta)}(\si)\right)\tilde\zeta\right]
\,ds\,d\rho_1\right|\le C_{A}^{(16)}\delta^{1/2-(4\ga_A+2\ep_2+2\ep_3)}(u-t)\|G\|_{1}T\bbE[\tilde\zeta].\nonumber
\end{eqnarray}
For $j=1,\ldots,d$ we let
\[
V_{j}(\bby,\bby',\bbl):=\sum\limits_{i,k=1}^d\left(
H_0'\!'(|\bbl|)-H_0'(|\bbl|)\right)\partial_{y_i,y_j,y_k}^3
R(\bby-\bby',|\bbl|)\hat{l}_{i}\hat{l}_{k}
\]
\[
+\sum\limits_{i=1}^d
H_0'(|\bbl|)|\bbl|^{-1}\partial_{y_i,y_i,y_j}^3
R(\bby-\bby',|\bbl|),
\]
and also
\begin{equation}\label{70605b}
\Lambda(t,\bby,\bby',\bbl;\pi):=\Theta(t,\bby,\bbl;\pi)\Theta(t,\bby',\bbl;\pi),\quad
t\ge0,\,\bby,\bby'\in\R^d,\,\bbl\in\R^d_*,\,\pi\in{\cal C},
\end{equation}
$
P:=\left(\bmL^{(\delta)}(\si,s),\bmL^{(\delta)}(\si,\rho_1)
,\bl^{(\delta)}(\si)\right)
$, $
P_\delta:=\left(\delta^{-1}\bmL^{(\delta)}(\si,s),\delta^{-1}\bmL^{(\delta)}(\si,\rho_1)
,\bl^{(\delta)}(\si)\right)
$ and
$$
\overline{\Theta}(s):=
\Theta(s,\by^{(\delta)}(s),\bl^{(\delta)}(s);\by^{(\delta)}(\cdot),\bl^{(\delta)}(\cdot)).
$$
Applying Lemma  \ref{lm3} and part ii) of Lemma \ref{mix1}, as in (\ref{60103a})
and (\ref{60103}), we conclude that the difference between the second term on the left hand side of
(\ref{52701}) and
\begin{equation}
\label{53001}
 \frac{1}{\delta^{2}}\sum\limits_{j=1}^d
\int\limits_t^u\int\limits_\si^s (s-\rho_1)
\bbE\left[\partial_{j}G(\bl^{(\delta)}(\si))
\Lambda(\si,P)V_{j}\left(P_\delta\right)\,\tilde\zeta\right]
\,ds\,d\rho_1,
\end{equation}
can be estimated by
$C_{A}^{(17)}\delta^{\gamma_A^{(1)}}(u-t)\|
G\|_{1}\bbE[\tilde\zeta]$ for some $\gamma_A^{(1)}>0$.
Using the fact that
 \begin{equation}\label{051210}
|\bl^{(\delta)}(\rho)-\bl^{(\delta)}(\si)|\le C_A^{(22)} \delta^{1/2-\ga_A},\quad \rho\in[\si,s],
\end{equation}
estimate \eqref{80704} and Lemma \ref{lm3} we can argue that
$$
\left|\Lambda\left(\si,P\right)-\overline{\Theta}^2(s)\right|
\leq C_{A}^{(18)}(\delta^{1/2-\ga_A-\ep_1}+\delta^{1/2-2(\ga_A+\ep_2+\ep_3)}T).
$$
 We conclude therefore that the magnitude of the difference between
the expression in (\ref{53001})
and \begin{equation}
\label{53101} \frac{1}{\delta^{2}}\sum\limits_{j=1}^d
\int\limits_{t}^u
\bbE\left[\partial_{j}G(\bl^{(\delta)}(\si))\overline{\Theta}^2(s)
\left( \int\limits_\si^s
(s-\rho_1)V_{j}(P_\delta)\,d\rho_1\right)\,\tilde\zeta\right]\,ds,
\end{equation}
can be estimated by
$C_{A}^{(19)}\delta^{\gamma_A^{(2)}}(u-t)\|G\|_{1}T\bbE[\tilde\zeta]$
for some $\gamma_A^{(2)}>0$.
Using shorthand notation $Q(\si):=H_0'(|\bl^{(\delta)}(\si)|)\,\hat{\bl}^{(\delta)}(\si)$
we can write  the integral from $\si$ to $s$ appearing above as being equal to
\begin{eqnarray*}
&&  \frac{1}{\delta^{2}}
\int\limits_{s-\delta^{1-\ga_A}}^s
 (s-\rho_1) \left[\sum\limits_{i,k=1}^d\left(
H_0'\!'(|\bl^{(\delta)}(\si)|)-H_0'(|\bl^{(\delta)}(\si)|)\right)\partial_{y_i,y_j,y_k}^3
R\left(\frac{s-\rho_1}{\delta}\,Q(\si),|\bl^{(\delta)}(\si)|\right)
\right.
\nonumber\\
&&\left.\times\hat{l}_{i}^{(\delta)}(\si)\hat{l}_{k}^{(\delta)}(\si)+H_0'(|\bl^{(\delta)}(\si)|)
|\bl^{(\delta)}(\si)|^{-1}\sum\limits_{i=1}^d\partial_{y_i,y_i,y_j}^3
 R\left(\frac{s-\rho_1}{\delta}\,Q(\si),|\bl^{(\delta)}(\si)|\,\right)
\right]\,d\rho_1,\nonumber\\
\end{eqnarray*}
which upon the change of variables $\rho_1:=(s-\rho_1)/\delta$ is equal to
\begin{eqnarray}&&
\int\limits_{0}^{\delta^{-\ga_A}}
 \rho_1 \left[\sum\limits_{i,k=1}^d\left(
H_0'\!'(|\bl^{(\delta)}(\si)|)-H_0'(|\bl^{(\delta)}(\si)|)\right)\partial_{y_i,y_j,y_k}^3
R\left(\rho_1\,Q(\si),|\bl^{(\delta)}(\si)|\right)\hat{l}_{i}^{(\delta)}(\si)\hat{l}_{k}^{(\delta)}(\si)
\right.\nonumber
\\
\label{70606}&&\left.+
H_0'(|\bl^{(\delta)}(\si)|)|\bl^{(\delta)}(\si)|^{-1}\sum\limits_{i=1}^d\partial_{y_i,y_i,y_j}^3
 R\left(\rho_1\,Q(\si),|\bl^{(\delta)}(\si)|\,\right)
\right]\,d\rho_1.
\end{eqnarray}
Using the fact that
\[
\sum\limits_{k=1}^d\partial_{y_i,y_j,y_k}^3R\left(\rho_1\,Q(\si),
|\bl^{(\delta)}(\si)|\right)
 \hat{l}_{k}^{(\delta)}(\si)
 =\frac{1}{H_0'(|\bl^{(\delta)}(\si)|)}\,\frac{d}{d\rho_1}\left[
\partial_{y_i,y_j}^2R\left(\rho_1\,Q(\si),|\bl^{(\delta)}(\si)|\right)\right]
\]
we obtain, upon  integrating by parts  in the first
term on the  right hand side of (\ref{70606}), that this
expression equals
\begin{eqnarray}
\label{70607}
&&H_0'(|\bl^{(\delta)}(\si)|)^{-1}\left(
H_0'\!'(|\bl^{(\delta)}(\si)|)-H_0'(|\bl^{(\delta)}(\si)|)\right)\sum\limits_{i=1}^d
\left[\vphantom{\int\limits_0^1}\delta^{-\ga_A}\partial_{y_i,y_j}^2
R\left(\delta^{-\ga_A}\,Q(\si),|\bl^{(\delta)}(\si)|\right)
\hat{l}_{i}^{(\delta)}(\si)\right.
\nonumber\\
&&\left.-\int\limits_0^{\delta^{-\ga_A}}
\partial_{y_i,y_j}^2R\left(\rho_1\,Q(\si),|\bl^{(\delta)}(\si)|\right)\hat{l}_{i}^{(\delta)}(\si)
\,d\rho_1\right]\\
&&+
H_0'(|\bl^{(\delta)}(\si)|)|\bl^{(\delta)}(\si)|^{-1}\int\limits_0^{\delta^{-\ga_A}}
\rho_1  \partial_{y_i,y_i,y_j}^3R\left(\rho_1\,Q(\si),|\bl^{(\delta)}(\si)|\right)\,d\rho_1.\nonumber\\
\end{eqnarray}
Note that $\nabla R(\bze)=\bze$  and
\[
\sum\limits_{i=1}^d\partial_{y_i,y_j}^2R\left(\rho_1\,Q(\si),|\bl^{(\delta)}(\si)|\right)
\hat{l}_{i}^{(\delta)}(\si)
=\frac{1}{H_0'(|\bl^{(\delta)}(\si)|)}\,\frac{d}{d\rho_1}\partial_{y_j}
R\left(\rho_1\,Q(\si),|\bl^{(\delta)}(\si)|\right).
\]
We obtain therefore that the expression in
\eqref{70607} equals
\begin{eqnarray}
&&
H_0'(|\bl^{(\delta)}(\si)|)^{-1}\left(
H_0'\!'(|\bl^{(\delta)}(\si)|)-H_0'(|\bl^{(\delta)}(\si)|)\right)\left[\sum\limits_{i=1}^d
\vphantom{\int\limits_0^1}\delta^{-\ga_A}\partial_{y_i,y_j}^2
R\left(\delta^{-\ga_A}\,Q(\si),|\bl^{(\delta)}(\si)|\right)
\hat{l}_{i}^{(\delta)}(\si)\right.\nonumber\\
&&\!\!\!\left.-H_0'(|\bl^{(\delta)}(\si)|)^{-1}\partial_{y_j}
R\left(\delta^{-\ga_A}\,Q(\si),|\bl^{(\delta)}(\si)|\right)
\vphantom{\int\limits_0^1}\right]\label{031004}\\
&&+
H_0'(|\bl^{(\delta)}(\si)|)|\bl^{(\delta)}(\si)|^{-1}\sum\limits_{i=1}^d\int\limits_0^{\delta^{-\ga_A}}
\rho_1  \partial_{y_i,y_i,y_j}^3R\left(\rho_1\,Q(\si),|\bl^{(\delta)}(\si)|
\right)\,d\rho_1.\nonumber
\end{eqnarray}
Recalling assumption \eqref{53102-intro} we conclude that the expressions corresponding to
the first two terms appearing in \eqref{031004}
are of order of magnitude $O(\delta^{\gamma_A^{(3)}})$
for some $\gamma_A^{(3)}>0$. Summarizing  work done in this section, we have shown that
\begin{equation}\label{80305}
  \left|\bE\left\{\left[I^{(1)}-\sum\limits_{j=1}^d
\int\limits_{t}^u
C_j(\bl^{(\delta)}(\si))\overline{\Theta}^2(s)\partial_{j}G(\bl^{(\delta)}(\si))\,ds
\right]\,\tilde\zeta\right\}\right|\leq
 C_{A}^{(20)}\delta^{\ga_A^{(4)}}(u-t)\|
  G\|_{1}T^2\bE\tilde\zeta
\end{equation}
for some constants $C_{A}^{(20)},\ga_A^{(4)}>0$ and (cf. \eqref{011210})
$$
C_j(\bbl):=E_j(\hat\bbl,|\bbl|)+
\frac{\partial_{y_j}R_1\left(\bze,|\bbl|\right)}{H_0'(|\bbl|)},
$$
$$
E_j(\hat\bbl,k):=-\frac{H_0'(k)}{k}\,\sum\limits_{i=1}^d
\int\limits_0^{+\infty}
\rho_1  \partial_{y_i,y_i,y_j}^3R\left(\rho_1\,H_0'(k)\hat\bbl,k\right)\,d\rho_1,\quad\,j=1,\ldots,d.
$$

\subsubsection{The terms $\bbE[I^{(2)}\tilde\zeta]$ and $\bbE[I^{(3)}\tilde\zeta]$}
 The calculations concerning these terms essentially follow the
respective steps performed in the previous section
so we only highlight their main points. First, we note
that the difference between $\bbE[I^{(2)}\tilde\zeta]$
and
\begin{equation}
\label{70611}
\frac{1}{\delta}\sum\limits_{i,j=1}^d
\int\limits_t^u\int\limits_\si^s\bbE\left[\partial_jG(\bl^{(\delta)}(\si))
\partial_{\ell_i}F_{j,\delta}\left(s,\frac{\by^{(\delta)}(s)}{\delta},\bl^{(\delta)}(\si)\right)
 F_{i,\delta}\left(\rho,\frac{\by^{(\delta)}(\rho)}{\delta},\bl^{(\delta)}(\si)\right)\tilde\zeta\right]
ds\,d\rho
\end{equation}
 is less than, or equal to
$C_A^{(21)}\delta^{\gamma_A^{(5)}}(u-t)\|G\|_{1}\bbE[\tilde\zeta]$, cf. \eqref{051210}.
Next we note that (\ref{70611}) equals
\begin{eqnarray}
\label{60104}
&&\frac{1}{\delta}
\sum\limits_{i,j=1}^d\int\limits_t^u\int\limits_\si^s \bbE\left[
\partial_{j}G(\bl^{(\delta)}(\si))
\partial_{\ell_i}F_{j,\delta}\left(s,\frac{\bmL^{(\delta)}(\si,s)}{\delta},\bl^{(\delta)}(\sigma)\right)
F_{i,\delta}\left(\rho,\frac{\bmL^{(\delta)}(\si,\rho)}{\delta},\bl^{(\delta)}(\si)\right)\tilde\zeta\right]
ds\,d\rho\nonumber\\
&&+
\frac{1}{\delta^2}\sum\limits_{i,j,k=1}^d\,
\int\limits_t^u\int\limits_\si^s\int\limits_0^1 \bbE\left[
\partial_{j}G(\bl^{(\delta)}(\si))
\partial_{\ell_i}\partial_{y_k}F_{j,\delta}\left(s,
\frac{\bmR^{(\delta)}(v,\si,s)}{\delta},\bl^{(\delta)}(\si)\right)\right.\\
&&\times\left.
F_{i,\delta}\left(\rho,\frac{\bmL^{(\delta)}(\si,\rho)}{\delta},\bl^{(\delta)}(\si)\right)
(y_{k}^{(\delta)}(s)-L_{k}^{(\delta)}(\si,s))\tilde\zeta\right] ds\,d\rho\,dv
\nonumber\\
&&
+
\frac{1}{\delta^2}\sum\limits_{i,j,k=1}^d\,
\int\limits_t^u\int\limits_\si^s\int\limits_0^1 \bbE\left[
\partial_{j}G(\bl^{(\delta)}(\si))
\partial_{\ell_i}F_{j,\delta}\left(s,\frac{\by^{(\delta)}(s)}{\delta},\bl^{(\delta)}(\si)\right)\right.\nonumber\\
&&\times\left.
\partial_{y_k}F_{i,\delta}\left(\rho,\frac{\bmR^{(\delta)}(v,\si,\rho)}{\delta},\bl^{(\delta)}(\si)\right)
(y_{k}^{(\delta)}(\rho)-L_{k}^{(\delta)}(\si,\rho))\tilde\zeta\right]
ds\,d\rho\,dv.\nonumber
\end{eqnarray}
A straightforward argument using Lemma \ref{lm3} and (\ref{80704})
shows that both the second and third terms of (\ref{60104}) can be
estimated by $C_A^{(23)}\delta^{1/2-(3\ga_A+2\ep_2+2\ep_3)}(u-t)\|
G\|_{1}T^2\bbE[\tilde\zeta]$. The first term, on the other hand, can
be handled with the help of part ii) of Lemma \ref{mix1} in the
same fashion as we have dealt with the term $J^{(2)}_{1,2,1}$,
given by (\ref{70612}) of Section \ref{seca41}, and we obtain that
\begin{equation}
\label{80306}
\left|\bbE\left\{\left[I^{(2)}-\sum\limits_{j=1}^d
\int\limits_{t}^u
\left(D_j(|\bl^{(\delta)}(\si)|)\overline{\Theta}^2(s)+
J_j(s;\by^{(\delta)}(\cdot),\bl^{(\delta)}(\cdot))\overline{\Theta}(s)\right)\partial_{j}G(\bl^{(\delta)}(\si))\,ds
\right]\tilde\zeta\right\}\right|
\end{equation}
$$
\leq C_A^{(24)}\delta^{\gamma_A^{(6)}}(u-t)\|
G\|_{1}T\bbE[\tilde\zeta].
$$
Here
\begin{equation}\label{061210}
D_j(l):=\frac{\partial_{y_j}R_2(\bze,l)}{H_0'(l)},
\qquad R_2(\bby,l):=\mathbb E[\partial_lH_1(\bby,l)H_1(\bze,l)],
\end{equation}
$$
J_j(s;\by^{(\delta)}(\cdot),\bl^{(\delta)}(\cdot)):=-
\sum\limits_{i=1}^d \overline{\Theta}_i(s)D_{i,j}(\hat\bl^{(\delta)}(\si),|\bl^{(\delta)}(\si)|),
$$
$$
\overline{\Theta}_i(s):=\partial_{l_i}\Theta(s,\by^{(\delta)}(s),\bl^{(\delta)}(s);\by^{(\delta)}(\cdot),
\bl^{(\delta)}(\cdot)).
$$
Finally, concerning the limit of $\bbE[I^{(3)}\tilde\zeta]$, another application of \eqref{80704}
yields
\begin{equation}
\label{60201} \left|\bbE[I^{(3)}\tilde\zeta] -{\cal I}\right|\leq C_A^{(25)}\delta^{\gamma_A^{(7)}}(u-t)\|
G\|_{1}\bbE[\tilde\zeta],
 \end{equation}
 where
\begin{eqnarray*}
&& {\cal I}:=\frac{1}{\delta}
\int\limits_t^u\int\limits_\si^s\bbE\left\{
\partial^2_{i,j}G(\bl^{(\delta)}(\si))\vphantom{\int} F_{j,\delta}\left(s,\frac{\bmL^{(\delta)}(\si,s)}{\delta},
\bl^{(\delta)}(\si)\right)
F_{i,\delta}\left(\rho,\frac{\bmL^{(\delta)}(\si,\rho)}{\delta},\bl^{(\delta)}(\si)\right)\tilde\zeta\right\}
ds\,d\rho.
\end{eqnarray*}
Then, we can use part ii) of Lemma \ref{mix1} in order to obtain
\begin{equation}\label{021210}
\left|{\cal I}-\sum\limits_{i,j=1}^d
\int\limits_{t}^u
D_{i,j}(\hat\bl^{(\delta)}(\si),|\bl^{(\delta)}(\si)|)
\overline{\Theta}^2(s)\partial^2_{i,j}G(\bl^{(\delta)}(\si))\,ds\right|\leq C_A^{(26)}
\delta^{\gamma_A^{(8)}}(u-t)\|
G\|_{2}T\bbE[\tilde\zeta].
\end{equation}
Next we replace the argument $\si$,  in formulas \eqref{80305}, \eqref{80306} and
\eqref{60201}, by $s$. This can be done thanks to estimate \eqref{051210} 
and the assumption on the regularity of the random field
$H_1(\cdot,\cdot)$. In order to make this approximation work we will
be forced to use the third derivative of $G(\cdot)$.

Finally (cf. \eqref{011210}, \eqref{061210}) note that
$$
\nabla_{\bby}R_1(\bze,l)+\nabla_{\bby}R_2(\bze,l)=\nabla_{\bby\,\left|\vphantom{\int_0^1}\right.\bby=\bze}
\mathbb E[\partial_l H_1(\bby,l)H_1(\bby,l)]=\bze.
$$
Hence  we conclude that the assertion of the lemma holds for any function $G\in C^3(\R^d_*)$
satisfying $\|G\|_3<+\infty$.
Generalization to an arbitrary $G\in C^{1,1,3}_b([0,+\infty)\times\R^{2d}_*)$
is fairly standard. Let $r_0$ be any integer and consider $s_k:=t+kr_0^{-1}(u-t)$, $k=0,\ldots,r_0$.
Then
\begin{equation}\label{1070804}
\bbE\left\{[G(u,\by^{(\delta)}(u),\bl^{(\delta)}(u))-
G(t,\by^{(\delta)}(t),\bl^{(\delta)}(t))]\tilde\zeta\right\}
\end{equation}
$$
=
\sum\limits_{k=0}^{r_0-1}\bbE\left\{[G(s_{k+1},\by^{(\delta)}(s_{k+1}),\bl^{(\delta)}(s_{k+1}))-
G(s_{k},\by^{(\delta)}(s_{k}),\bl^{(\delta)}(s_{k}))]\tilde\zeta\right\}.
$$
$$
=\sum\limits_{k=0}^{r_0-1}\bbE\left\{[G(s_{k},\by^{(\delta)}(s_{k}),\bl^{(\delta)}(s_{k+1}))-
G(s_{k},\by^{(\delta)}(s_{k}),\bl^{(\delta)}(s_{k}))]\tilde\zeta\right\}
$$
$$+
\sum\limits_{k=0}^{r_0-1}\bbE\left\{[G(s_{k+1},\by^{(\delta)}(s_{k+1}),\bl^{(\delta)}(s_{k}))-
G(s_{k},\by^{(\delta)}(s_{k}),\bl^{(\delta)}(s_{k}))]\tilde\zeta\right\}
$$
Using the already proven part of the lemma we obtain
\begin{equation}\label{2070804}
\left|\sum\limits_{k=0}^{r_0-1}\bbE\left\{[\widehat{ N}_{s_{k+1}}(G(s_{k},\by^{(\delta)}(s_{k}),\cdot\,))-
\widehat{ N}_{s_k}(G(s_{k},\by^{(\delta)}(s_{k}),\cdot\,))]\tilde\zeta\right\}\right|
\end{equation}
$$
\le C_A^{(27)}\delta^{\gamma_A^{(9)}}
(u-t)\|G\|_{1,1,3}T^2\bbE\tilde\zeta.
$$
On the other hand, the second term on the right hand side of \eqref{1070804}
equals
\begin{equation}\label{3070804}
\sum\limits_{k=0}^{r_0-1}\bbE\left\{\int\limits_{s_k}^{s_{k+1}}\left\{
\partial_\rho+\left[H_0'(|\bl^{(\delta)}(\rho)|)+
\sqrt{\delta}\partial_l H_1\left(\frac{\by^{(\delta)}(\rho)}{\delta},|\bl^{(\delta)}(\rho)|\right)\right]\,
\hat\bl^{(\delta)}(\rho)\cdot\nabla_\bby\right\} \right.
\end{equation}
$$
\vphantom{\int\limits_0^1}\times\left.
G(\rho,\by^{(\delta)}(\rho),\bl^{(\delta)}(s_{k}))
\tilde\zeta\,d\rho\right\}
$$
The conclusion of the lemma for an arbitrary function $G\in C^{1,1,3}_b([0,+\infty)\times\R^{2d}_*)$
is an easy consequence of \eqref{1070804}--\eqref{3070804} upon passing to the limit with $r_0\to+\infty$.
\endproof

\end{appendix}


\begin{thebibliography}{99}

\bibitem{bakoryz}
G. Bal, T. Komorowski and L. Ryzhik, {Self-averaging of Wigner
Transforms in Random Media}, Comm. Math. Phys., {\bf 242}, 2003, 81--135.

\bibitem{DGL}
D. D{\"u}rr, S. Goldstein and J. Lebowitz, Asymptotic motion of a
classical particle in a random potential in two dimensions: Landau
model,  Comm. Math. Phys., {\bf 113}, 1987, 209--230.

\bibitem{Erdos-Yau}
{L.~Erd{\"o}s and H.~T. Yau}, {Linear Boltzmann equation as the weak
  coupling limit of a random Schr\"odinger Equation}, Comm. Pure Appl. Math.,
  {\bf 53}, 2000, 667--735.

\bibitem{ESY}
L. Erd{\"o}s, M. Salmhofer and H.T. Yau, in preparation, 2004.


\bibitem{GMMP}
{ P.~G\'erard, P.~A. Markowich, N.~J. Mauser, and F.~Poupaud}, {\em
{Homogenization limits and Wigner transforms}}, Comm. Pure
Appl. Math., {\bf 50}, 1997, 323--380.


\bibitem{giksk} I.I. Gikhman and  A.V. Skorochod, {\em Theory of stochastic processes, vol. 3},
Springer Verlag, 1974, Berlin.

\bibitem{gilbarg} D. Gilbarg and N.S. Trudinger, {\em Elliptic partial differential equations
of second order},
Springer Verlag, 1998, Berlin.


\bibitem{kp1}
H. Kesten and G. Papanicolaou. {A Limit Theorem For Turbulent Diffusion},
Comm. Math. Phys., {\bf  65}, 1979, 97--128.

\bibitem{KP}
H. Kesten, G. C. Papanicolaou,
A Limit Theorem for Stochastic Acceleration,  Comm. Math. Phys.,
{\bf 78}, 1980, 19--63.


\bibitem{kustr} S. Kusuoka and D. Stroock,
{Appliactions of the Malliavin calculus, Part II},
J. Fac. Sci. Univ. Tokyo, Sect. IA, Math, \textbf{32}, 1985, 1--76.

\bibitem{LP}
{P.-L. Lions and T.~Paul}, {\em {Sur les mesures de Wigner}},
Rev. Mat.  Iberoamericana, {\bf 9}, 1993, 553--618.


\bibitem{RPK-WM}
L. Ryzhik, G. Papanicolaou, J. Keller, Transport equations for elastic
and other waves in random media, Wave Motion, {\bf 24}, 1996, 327-370.


\bibitem{stroock} D. Strook,  {\em
An Introduction to the Analysis of Paths on a Riemannian Manifold},
Math. Surv. and Monographs v. \textbf{74}, 2000.

\bibitem{stroock-varadhan} D. Strook, S. R. S. Varadhan {\em
Multidimensional Diffusion Processes}, Berlin, Heidelberg, New York:
Springer-Verlag, 1979.

\bibitem{Taylor} M. Taylor,
{\em Partial differential equations, vol. 2}, New York,
Springer-Verlag, 1996.
\end{thebibliography}
\end{document}